\documentclass[aps,reprint,10pt,showkeys,prx,superscriptaddress,twocolumn,longbibliography,nofootinbib]{revtex4-2}

\usepackage[english]{babel}
\usepackage[utf8]{inputenc}
\usepackage{amsmath,dsfont,mathtools}
\usepackage{tikz}
\usepackage{multirow}
\usepackage{amssymb}
\usepackage{graphicx}
\usepackage{pgf}
\usepackage{algorithm}
\usepackage{dcolumn}
\usepackage{bm}
\usepackage{makecell}
\usepackage[normalem]{ulem}
\usepackage{color}
\graphicspath{{figures/}}
\usepackage{adjustbox}
\usepackage{physics}
\usepackage{amsfonts}
\usepackage{soul} 
\usepackage{overpic}
\usepackage{lmodern}
\usepackage{fix-cm}
\usepackage{nowidow}
\usepackage{amsthm}
\usepackage{comment}
\usepackage[noend]{algpseudocode}
\usepackage{thmtools}
\usepackage{thm-restate}
\usepackage{xcolor}
\usepackage{mathdots}
\usepackage{yhmath}
\usepackage{ragged2e}
\usepackage{silence}
\usepackage[scr=rsfso]{mathalfa}
\usepackage{stmaryrd} 
\usepackage{booktabs}

\usetikzlibrary{quantikz2}
\usetikzlibrary{fadings}
\usetikzlibrary{patterns}
\usetikzlibrary{shadows.blur}
\usetikzlibrary{shapes}
\theoremstyle{definition}
\newtheorem{definition}{Definition}[section]
\theoremstyle{plain}
\newtheorem{lemma}{Lemma}[section]
\newtheorem{proposition}{Proposition}[section]

\newtheorem{theorem}{Theorem}[section]

\newcommand{\poly}{\ensuremath{\operatorname{poly}}}

\newcommand{\Q}{\ensuremath{\mathcal{Q}}}

\newcommand{\g}{{\ensuremath{\mathfrak{g}}}}
\newcommand{\G}{\mathcal{G}}
\newcommand{\risk}{\mathcal{R}}

\usepackage[table]{xcolor}

\newcommand{\R}{\mathbb{R}}

\newcommand{\btheta}{{\boldsymbol{\theta}}}

\newcommand{\bx}{\ensuremath{\boldsymbol{x}}}
\newcommand{\br}{\ensuremath{\boldsymbol{r}}}

\newcommand{\bzero}{\ensuremath{\boldsymbol{0}}}
\newcommand{\bi}{\boldsymbol{i}}

\newcommand{\bj}{\boldsymbol{j}}

\newcommand{\przero}{\ket{0}\!\!\bra{0}}
\newcommand{\prone}{\ket{1}\!\!\bra{1}}
\newcommand{\bprzero}{\ket{\bzero}\!\!\bra{\bzero}}

\newcommand{\E}{\mathbb{E}}

\newcommand{\pr}{\operatorname{Pr}_\btheta}
\newcommand{\opr}{\operatorname{Pr}}
\newcommand{\noi}{\bar{\imath}}
\newcommand{\noj}{\bar{\jmath}}
\newcommand{\cl}{\mathcal{C}\!\ell}
\newcommand{\ccl}{\boldsymbol{c}_{\mathcal{C}\!\ell}}
\newcommand{\cq}{\boldsymbol{c}_{\mathcal{Q}}}
\newcommand{\Var}{\operatorname{\mathds{V}\!ar}}
\newcommand{\1}{\mathds{1}}
\newcommand{\F}{\mathbb{F}}

\definecolor{customblue}{HTML}{17178B}

\newcommand{\customauthors}[1]{
    \bgroup\def\and{, }\def\And{, }\def\AND{, }
    \author{#1}
    \egroup
}

\usepackage[colorlinks=true,citecolor=blue,linkcolor=blue]{hyperref}
\newcommand{\figref}[2][]{\hyperref[#2]{\ref*{#2}#1}}
\usepackage{cleveref}

\begin{document}
\preprint{APS/123-QED}
\makeatletter
\let\orig@addcontentsline\addcontentsline
\def\addcontentsline#1#2#3{}
\makeatother
 
\title{The Born Ultimatum: Conditions for Classical Surrogation of\texorpdfstring{\\}{ } Quantum Generative Models with Correlators.}

\author{Mario Herrero-Gonz\'alez}
\email{mario.herrero@ed.ac.uk}
\affiliation{School of Informatics, University of Edinburgh, United Kingdom}
\affiliation{Laboratoire d’Informatique de Paris 6, CNRS, Sorbonne Université, France}

\author{Brian Coyle}
\affiliation{School of Informatics, University of Edinburgh, United Kingdom}
\affiliation{Fujitsu Research of Europe Ltd., Slough SL1 2BE, United Kingdom}

\author{Kieran McDowall}
\affiliation{National Quantum Computing Centre, United Kingdom}

\author{Ross Grassie}
\affiliation{School of Informatics, University of Edinburgh, United Kingdom}

\author{Sjoerd Beentjes}
\altaffiliation{Senior authors, equal contribution}
\affiliation{School of Mathematics and Maxwell Institute for Mathematical Sciences, University of Edinburgh, United Kingdom}

\author{Ava Khamseh}
\altaffiliation{Senior authors, equal contribution}
\affiliation{School of Informatics, University of Edinburgh, United Kingdom}
\affiliation{MRC Human Genetics Unit, Institute of Genetics \& Cancer, University of Edinburgh, United Kingdom}

\author{Elham Kashefi}
\altaffiliation{Senior authors, equal contribution}
\affiliation{School of Informatics, University of Edinburgh, United Kingdom}
\affiliation{Laboratoire d’Informatique de Paris 6, CNRS, Sorbonne Université, France}

\hyphenpenalty=10000
\exhyphenpenalty=10000
\begin{abstract}
Quantum Circuit Born Machines (QCBMs) are powerful quantum generative models that sample according to the Born rule, with complexity-theoretic evidence suggesting potential quantum advantages for generative tasks. Here, we identify QCBMs as a quantum Fourier model independently of the loss function. This allows us to apply known dequantization conditions when the optimal quantum distribution is available. However, realizing this distribution is hindered by trainability issues such as vanishing gradients on quantum hardware. Recent train-classical, deploy-quantum approaches propose training classical surrogates of QCBMs and using quantum devices only for inference. We analyze the limitations of these methods arising from deployment discrepancies between classically trained and quantumly deployed parameters. Using the Fourier decomposition of the Born rule in terms of correlators, we quantify this discrepancy analytically. Approximating the decomposition via distribution truncation and classical surrogation provides concrete examples of such discrepancies, which we demonstrate numerically. We study this effect using tensor-networks and Pauli-propagation-based classical surrogates. Our study examines the use of IQP circuits,  matchcircuits, Heisenberg-chain circuits, and Haldane-chain circuits for the QCBM ansatz. In doing so, we derive closed-form expressions for Pauli propagation in IQP circuits and the dynamical Lie algebra of the Haldane chain, which may be of independent interest.
\end{abstract}

\maketitle
\section{Introduction}\label{sec:intro}
Quantum machine learning has received significant attention in recent years due to its relevance to various applications and theoretical results regarding its ability to deliver practical applications~\cite{cerezo_challenges_2022, gil-fuster_relation_2025}. Although a growing body of literature is shrinking the space where we may find quantum advantage in quantumly minimized solutions~\cite{bermejo_quantum_2024, thanasilp_exponential_2024}, one class of problems that seems promising at the inference stage is generative modeling. These tasks require learning from target data, which normally under-represents a complex true underlying distribution. By aligning the model with the target, the model can produce new data points and generalize to the underlying case.

One popular quantum generative model is the quantum circuit Born machine (QCBM). In this setting, one trains a parameterized quantum circuit (PQC) so that its output probability distribution matches the true distribution better than its classical counterpart. Previous work shows that this specific quantum model with  can be competitive in a data-limited regime~\cite{hibat-allah_framework_2024} when compared to classical generative models. QCBMs have also been directly compared with restricted Boltzmann machines (RBM) and have shown improved accuracy in the production of synthetic data~\cite{coyle_quantum_2021} when measured by the same standard with an equal number of parameters. However, it is unknown what are the minimal quantum or classical resources for training such models in practice. In this work, we address this question.

\begin{figure*}[t!]
    \centering
    \begin{adjustbox}{width=\textwidth, trim=15pt 30pt 1pt 0pt, clip}
        \input{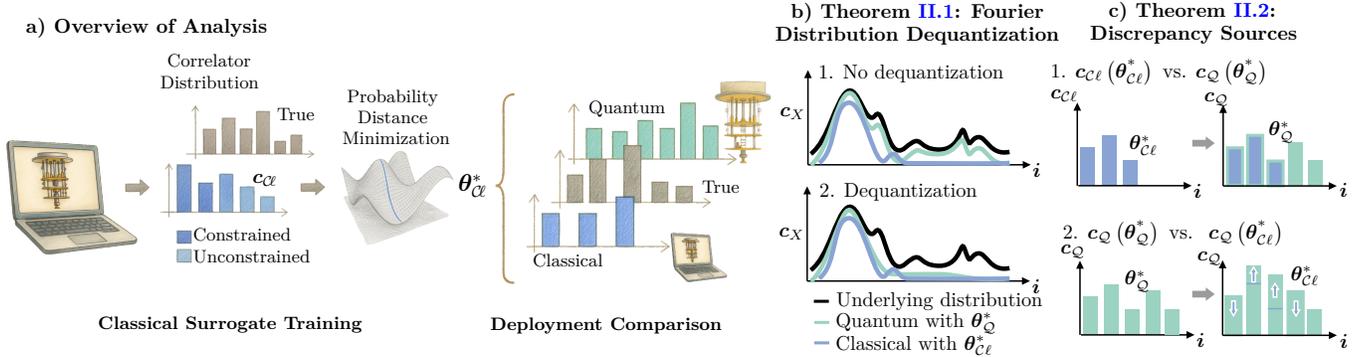}
    \end{adjustbox}
    \caption{ \label{fig:main_overview_dequant:left}\justifying
        \textbf{a)} We adopt a \textbf{train-on-classical, deploy-on-quantum}~\cite{recio-armengol_train_2025} approach for quantum circuit Born machines. Training is performed using the Fourier decomposition in terms of correlators of the QCBM’s probability distribution. $\boldsymbol{c}_X(\boldsymbol{\theta}^*_Y)$ are the set of correlators produced by model $X \in \{\cl, \mathcal{Q}\}$ using optimal parameters ($\boldsymbol{\theta}^*_Y$) generated by training model, $Y \in \{\cl, \mathcal{Q}\}$.
        Classical surrogates, $\ccl$, are used to approximate a `\emph{constrained}' subset of correlators within a truncated version of the Fourier decomposition to train the model. Once the optimal classical parameters $\boldsymbol \theta^*_{\cl}$ are obtained, they are deployed on the quantum model, $\boldsymbol c_{\mathcal{Q}}(\btheta^*_{\cl})$. A quantum advantage arises when the distribution induced by the quantum deployment approximates the underlying target distribution better than the classical counterpart and we provide conditions under such an advantage holds. \textbf{b) Deployment-Dequantization conditions:} When quantum training is possible to find $\btheta^*_{\mathcal{Q}}$ \emph{and} if the learned quantum distribution $\boldsymbol c_{\mathcal{Q}}(\btheta^*_{\mathcal{Q}})$: \textbf{b.1)} has significant weight across the full frequency support, this prevents alignment with any classical counterpart, $ c_{\cl}$; or \textbf{b.2)} is less expressive and concentrates only on a restricted support, the classical distribution can align with it. \textbf{c)} Sources of discrepancy in deploying the ideal quantum and classical parameters, $\btheta^*_{\mathcal{Q}}, \btheta^*_{\cl}$ respectively: \textbf{c.1)} the quantum model, $\boldsymbol c_{\mathcal{Q}}(\btheta^*_{\mathcal{Q}})$, can represent a broader range of frequencies, even if not all are necessarily useful (cf.~\textbf{b.2}); \textbf{c.2)} using optimal classical - $\btheta^*_{\cl}$ - rather than quantum parameters - $\btheta^*_{\mathcal{Q}}$ - decreases the overall optimality, with the extent of this reduction determined by the inductive bias of the ansatz. We elabotate on this and give precise statements in Sec.~\ref{sec:comparison}.
        }
    \label{fig:main_fig_deployment}
\end{figure*}

Our main objective is then to isolate the contribution of correlations of different order to the probability distribution of the QCBM. This framework allows us to naturally obtain a relation between the correlations that are present in the data (\emph{data complexity}) and the \emph{level of statistical generative capacity} which is the maximum number of correlators the learning model can control up to a certain accuracy. The second objective is to determine practically whether the required level of statistical generative capacity can be achieved with a classical surrogate of the quantum model rather than with the quantum model itself. Consequently, this allows us to benchmark where the classical limit is.
Our two objectives and related key concepts are depicted in Fig.~\ref{fig:main_fig_deployment}. Implicitly, we concentrate on deriving insights into the trainability, expressivity, and simulability of the QCBM through the analysis of correlations.

Regarding classical methods, some of the authors previously employed Restricted Boltzmann Machines (RBMs) to quantify higher-order variable dependencies~\cite{cossu_machine_2019,beentjes_higher-order_2020}, providing analytical formulations and statistical criteria to enhance accuracy~\cite{cossu_machine_2019}. Despite success with simulated Ising models, RBMs faced severe computational limitations for large or feature-rich datasets. To mitigate this, \emph{Stator}~\cite{jansma_high_2025} is a model that quantifies higher-order interactions directly, leveraging structure learning algorithms to boost statistical power, albeit at significant computational cost~\cite{beentjes_higher-order_2020,kuipers_efficient_2022}. Motivated by these classical challenges, we explore QCBMs as alternative models for effectively sampling new datapoints with specific constrained correlations.

\begin{table}[]
\begin{tabular}{cc}
\hline
\multicolumn{1}{l}{}                              & \textbf{Analytical Result} \\ \hline
\multicolumn{1}{l}{\textbf{Quantum Circuits}}     & \multicolumn{1}{l}{}       \\ \hline
\multirow{2}{*}{Haldane Ansatz}                   & Closed for for algebra     \\
                                                  & (Eq.~\ref{eqn:Haldane_dla})                   \\
\multirow{2}{*}{Matchcircuit Ansatz}              & Probability variance       \\
                                                  & (Eq.~\ref{eq:var_prob}, Fig.~\ref{fig:variancealgebra})           \\ \hline
\multicolumn{1}{l}{\textbf{Classical Surrogates}} & \multicolumn{1}{l}{}       \\ \hline
\multirow{2}{*}{Tensor Network Surrogate}         & Correlator variance        \\
                                                  & (Eq.~\ref{eq:tn_var}, Fig.~\ref{fig:variancealgebra})           \\
\multirow{2}{*}{Pauli Propagation Surrogate}      & Closed form for            \\
                                                  & IQP circuit (Eq.~\ref{eqn:iqp_pps_main_result})      \\ \hline
\end{tabular}
\caption{\justifying \textbf{Complementary analytical results to the main theorems}. The results on the ansatz\"{e} serve as a step forward to understand and identify problems when choosing circuit architectures. The TN surrogate variance allows us to show that we can still hit vanishing approximated correlators when the complexity of them is too large. In the PPS in the IQP case we manage to find a closed efficient form to faster compute the surrogated correlators than in the vanilla version of PPS.}
\end{table}

The outline of the paper is as follows. In the remaining of this section we formally present the QCBM, its approximations, the datasets and different ansatz\"{e}. In Section~\ref{sec:comparison} give the main theorems regarding the dequantization conditions and the sources of discrepancy. In Section~\ref{sec:concentration} we argue that in general it is not possible to always obtain the optimal quantum distribution from uniquely quantum training and show how as we increase the complexity of the surrogates we can also have vanishing correlators. In Section~\ref{sec:PPS} we give a closed form for the PPS of the IQP circuit.
In Section~\ref{sec:numerics} we study the truncation, generalization, scaling and surrogate training in the practical setting.

Lastly, in Section~\ref{sec:conclusion}, we briefly summarize the main findings and in Section~\ref{sec:discussion} we cover various ideas for future work which can build from this approach.
\subsection{The Quantum Circuit Born Machine}\label{sec:qcbm}

\begin{figure*}[htbp]
   \resizebox{\linewidth}{!}{\input{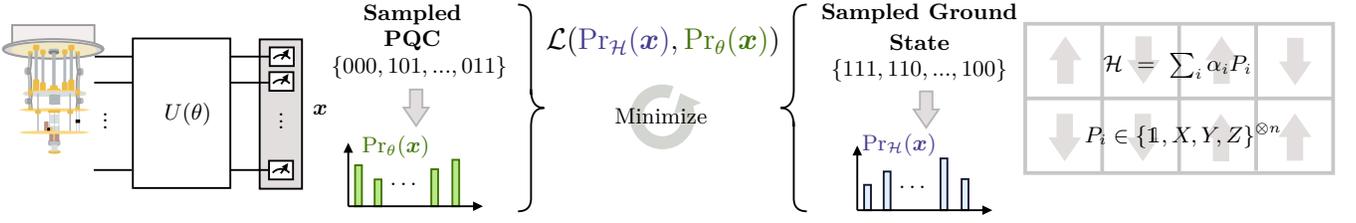}}
    \caption{\justifying\textbf{The Quantum Circuit Born Machine (QCBM).} In its original version the PQC circuit is sampled to obtain the discrete binary bitsring $\bx$. After sampling sufficiently many times one can build an estimate of $\operatorname{Pr}_{\btheta}$. The same is done for the Hamiltonain $\mathcal{H}$ and $\operatorname{Pr}_{\mathcal{H}}$ or a different data source. Both probabilities are compared via the gradient of the loss function of choice $\mathcal{L}$ which informs parameter updates. This procedure is repeated until $\mathcal{L}$ reaches a predefined convergence criterion, such as a small gradient norm, a plateau in loss values, or a target accuracy threshold. 
    }    \label{fig:qcbm_scheme}
\end{figure*}

We now specify the QCBM that is tasked with reproducing the probability distributions $\operatorname{Pr}_{\mathcal{H}}(\boldsymbol{x})$ of the systems in Sec.\ref{sec:data}. The QCBM is a quantum generative model that parameterizes a probability distribution over bitstrings using a quantum circuit. Given an $n$-qubit quantum system, the circuit is assumed to initialize all qubits in the computational basis state $\ket{\bzero} \equiv |0\rangle^{\otimes n} $.
A parameterized unitary transformation $ U(\boldsymbol{\theta}) $ is applied, leading to the preparation of the quantum state
\begin{equation}
    |\psi(\boldsymbol{\theta})\rangle = U(\boldsymbol{\theta}) |\bzero \rangle = \sum_{\bx} \alpha_{\bx}(\btheta)\ket{\bx}
\end{equation}
where $ \boldsymbol{\theta} $ denotes the set of tunable parameters. In the above, we have an orthonormal basis $\{\ket{\bx}\}$ with $\bx\in\mathbb{Z}_2^{n}$ and coefficients $\alpha_{\bx}(\btheta)\in \mathbb{C}$ .The unitary operator$U(\boldsymbol{\theta})$ is commonly composed of layers of parameterized Pauli rotations and entangling gates. The complexity of the target distribution influences the number of qubit connections required and, consequently, the depth of the circuit needed to approximate the distribution effectively in a non-trivial manner.

Once the quantum state is prepared, measurements are performed in the canonical $ Z $-basis, producing the classical bitstring $ \boldsymbol{x} \in \mathbb{Z}_2^n $ with probability given by the Born rule,
\begin{equation}\label{eq:born_rule}
\operatorname{Pr}_{\boldsymbol{\theta}}(\boldsymbol{x}) =\Tr(\ketbra{\bx}{\bx} \rho_{\boldsymbol{\theta}}) = |\langle \boldsymbol{x} | \psi(\boldsymbol{\theta}) \rangle|^2.
\end{equation}

With $\rho_{\boldsymbol{\theta}} := \ketbra{\psi(\boldsymbol{\theta})}{\psi(\boldsymbol{\theta})}$ if the QCBM ansatz is pure. These measurement outcomes define the modeled probability distribution $ \Pr_{\boldsymbol{\theta}}(\boldsymbol{x}) $, which can be compared to the target distribution through a loss function $\mathcal{L}$. The QCBM framework provides a way to represent distributions in terms of quantum states, with the quantum circuit serving as the generative mechanism. Fig.~\ref{fig:qcbm_scheme} provides a schematic of the  QCBM framework.

Our first contribution is the following observation. Specifically, we can write the probability given by the Born rule in terms of expectation values - \emph{correlators}, which in general are defined as follows, given a distribution $\operatorname{Pr}$:
\begin{equation}\label{eq:exact_corr}
    \langle S_{\boldsymbol{i}} \rangle \;=\; \underset{\operatorname{Pr}}{\mathbb{E}} \!\left[\prod_{i\in\boldsymbol{i}} \bx_i\right]
\end{equation}
where $x_i^{(j)} \in \mathbb{Z}_2$ denotes the observed value of variable $i$ in sample $j$. 
To distinguish between terminology, when the random variables, $\bx$, are generated from \emph{data}, i.e. under the distribution $\operatorname{Pr}_{\textnormal{Data}}$, we refer to $\langle S_{\boldsymbol{i}} \rangle$ as as correla\emph{tions} or moments, while when they are generated from the model (parameterized by $\boldsymbol{\theta}$), $\operatorname{Pr}_{\boldsymbol{\theta}}$, we refer to $\langle S_{\boldsymbol{i}} \rangle$ as correlators.

Lemma~\ref{lemma:decomposition_full} forms the central object of this work, from which the subsequent results are derived.
\begin{lemma}\label{lemma:decomposition_full}
The probability of any bitstring $\bx = (x_1,x_2,...,x_n)\in\mathbb{Z}_2^n$ can be expressed in terms of $Z$-basis correlators, namely
    \begin{equation} \label{eq:decomposition_pr}
        \pr(\bx) = \frac{1}{2^n} \sum_{\bi\in 2^{[n]}} (-1)^{\sum_{i\in {\bi}} x_i}\left\langle Z_{\bi}\right\rangle_{\btheta},
    \end{equation}
    where
    \begin{equation}
        Z_{\bi} =  \left(\bigotimes_{i\in {\bi}} Z_i\right)\! \otimes\! \1_{\boldsymbol{\noi}}
        \,\text{ with }\, \boldsymbol{\noi} = \{ i \notin {\bi}\},
    \end{equation}
\end{lemma}
where now the different $\langle Z_{\bi}\rangle_{\btheta}$ are the correlators of the QCBM. This result is the Fourier transform of the probability distribution for a given bitstring. Equivalently, we can rewrite the correlators in terms of bitstring probabilities.
\begin{lemma}\label{lem:exp_from_probs}
     A $Z$-basis $k$-order correlator acting on a subset of ${\bi}\subseteq \{1,2,...,n\}$ qubits can be written in terms of the bitstring probabilities,  
    \begin{equation}\label{eq:exp_from_probs}
        \left\langle Z_{\bi} \right\rangle_{\btheta} = \sum_{\bx} (-1)^{\sum_{i \in{\bi}}x_i}\operatorname{Pr}_\btheta(\bx).
    \end{equation}
\end{lemma}
\noindent
See Appendix~\ref{app:general} for the proofs of the above Lemmas. The correlators can be estimated statistically via Monte Carlo averaging over \(m\) samples:
\begin{equation}\label{eq:empirical_corr}
    \langle S_{\boldsymbol{i}} \rangle \;=\; \mathbb{E} \!\left[\prod_{i\in\boldsymbol{i}} x_i\right]
    \;\approx\; \frac{1}{m}\,\sum_{j=1}^{m}\,\prod_{i\in\boldsymbol{i}} x_i^{(j)},
\end{equation}
with variance (assuming a pairwise correlation between samples, \(\gamma, \gamma = 0\) implies independence of samples):
\begin{equation}
    \Var(\langle S_{\boldsymbol{i}} \rangle) 
    \;\approx\;\frac{1}{m}\,\langle S^*_{\boldsymbol{i}}\rangle\,(1-\langle S^*_{\boldsymbol{i}}\rangle)\,\bigl[1+(m-1)\gamma\bigr].
\end{equation}
assuming each \(x_i^{(j)}\) is a Bernoulli random variable with success probability, \(\langle S^*_{\boldsymbol{i}}\rangle\). In essence, correlators/correlations give the probability of all of the bits within the sub-bitstring to be equal to one. Therefore, when calculating the estimator, a given sampled bitstring no longer contributes to the sample average from the moment one of the bits is equal to zero.

While quantum effects may enhance expressivity of the overall distribution, the practical evaluation of correlators poses significant computational challenges. Evaluating all correlations up to order $k$ involves summing over $\sum_{p=1}^k \binom{n}{p}$ subsets, leading to a complexity
\begin{equation}
    \mathcal{O}\!\left(m\,\sum_{p=1}^k \binom{n}{p}\right),
\end{equation}
which remains polynomial in $n$ for low enough values of $k$. For example, evaluating all correlators from the QCBM would require computing all single qubit Pauli-Z correlators, $\{\langle Z_i\rangle | i \in \{1, \dots, n\}\}$, then all pairs of two qubit terms, $\{\langle Z_iZ_j\rangle | i, j \in \{1, \dots, n\}, i\neq j\}$, and so on. Therefore, considering all combinations up to
$n$ features grows exponentially ($2^n-1$) and thus the estimation or Eq.~\ref{eq:decomposition_pr} scales as \(\mathcal{O}(m\,2^n)\). In high-dimensional settings, practical classical strategies often involve reducing the number of correlations considered, such as by imposing sparsity through probabilistic sparse principal components analysis frameworks~\cite{zou_sparse_2006}, or constraining interactions via low-rank latent representations as in variational autoencoders~\cite{kingma_auto-encoding_2022}. This is precisely the approach we take in this work.

\subsection{Approximate Born Rule via Distribution Truncation}
As mentioned above, it is clear that efficiently training over all $2^n$ correlators is intractable as the system size grows. As such, one may \emph{restrict} or \emph{constrain} the possible correlators to some subset, and as a result, we form an approximation, \(\widetilde{\pr}(\bx)\) to the true probability distribution. The more correlators we include, the better the approximation. Among the many possible truncation approximation strategies, we restrict our attention to two: \emph{k-order} and \emph{random-sampling} truncation. We give definitions for these two cases below. 

\begin{definition} \label{defn:k-order_truncation}
Given a probability distribution, $\Pr(\bx)$, according to the the Born rule, $\Pr(\bx)\sim \Tr(\ketbra{\bx}{\bx}\rho) = |\braket{\bx}{\psi}|^2$, the summation in Lemma~\ref{lemma:decomposition_full} can be split into contributions up to order $k$, and we define the \emph{k-order} truncation approximation to $\Pr(\bx)$ as:
\begin{equation}\label{eq:truncated_prob}
    \operatorname{Pr}_{\btheta}^{(k)}({\bx}) = \frac{1}{2^n} \sum^{k}_{p=0}\,\sum_{\substack{{\bi}\subseteq\{1,...,n\}:\\ |{\bi}|=p}}\!\!\!\! (-1)^{\sum_{i\in {\bi}} x_i}\big\langle Z_{\bi}\big\rangle_{\btheta}.
\end{equation}    
\end{definition}

\begin{definition} \label{defn:random-sampling_truncation}
    Given a probability distribution, $\Pr(\bx)$, according to the the Born rule, $\Pr(\bx)\sim \Tr(\ketbra{\bx}{\bx}\rho) = |\braket{\bx}{\psi}|^2$, one can consider a random subset of correlators $\Omega \subseteq 2^{[n]}$, and define the \emph{Random Fourier Correlator} (RFC) truncation approximation to $\Pr(\bx)$ as:

    \begin{equation}\label{eq:random_corr_prob}
        \operatorname{Pr}_{\btheta}^{\Omega}({\bx}) = \frac{1}{2^n} \sum_{{\bi}\in \Omega} (-1)^{\sum_{i\in {\bi}} x_i}\big\langle Z_{\bi}\big\rangle_{\btheta}.
    \end{equation}
\end{definition}

\begin{figure}[htbp]
     \centering
     \begin{adjustbox}{width=\linewidth}
     \input{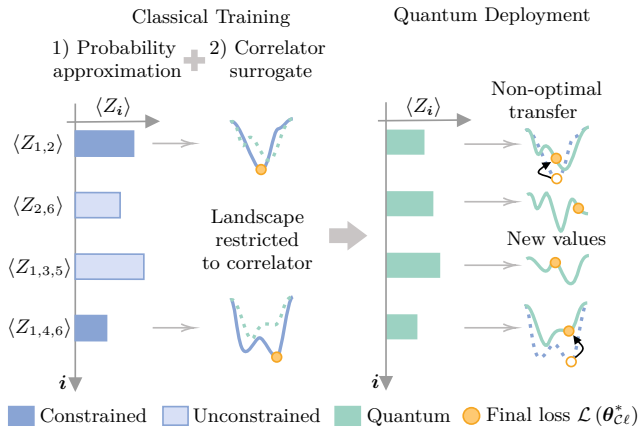}
   \end{adjustbox}
     \caption{\justifying\textbf{Approximations summary.} In the first approximation we select the set of constrained correlators to train. In the second approximation we choose the correlator surrogate. The restricted landscape to the correlator $\mathcal{L}(\langle Z_{\boldsymbol{i}}\rangle)$ changes with respect to the original quantum model one. When deploying, after the classical training, the previous optimum value in the landscape is no longer optimum for the constrained surrogated correlators. Furthermore, new values arise from the unconstrained correlators.
     }
     \label{fig:approximations}
 \end{figure}

Normalization is preserved as long as the empty set is included, since any approximation then builds on the trivial distribution $\Pr^{\varnothing}_\btheta = 2^{-n}$ for all $\bx$. This follows because, when summing over all bitstrings, each correlator contributes equally with positive and negative terms (see Eq.~\ref{eq:matrix_form}). However, the individual approximated values $\Pr^{\Omega}(\bx)$ or $\Pr^{(k)}(\bx)$ need not remain non-negative, as the correlator expectation values lie in $[-1,1]$. These approximations therefore yield pseudo-probabilities, which restricts us to losses that can accommodate negative domains.

Different truncation strategies can be devised. For instance, one may combine the $k$-order and RFC sampling schemes by randomly selecting observables within prescribed orders. Alternatively, Ref.~\cite{huang_generative_2025} proposes partitioning the $n$ qubits into $\alpha$ independent subsets of size $m$ (so that $n=\alpha m)$, which are trained separately. In our framework, this setting translates to having $\alpha\left(2^m-1\right)$ constrained correlators excluding the identity term, and the maximum order of correlators that can be trained is $m$.

\subsection{Approximate Correlator Evaluation via Classical Surrogates} \label{subsec:surrogates}
Classical surrogates were introduced in~\cite{schreiber_classical_2023} as models that can efficiently learn the input-output relation from a quantum system. Here, we surrogate the minimization procedure by first surrogating via the approximation of the full Fourier development of the probability distribution as in Eq.~\ref{eq:truncated_prob} and Eq.~\ref{eq:random_corr_prob}; and second by additionally surrogating the correlators within this development. In Ref.~\cite{recio-armengol_iqpopt_2025,recio-armengol_train_2025} the MMD loss is used to approximate the correlators using the Bernoulli distribution in indexes and an approximation of a thousand random terms in the canonical basis. To expand this analysis, in this work we use Tensor Networks and Pauli Propagation jointly with the $k$-order truncation and random correlator sampling for any type of loss with positive domain. Other methods such as $\mathfrak{g}$-sim~\cite{goh_lie-algebraic_2025}, neural quantum states~\cite{rende_foundation_2025} or neural network encoded angles~\cite{miao_neural-network-encoded_2024} among others exist. Moreover, we treat the surrogates in their vanilla version, although there exist techniques which adapt interactively the surrogate with calls to the quantum computer~\cite{oleary_efficient_2025,luo_quack_2024}.
 
\subsubsection{Tensor Network Correlator Surrogates}  \label{subsec:tensor_networks}
Tensor Networks have been used for sampling problems~\cite{ferris_perfect_2012} as well as for the Born Machine~\cite{rudolph_synergy_2023} where it is numerically proven that they provide useful initialisations. The strength of tensor networks stems from their capacity to capture intricate correlations in a resource-efficient manner, enabling the study of dynamics within vector spaces that would otherwise be computationally prohibitive. This is done by effectively expanding a state into smaller elements and reducing the dimensionality of the smaller sized ones via the bond dimension. As specified in the Results Subsec.~\ref{subsec:tensor_networks} the computational resources needed for the Tensor Network simulation increase with such bond dimension.

We will firstly focus on the analytical side of tensor networks surrogating QCBM instances, specifically on how the $\ccl(\btheta)$ vanish depending on the computational resources. Later, in Section~\ref{sec:train_surr}, we evaluate the overall training in practice.
In this framework, large tensors representing a whole quantum state are described by a network of smaller, interconnected tensors. Each tensor represents local degrees of freedom, and the connections (bonds) encode dependencies between different parts of the system.

The local dimension $\ell$ refers to the number of independent states or levels a single site or subsystem in a larger system can occupy. For a single qubit in a quantum circuit, $\ell=2$. The bond dimension $\chi$ is the size of the internal indices (or virtual indices) that connect tensors in the network, and controls the area-law for the entanglement entropy within the tensor network~\cite{orus_practical_2014}. A larger bond dimension allows the network to represent states with more complex entanglement, making it more expressive but also requires additional resources.

We can describe a quantum state with the tensor network diagram
\begin{equation} \label{eq:mps}
    \ket{\psi} = U\ket{\bzero} = 
        \raisebox{-1.4em}{\resizebox{.22\textwidth}{!}{\tikzset{every picture/.style={line width=0.75pt}} 

\begin{tikzpicture}[x=0.7pt,y=0.6pt,yscale=-1,xscale=1]

\draw    (38.28,42.39) -- (38.28,29.46) ;
\draw  [color={rgb, 255:red, 84; green, 142; blue, 161 }  ,draw opacity=1 ][fill={rgb, 255:red, 186; green, 235; blue, 252 }  ,fill opacity=1 ] (20.81,42.65) -- (56,42.65) -- (56,67.64) -- (20.81,67.64) -- cycle ;
\draw    (38.9,80.91) -- (38.9,67.98) ;
\draw    (56,55.15) -- (72.23,55.15) ;
\draw    (90.56,42.39) -- (90.56,29.46) ;
\draw  [color={rgb, 255:red, 84; green, 142; blue, 161 }  ,draw opacity=1 ][fill={rgb, 255:red, 186; green, 235; blue, 252 }  ,fill opacity=1 ] (73.09,42.65) -- (108.28,42.65) -- (108.28,67.64) -- (73.09,67.64) -- cycle ;
\draw    (91.18,80.91) -- (91.18,67.98) ;
\draw    (108.28,55.15) -- (124.51,55.15) ;
\draw  [dash pattern={on 4.5pt off 4.5pt}]  (124.51,55.15) -- (175.93,55.15) ;
\draw    (175.93,55.15) -- (192.17,55.15) ;
\draw  [color={rgb, 255:red, 84; green, 142; blue, 161 }  ,draw opacity=1 ][fill={rgb, 255:red, 186; green, 235; blue, 252 }  ,fill opacity=1 ] (192.42,42.65) -- (227.6,42.65) -- (227.6,67.64) -- (192.42,67.64) -- cycle ;
\draw    (210.5,80.91) -- (210.5,67.98) ;
\draw    (209.88,42.39) -- (209.88,29.46) ;
\draw  [color={rgb, 255:red, 155; green, 155; blue, 155 }  ,draw opacity=1 ][fill={rgb, 255:red, 204; green, 204; blue, 204 }  ,fill opacity=1 ] (79.25,4.21) -- (102,4.21) -- (102,29.2) -- (79.25,29.2) -- cycle ;
\draw  [color={rgb, 255:red, 155; green, 155; blue, 155 }  ,draw opacity=1 ][fill={rgb, 255:red, 204; green, 204; blue, 204 }  ,fill opacity=1 ] (26.97,4.21) -- (49.72,4.21) -- (49.72,29.2) -- (26.97,29.2) -- cycle ;
\draw  [color={rgb, 255:red, 155; green, 155; blue, 155 }  ,draw opacity=1 ][fill={rgb, 255:red, 204; green, 204; blue, 204 }  ,fill opacity=1 ] (198.57,4.21) -- (221.32,4.21) -- (221.32,29.2) -- (198.57,29.2) -- cycle ;
\draw  [dash pattern={on 4.5pt off 4.5pt}]  (227.84,55.15) -- (243.75,55.15) ;
\draw  [dash pattern={on 4.5pt off 4.5pt}]  (20.32,55.15) -- (4.7,55.15) ;

\draw (29.61,6.34) node [anchor=north west][inner sep=0.75pt]    {$| 0\rangle $};
\draw (81.67,6.53) node [anchor=north west][inner sep=0.75pt]    {$| 0\rangle $};
\draw (200.97,6.53) node [anchor=north west][inner sep=0.75pt]    {$| 0\rangle $};
\draw (29.75,47.66) node [anchor=north west][inner sep=0.75pt]    {$U_{1}$};
\draw (82.03,47.66) node [anchor=north west][inner sep=0.75pt]    {$U_{2}$};
\draw (201.47,47.66) node [anchor=north west][inner sep=0.75pt]    {$U_{n}$};
\draw (115.2,34.08) node [anchor=north west][inner sep=0.75pt]    {$\chi $};
\draw (42.62,70.36) node [anchor=north west][inner sep=0.75pt]    {$\ell$};
\end{tikzpicture}}}.
\end{equation}
This description corresponds to a Matrix Product State (MPS) which can be unitarily embedded~\cite{haferkamp_emergent_2021,haag_typical_2023,perez-garcia_matrix_2007,gross_quantum_2010}. The unitaries $U_1, U_2, \dots\, U_n \in U(\ell\chi)$ are matrices in $\mathbb{C}^{\ell}\otimes \mathbb{C}^\chi$ and the normalization of $\ket{\psi}$ depends on the boundary conditions, here we use periodic boundary conditions. The state vector $\ket{0}\in \mathbb{C}^\ell$ is fed to the unitaries with a leg accounting for the local dimension $\ell$. 

\subsubsection{Pauli Propagation Correlator Surrogates (PPS)} 
The second method we will use for surrogating correlators is via Pauli Propagation. Pauli Propagation Surrogates (PPS) have been thoroughly explored for the calculation of expectation values. This approach is based on developing the measurement outcome in terms of commuting and non-commuting terms and truncating the non-commuting terms which go over a threshold. To evaluate the potential of PPS in approximating generative tasks, we take such framework thoroughly developed in~\cite{angrisani_classically_2024,bermejo_quantum_2024,begusic_fast_2024,begusic_simulating_2023,fontana_classical_2023,nemkov_fourier_2023,martinez_efficient_2025} to estimate the truncated probability shown in Eq.~\ref{eq:truncated_prob}, as the expectation values are the correlators. For example, when classifying the phase of different quantum states, only a limited subset of correlations may be sufficient to distinguish between phases, allowing for efficient feature extraction and classification. However, near phase transition points, the system’s critical behavior may require a better informed ansatz to capture the growing correlation lengths accurately. This can be observed through the increment in precision when augmenting the number of samples in~\cite{bermejo_quantum_2024}. 

Pauli propagation tracks how Pauli operators evolve through the circuit in the Heisenberg picture, allowing one to compute expectation values without explicit state evolution. Each unitary layer $U$ acts as $ P_i \mapsto U^{\dagger} P_i U=\sum_j \alpha_{i j}(\theta) P_j,$ where the coefficients $\alpha_{i j}(\boldsymbol{\theta})$ depend on the circuit parameters and exhibit trigonometric structure, depending on products of sine, cosine and ones from the underlying gate rotations. Iterating this rule across layers yields the propagated observable $ O^{\prime}=\sum_j \beta_j(\theta) P_j,$ where the amplitudes $\beta_j(\theta)$ encode how operator weight is redistributed among Pauli strings. To control the exponential growth of terms, one introduces a weight truncation, retaining only Pauli strings with at most $w$ non-identity operators, $O^{(\leq w)}=\sum_{\left|P_j\right| \leq w} \beta_j(\theta) P_j.$

This procedure defines a systematic approximation hierarchy that captures dependencies up to weight $w$ while maintaining the explicit functional dependence of each propagated term on the circuit parameters.

In both cases, the separation between the classically surrogated correlator and the quantum correlator is controlled by a \emph{complexity cuttof}. In the Tensor Networks case it is the bond dimension and for the Pauli Propagation Surrogate the weight truncation.

\subsection{Data}\label{sec:data}
In the analysis of discrete binary datasets, each observation is a bitstring $\boldsymbol{x} = (x_1, x_2, \ldots, x_n)$ drawn from the $n$-dimensional binary hypercube $\{0,1\}^n$, which can be naturally identified with the vector space $\mathbb{Z}_2^n$ (with $\mathbb{Z}_2 = \mathbb{Z}/2\mathbb{Z}$). Beyond pairwise interactions, understanding higher-order dependencies is essential for capturing complex relationships among features. Higher-order correlations quantify how multiple features interact simultaneously, revealing patterns not discernible from lower-order statistics.
Motivated by the fact that the majority of hard classical problems can be translated into a Hamiltonian form~\cite{lucas_ising_2014}, we will use a similar approach as in~\cite{bermejo_quantum_2024,angrisani_classically_2024}. For a quantum inspired approach, the aim is to learn the distribution of ground states. Appendix~\ref{app:data_obtention} details the procedure for obtaining these ground states.

In this work, we consider various Hamiltonians, decomposed generally as a sum of Pauli operators, \(\mathcal{H} = \sum_i \alpha_i P_i\), where $\alpha_i$ are real, non-negative coefficients representing interaction strengths. The commutation and anti-commutation relations within the Pauli group play a crucial role in determining the system’s properties and symmetries. Having ground states of Hamiltonian systems as the target distribution to learn, allows us to naturally map the circuit architecture intuition to that of the underlying Hamiltonian through Lie algebraic arguments~\cite{wiersema_classification_2024,goh_lie-algebraic_2025} as seen in Subsec.~\ref{subsec:generalization}.

The correlations in the ground state are determined by the phase diagram of the Hamiltonian parameters. Closer to a phase transition, higher-order correlations become more significant~\cite{vidal_entanglement_2003}. When there is a gap between the ground state and the first excited state, and the parameters are far from a phase transition, correlations are short-ranged. 

\begin{figure}[htbp]
\centering
    \resizebox{\linewidth}{!}{\resizebox{\textwidth}{!}{
\begin{tikzpicture}[remember picture]
    \node[inner sep=0pt, xshift=8.3cm] (plot){ \resizebox{.85\textwidth}{!}{\input{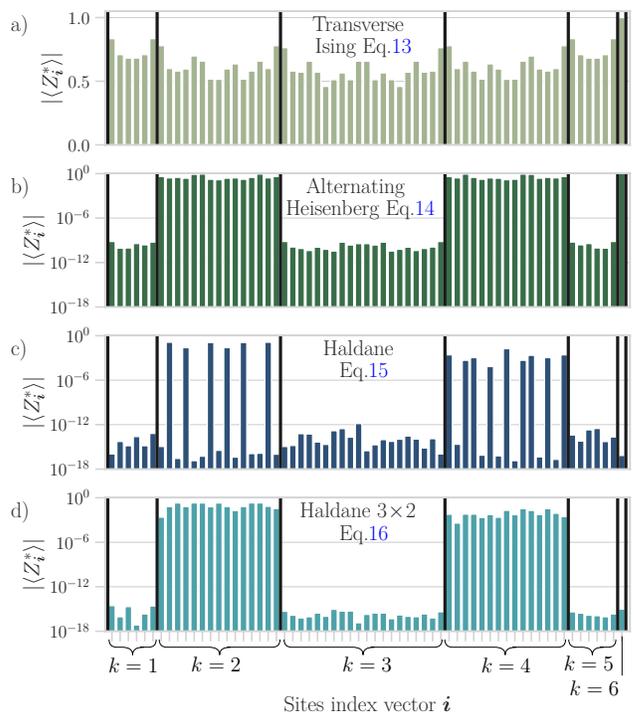}}};

\begin{scope}[overlay] 
\node at (9.355,-7.24) { 
\begin{tikzpicture}[x=0.75pt,y=0.75pt,yscale=-1,xscale=1]

\draw   (80.8,30.89) .. controls (80.8,35.56) and (83.13,37.89) .. (87.8,37.89) -- (92.15,37.89) .. controls (98.82,37.89) and (102.15,40.22) .. (102.15,44.89) .. controls (102.15,40.22) and (105.48,37.89) .. (112.15,37.89)(109.15,37.89) -- (116.5,37.89) .. controls (121.17,37.89) and (123.5,35.56) .. (123.5,30.89) ;
\draw   (126.3,30.89) .. controls (126.3,35.56) and (128.63,37.89) .. (133.3,37.89) -- (171.05,37.89) .. controls (177.72,37.89) and (181.05,40.22) .. (181.05,44.89) .. controls (181.05,40.22) and (184.38,37.89) .. (191.05,37.89)(188.05,37.89) -- (228.8,37.89) .. controls (233.47,37.89) and (235.8,35.56) .. (235.8,30.89) ;
\draw   (239.51,30.89) .. controls (239.51,35.56) and (241.84,37.89) .. (246.51,37.89) -- (301.16,37.89) .. controls (307.83,37.89) and (311.16,40.22) .. (311.16,44.89) .. controls (311.16,40.22) and (314.49,37.89) .. (321.16,37.89)(318.16,37.89) -- (375.8,37.89) .. controls (380.47,37.89) and (382.8,35.56) .. (382.8,30.89) ;
\draw   (497.7,30.89) .. controls (497.7,35.56) and (500.03,37.89) .. (504.7,37.89) -- (509.05,37.89) .. controls (515.72,37.89) and (519.05,40.22) .. (519.05,44.89) .. controls (519.05,40.22) and (522.38,37.89) .. (529.05,37.89)(526.05,37.89) -- (533.4,37.89) .. controls (538.07,37.89) and (540.4,35.56) .. (540.4,30.89) ;
\draw    (546,28) -- (546,63.6) ;
\draw   (385.3,30.89) .. controls (385.3,35.56) and (387.63,37.89) .. (392.3,37.89) -- (430.05,37.89) .. controls (436.72,37.89) and (440.05,40.22) .. (440.05,44.89) .. controls (440.05,40.22) and (443.38,37.89) .. (450.05,37.89)(447.05,37.89) -- (487.8,37.89) .. controls (492.47,37.89) and (494.8,35.56) .. (494.8,30.89) ;

\draw (79,46.5) node [anchor=north west][inner sep=0.75pt]  [font=\Large]  {$k=1$};
\draw (154,45.5) node [anchor=north west][inner sep=0.75pt]  [font=\Large]  {$k=2$};
\draw (291,45.5) node [anchor=north west][inner sep=0.75pt]  [font=\Large]  {$k=3$};
\draw (417,46.5) node [anchor=north west][inner sep=0.75pt]  [font=\Large]  {$k=4$};
\draw (492,44.5) node [anchor=north west][inner sep=0.75pt]  [font=\Large]  {$k=5$};
\draw (497,67) node [anchor=north west][inner sep=0.75pt]  [font=\Large]  {$k=6$};

\end{tikzpicture}};
    \end{scope}
\end{tikzpicture}
}}
    \caption{\justifying\textbf{$Z$-basis $k$-order correlations for ground state} of $n=6$ qubits in: a) the one-dimensional Ising Hamiltonian, with $J=0.7$ and $h=0.33$; b) the Alternating Heisenberg Hamiltonian with $J_{\rm even}=1.4$ and $J_{\rm odd} = 0.6 $; then for the correlations of the Haldane Hamiltonian with parameters $J=1$, $h_1=0.7$ and $h_2=0.33$ we have c) the one-dimensional chain with $n=6$; and d) the rectangular $y$-periodic lattice with $n_x=3$ and $n_y=2$.
    }    \label{fig:correlations_hamiltonians}
\end{figure}

We focus on one- and two-dimensional systems. In 1D, we have: 

The \textit{transverse field Ising model (TFIM)}~\cite{syljuasen_concurrence_2004} with $X\!X$ coupling strength, $J$ and transverse field, $h$:
        \begin{equation}\label{eq:ising_1d}
            \mathcal{H} = -J \sum_{i=1}^{n-1} X_i X_{i+1} \;-\; h \sum_{i=1}^n Z_i,
        \end{equation}
The \emph{Heisenberg bond-alternating XXX model}~\cite{bermejo_quantum_2024,kokcu_fixed_2022}, with nearest-neighbor coupling $J_i \in\{J_{\mathrm{odd}}, J_{\mathrm{even}}\}$:
    \begin{equation}\label{eqn:heisenberg}
        \mathcal{H} = \sum_{i=1}^{n-1} J_i \,\bigl(X_iX_{i+1} + Y_iY_{i+1} + Z_iZ_{i+1}\bigr),
    \end{equation}
    
A \textit{Haldane chain model}~\cite{bermejo_quantum_2024,haldane_nonlinear_1983} with: next-nearest-neighbor interactions, $Z_i X_{i+1} Z_{i+2}$ with strength, $J$; transverse field, $h_1$; and nearest-neighbor $XX$ coupling, $h_2$:
\begin{equation}\label{eq:haldane_1d}
    \mathcal{H} = -J \sum_{i=1}^{n-2} Z_i X_{i+1} Z_{i+2} 
    \;-\; h_1 \sum_{i=1}^{n} X_i \;-\; h_2 \sum_{i=1}^{n-1} X_i X_{i+1},
\end{equation}

In the \textit{two-dimensional case}, we consider a lattice that is periodic in the $y$-direction, so that boundary sites in this direction are identified with sites on the opposite edge. The Hamiltonian is written as
    \begin{equation}\label{eq:haldane_2d}
    \mathcal{H} = -J \sum_{\langle i,j,k \rangle} Z_i X_j Z_k \;-\; h_1 \sum_{i} X_i \;-\; h_2 \sum_{\langle i,j \rangle} X_i X_j,
    \end{equation}
 Here, $\langle i,j,k\rangle$ indicates the relevant triplets of sites in the chosen lattice arrangement, and $\langle i,j\rangle$ denotes pairs of nearest neighbors. 

Given that we are treating with Pauli-based Hamiltonians, the correlations $\langle S_{\bi}\rangle$ are measured with respect to the $Z$ basis and hence can be noted as $Z$-basis correlations, $\langle Z_{\bi}\rangle$. For small systems, such as the ones in this paper, we have access to the exact probability function of the ground state, from which we can obtain the exact correlations denoted with an asterisk,  $\langle Z^*_{\bi}\rangle$. This exact computation enables us to isolate the effect shot noise error has in training and focus on the exact underlying distribution.

Fig.~\ref{fig:correlations_hamiltonians} displays the correlation spectra of the ground states for the different Hamiltonians. In the TFIM, correlations remain non-vanishing across the spectrum. The alternating Heisenberg chain exhibits three dominant orders, $k=2,4,$ and $6$, with correlations spanning all index sets within these orders, while moderate contributions persist at other orders. For the one-dimensional Haldane chain, only a subset of the $k=2$ and $k=4$ correlations deviate significantly from zero. In contrast, in the two-dimensional case, all correlations at $k=2$ and $k=4$ remain appreciable.

Correlations are ubiquitous in real-world data, ranging from gene-expression matrices to fluid flows and market dynamics. While higher-order interactions often carry important information, these systems are typically subject to noise, external influences, and limited sampling, which makes reliable estimation of such moments challenging. Nonetheless, various state-of-the-art classical generative algorithms can offer effective solutions despite numerous complexity-theoretical challenges~\cite{chen_learning_2022,manduchi_challenges_2024}. In contrast to Hamiltonians, such correlations do not necessarily exhibit a natural structure that can be directly mapped onto a model like the quantum circuit Born machine. Nevertheless, we will explore training the QCBM in this setting to assess its ability to capture and represent these patterns.

To study a biological system with an expected large number of interdependent variables, we use single cell RNA sequencing (scRNA-seq) data, focusing on cell cycle-related genes. These are high-dimensional gene expression measurements across different phases of the cell cycle, which serve as a structured but complex dataset for testing probabilistic models. Specifically, we use astrocyte progenitor cells from mouse late embryonic (E18) brain~\cite{noauthor_transcriptional_nodate}, which were randomly subsampled to 11,950 cells in~\cite{jansma_high_2025} to identify reproducible higher-order gene interactions for cell cycle genes. Here, up to 37 binarised highly variable cell cycle genes are available, from which we can select a subset of them to train on.

\begin{figure}[htbp]
\centering
    \resizebox{\columnwidth}{!}{\resizebox{\textwidth}{!}{
\begin{tikzpicture}[remember picture]
    \node[inner sep=0pt, xshift=8.3cm] (plot){ \resizebox{.8\textwidth}{!}{\input{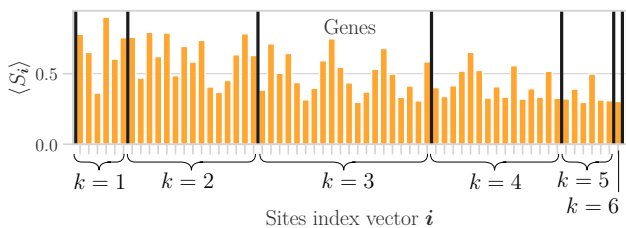}}};

\begin{scope}[overlay] 
\node at (8.94,-1.4) { 
\begin{tikzpicture}[x=0.75pt,y=0.75pt,yscale=-1,xscale=1]

\draw   (80.8,30.89) .. controls (80.8,35.56) and (83.13,37.89) .. (87.8,37.89) -- (92.15,37.89) .. controls (98.82,37.89) and (102.15,40.22) .. (102.15,44.89) .. controls (102.15,40.22) and (105.48,37.89) .. (112.15,37.89)(109.15,37.89) -- (116.5,37.89) .. controls (121.17,37.89) and (123.5,35.56) .. (123.5,30.89) ;
\draw   (126.3,30.89) .. controls (126.3,35.56) and (128.63,37.89) .. (133.3,37.89) -- (171.05,37.89) .. controls (177.72,37.89) and (181.05,40.22) .. (181.05,44.89) .. controls (181.05,40.22) and (184.38,37.89) .. (191.05,37.89)(188.05,37.89) -- (228.8,37.89) .. controls (233.47,37.89) and (235.8,35.56) .. (235.8,30.89) ;
\draw   (239.51,30.89) .. controls (239.51,35.56) and (241.84,37.89) .. (246.51,37.89) -- (301.16,37.89) .. controls (307.83,37.89) and (311.16,40.22) .. (311.16,44.89) .. controls (311.16,40.22) and (314.49,37.89) .. (321.16,37.89)(318.16,37.89) -- (375.8,37.89) .. controls (380.47,37.89) and (382.8,35.56) .. (382.8,30.89) ;
\draw   (497.7,30.89) .. controls (497.7,35.56) and (500.03,37.89) .. (504.7,37.89) -- (509.05,37.89) .. controls (515.72,37.89) and (519.05,40.22) .. (519.05,44.89) .. controls (519.05,40.22) and (522.38,37.89) .. (529.05,37.89)(526.05,37.89) -- (533.4,37.89) .. controls (538.07,37.89) and (540.4,35.56) .. (540.4,30.89) ;
\draw    (546,28) -- (546,63.6) ;
\draw   (385.3,30.89) .. controls (385.3,35.56) and (387.63,37.89) .. (392.3,37.89) -- (430.05,37.89) .. controls (436.72,37.89) and (440.05,40.22) .. (440.05,44.89) .. controls (440.05,40.22) and (443.38,37.89) .. (450.05,37.89)(447.05,37.89) -- (487.8,37.89) .. controls (492.47,37.89) and (494.8,35.56) .. (494.8,30.89) ;

\draw (79,46.5) node [anchor=north west][inner sep=0.75pt]  [font=\Large]  {$k=1$};
\draw (154,45.5) node [anchor=north west][inner sep=0.75pt]  [font=\Large]  {$k=2$};
\draw (291,45.5) node [anchor=north west][inner sep=0.75pt]  [font=\Large]  {$k=3$};
\draw (417,46.5) node [anchor=north west][inner sep=0.75pt]  [font=\Large]  {$k=4$};
\draw (492,44.5) node [anchor=north west][inner sep=0.75pt]  [font=\Large]  {$k=5$};
\draw (500,67) node [anchor=north west][inner sep=0.75pt]  [font=\Large]  {$k=6$};

\end{tikzpicture}};
    \end{scope}
\end{tikzpicture}
}}
    \caption{\justifying\textbf{$k$-order correlations in the scRNA-seq dataset} derived from the data probability distribution amongst $n=6$ genes with activated and not activated encoding. 
    }    \label{fig:correlations_bio}
\end{figure}

Fig.~\ref{fig:correlations_bio} shows the correlation distribution of the biological dataset, there is a general polynomial decay as the order increases with different peaks appearing.

\subsection{Quantum Circuit Born Machine Ansatz\"{e}} \label{subsec:qcbm_ansatze}
\subsubsection{Ansatz\"{e} and Lie Algebras}
Next, we discuss the possible choices of QCBM model we consider, i.e. the specific parameterization structure of the unitary operator, $U(\boldsymbol{\theta})$. We partition the possibilities into two categories, one which is generated from a specific combination of gates from a set $ U \in \mathcal{U}$ or from a set of generators, $G \in \mathcal{G}$ such that $U = e^{iG}$ - a precise ansatz structure. The second option generates ansatz\"{e} directly from a \emph{dynamical Lie algebra} (DLA), denoted as $\mathfrak{g}$. In the second approach, $U(\boldsymbol{\theta})$ will be generated using gates of the form $U = e^{i g}$ where $g \in \mathfrak{g}$. The DLA is an important concept in the study of parameterized quantum circuit ansatz\"{e} as it can be used to make statements about expressivity and vanishing gradients~\cite{ragone_lie_2024,diaz_showcasing_2023}.

These two cases can be related as $\mathfrak{g}$ is defined as the real span of the Lie closure of the elements of $i\mathcal{G}$. However, we make the distinction between these two cases due to the fact that the precise relationship between unitaries $\mathcal{U}$ or set of generators, $\mathcal{G}$ and the corresponding DLA, $\mathfrak{g}$ may be unknown (no closed form expression) or unwieldy in general. Secondly, as we primarily aim to learn Hamiltonians with a specific generator structure in this work, parameterizing a target ansatz via the DLA allows us to target the scenario where we do not have knowledge of the Hamiltonian terms but want to have a structured approach towards solving the problem. In this way we emphasize the fact that an algebra can be attained by different Hamiltonians.

\subsubsection{Instantaneous Quantum Polynomial (IQP) Circuit} \label{sssec:iqp_circuits_defn}
The Instantaneous Quantum Polynomial (IQP) circuit provides a model where samples drawn from its output distribution can exhibit a provable separation between classical and quantum computational capabilities~\cite{marshall_improved_2024,bremner_achieving_2017} (see Appendix~\ref{subsec:sampling_advantage} for details), while remaining efficiently verifiable. Its parameterized form is first introduced in~\cite{coyle_born_2020}
as,
\begin{equation}
U_{\text{IQP}}(\btheta) = H^{\otimes n} U_c(\boldsymbol{\theta}) H^{\otimes n},
\end{equation}
where $H$ denotes the Hadamard gate and $U_c(\btheta)$
is a product of mutually commuting gates, hence the term instantaneous. The number of such gates scales polynomially with the system size $n$. Although the parameterized IQP circuit is non-universal in its standard form, universality can be achieved by introducing ancillary qubits~\cite{kurkin_note_2025}.

\subsubsection{Matchcircuits} \label{sssec:matchcircuits_defn}
We start by analyzing the matchcircuits as they are one of the simplest known algebras that map directly to the TFIM. Matchcircuits are composed of matchgates, named for their connection to perfect matchings in graphs~\cite{valiant_quantum_2012,terhal_classical_2002,jozsa_matchgates_2008}. In general, parameterized matchcircuits are not efficiently simulable and depend on the weight of the correlator. Next, we see how higher-order correlators can be both significant and trainable for matchcircuits. 

Following~\cite{diaz_showcasing_2023} we start by considering a PQC based on matchgates (MG) where the generators of the rotation unitary operators $U(\btheta)$ are drawn from the set
\begin{equation}\label{eq:MG_set}
\G =\left\{X_i X_{i+1}\right\}_{i=1}^{n-1}\cup \left\{Z_i\right\}_{i=1}^n,
\end{equation}
which happens to have exact same terms as the one-dimensional Ising Hamiltonian in  Eq.~\ref{eq:ising_1d}. These generators give rise to a dynamical Lie algebra (DLA), which is expressed as

\begin{equation} \label{eq:MG_algebra}
\begin{aligned}
    \mathfrak{g}_{\text{MG}} &= \operatorname{span}_{\mathbb{R}} i\left\{Z_i, \widehat{X_i X_j}, \widehat{Y_i Y_j}, \widehat{X_i Y_j}, \widehat{Y_i X_j}\right\}_{1 \leq i < j \leq n}\\ &\simeq \mathfrak{so}(2n), 
\end{aligned}
\end{equation}
where the notation $ \widehat{A_i B_j} $ is defined as
\begin{equation}  \label{eqn:ab_hat_defn}
    \widehat{A_i B_j} = A_i Z_i Z_{i+1} \cdots Z_{j-1} B_j.
\end{equation}

\subsubsection{Heisenberg chain circuits}
Next, we take ansatz\"{e} which are generated directly from elements of the DLA, $\mathfrak{g}$. The first example is inspired from the  one-dimensional Heisenberg chain Hamiltonian, Eq.~\ref{eqn:heisenberg} and which has a Lie algebra (given in Kokcu et al.~\cite{kokcu_fixed_2022}) as:
\begin{multline} \label{eqn:heisenberg_dla}
    \mathfrak{g}_{\text { Heisenberg }}=\operatorname{span}(\\
    \{\text{Pauli strings with } N_X = a,  N_Y = b , N_Z=c \mid \\
     a+b, a+c, b+c \text { even, } a, b, c \geq 1\}  \backslash\{X^{\otimes n}, Y^{\otimes n}, Z ^{\otimes n}\})
\end{multline}
where $N_J$ is the number of terms in a Pauli/word string of type $J\in \{X, Y, Z\}$. This algebra
has exponential size $|\g_{\text{Heisenberg}}|=4^{n-1}-4$.  

\subsubsection{Haldane chain circuit}
Our second direct algebra-constructed ansatz is the Haldane chain circuit. As with the Heisenberg chain, this algebra is derived from the corresponding one-dimensional Haldane Hamiltonian, Eq.~\ref{eq:haldane_1d}. However, unlike the Heisenberg chain, the algebra for the Haldane chain was previously unknown, to the best of our knowledge. Our first result in this work is to provide an expression for the Lie algebra, before using it as a QCBM model ansatz. We do this by inspection numerically, by analyzing the terms generated from the nested commutants~\cite{ragone_lie_2024}, isolating each term individually and verifying it computationally. As such, we can only verify the expression up to a certain system size, so we leave is as a proposition for this work.

\begin{proposition}[Dynamical Lie algebra for the Haldane chain Hamiltonian (Eq.~\ref{eq:haldane_1d})]

The dynamical Lie algebra (DLA) for the one-dimensional Haldane chain, with Hamiltonian:
\begin{equation}\label{eq:haldane_1d_repeated_lem}
    \mathcal{H} = -J \sum_{i=1}^{n-2} Z_i X_{i+1} Z_{i+2} 
    \;-\; h_1 \sum_{i=1}^{n} X_i \;-\; h_2 \sum_{i=1}^{n-1} X_i X_{i+1},
\end{equation}
is given by:
    \begin{equation} \label{eqn:Haldane_dla}
    \begin{aligned}
        \g_{\text{Haldane}}\! = &\operatorname{span}\!\bigg(\bigg\{ 
        (\mathcal{YZ})_{\bj},
        (\mathcal{YZ})_{\bj}\mathcal{X}_{\boldsymbol{\noj}},
        \mathcal{X} \ \big|\ \bj = \bj_{\text{even}} \oplus \bj_{\text{odd}} : \\& \qquad \quad
         \left| \bj_{\text{even}} \right|\!,\!\left| \bj_{\text{odd}} \right|\! \in \! 2\mathbb{Z},  \, (\mathcal{YZ})_{\bj}\!\!=\!\!\{Y,Z\}_{\bj},\\ & \mathcal{X} \!\in\!\{X, I\}^{\otimes n},\mathcal{X}_{\boldsymbol{\noj}}\! \in\! \{X, I\}_{\boldsymbol{\noj}}
        \bigg\} \!\! \setminus \!\!
        \bigl\{\!X_o,
        X_e,
        X^{\otimes n}\!\bigr\}\bigg) .
    \end{aligned}
    \end{equation}
\end{proposition}

where we define $X_o := (\1 X)^{\lfloor n/2 \rfloor}\1^{(n \bmod 2)}$ and $X_e := (X\1)^{\lfloor n/2 \rfloor}X^{(n \bmod 2)}$. The proof of this can be found in Appendix~\ref{app:lie_modules} along with further discussion. The Cartan decomposition~\cite{kokcu_fixed_2022} allows to prove that the size of the one-dimensional Haldane algebra is exponential, $|\g_{\text{Haldane}}|=4^{n-1}-4$, as with the Heisenberg algebra. As also detailed numerically in the same Appendix, the $y$-periodic structure of the lattice for the two-dimensional case still follows the exponential rule but increments the number of elements in the algebra with respect to the one-dimensional case. Finally, we note that intersection size between the Haldane and Heisenberg chain algebras although lower is also exponential, $|\g_{\text{Haldane}} \cap \g_{\text{Heisenberg}}| \approx 2^{n-2}-2$, which we also verify computationally.

\section{Quantum advantage and Dequantization in Deployment}\label{sec:comparison}
Now that we have defined the models we use, we can ask the fundamental question we target in this work. Specifically, how does the classical surrogate or direct quantum training relate to the actual deployment of trained parameters on quantum devices? It is critical to quantify such relationships as it is actually the realistic situation one will encounter in practice, in almost all cases. \emph{Deployment discrepancy} is the result of (potentially significant) deviation between the deployed model in inference (due to a range of factors, perhaps most notably quantum noise and imperfect device calibration). On the other hand, this discrepancy may also be a source of quantum advantage as classical surrogate training will not, in general, be able to reproduce fully quantum correlations. We address this question in the following sections by first deriving theorems relating these scenarios, and then numerically investigating them through concrete examples. We refer to the conditions (the `ultimatum') we derive in the rest of this work to be the \emph{Deployment-Dequantization} conditions, as seen in Figs.~\figref[b), c)]{fig:main_fig_deployment}.

As established in Eq.~\ref{eq:truncated_prob}, we deal with an approximation method to the actual QCBM probability distribution. In this approximation we use correlation sampling with the unique constraint of having to always include the void correlation, $\langle Z_{\varnothing}\rangle\equiv\langle\1\rangle = 1$ for normalization purposes. In the supervised setting, where one assigns labels to data depending on the features, the randomized sampling method of features is referred to as \emph{Random Fourier Feature (RFF)}~\cite{rahimi_random_2007,rahimi_weighted_2008,li_towards_2021,sutherland_error_2021,avron_random_2017} sampling. We mimic the logic presented in ~\cite{landman_classically_2022,sweke_potential_2025,sahebi_dequantization_2025}, to assess the problem of the unsupervised learning dequantization of quantum deployments hence the name \emph{Random Fourier Correlator (RFC)} sampling in Defintion~\ref{defn:random-sampling_truncation}. The analysis divides into two:
\begin{enumerate}
    \item We first consider the hypothetical scenario of \textbf{train-on-quantum}, \textbf{deploy-on-quantum} where the quantum sampler is ideally trained and,
    \item \textbf{Train-on-classical}, \textbf{deploy-on-quantum}, where the quantum sampler is deployed with the angles of the trained classical surrogate, which is the primary method we advocate for in this work.
\end{enumerate}

\subsection{Technical Terminology} \label{ssec:deployment_technical_term}

Formally, the objective of the learning task is to minimize the \emph{empirical risk}. The empirical risk is the sampled mean over all ground truth samples of the unknown \emph{true risk}, and we take the squared difference for one single bit-string as it contains all the information associated to the correlators,

\begin{equation}
    \begin{aligned}\label{eq:risk}
    \risk[\operatorname{Pr}(\boldsymbol x)] &= \E[\mathcal{L}(\operatorname{Pr}(\bx),\operatorname{Pr^*}\!\!\!_{\mathcal H}(\bx)]\\
    & = \E\left[\left(\operatorname{Pr^*}\!\!\!_{\mathcal H}(\bx)-\operatorname{Pr}(\bx)\right)^2\right].
    \end{aligned}
\end{equation}

Our aim in this section is to study this quantity for QCBM deployment. However, the techniques we use to do so have only previously been used in the context of continuous variables, $\bx \in \mathbb{R}^n$, for example in supervised regression problems. To adapt these to the discrete (Boolean) case of $\bx \in \mathbb{Z}^n_2$, as we use here, we must make the following adjustments.

First, the Fourier transform of the probability distribution can be rewritten in terms of the inner product between a \emph{correlator} vector $\boldsymbol c(\btheta)$ and a \emph{sign features} vector $\boldsymbol \varphi(\bx)$, 

\begin{equation}\label{eq:fourier_prod}
    \operatorname{Pr_{\btheta}}(\bx)=\langle \boldsymbol c(\btheta), \boldsymbol \varphi(\bx)\rangle,
\end{equation}
with 
\begin{equation}
    \boldsymbol c(\btheta) =\frac{1}{\sqrt{2^{n}}}\bigl[\langle Z_{\bi}\rangle\bigr]_{\bi \in\Omega}, \quad 
    \boldsymbol \varphi (\bx) =\frac{1}{\sqrt{2^{n}}}\bigl[(\!-1)^{\sum_{i\in \bi}x_i}\bigr]_{\bi \in\Omega}.
\end{equation}
The index sample set $\Omega$ defines the correlators used to surrogate. The sign features can then be used to build the \emph{parity Boolean kernel} between sign feature vectors of two different bit-strings $s:\mathbb{Z}_2^{n}\times\mathbb{Z}_2^{n} \to \mathbb{R}$:
\begin{equation}
    s\left(\bx, \bx^{\prime}\right)=\left\langle\boldsymbol\varphi(\bx), \boldsymbol\varphi\left(\bx^{\prime}\right)\right\rangle=\frac{1}{2^n} \sum_{\bi \in \Omega}(-1)^{\sum_{i \in \bi}\left(\bx_i\oplus \bx_i^{\prime}\right)}.
\end{equation}
Note that as $\bx_i \oplus \bx'_i = \bx_i - \bx'_i\; ({\rm mod}\,2)$, the kernel is \emph{shift invariant}. Also, if we have the whole power set of correlators $\Omega=2^{[n]}$, then the kernel reduces to the identity or delta-kernel $s(\bx,\bx')=\delta_{\bx,\bx'}/2^n$.

The associated Reproducing Kernel Hilbert Space (RKHS) $\mathscr{H}$ is then the completion of the span of the representers~\cite{schwartz_sous-espaces_1964},
\begin{equation}\label{eq:rkhs}
    \mathscr{H} = {\rm span}\{s(\cdot,{\bx}) : {\bx} \in \mathbb{Z}_2^n \}.
\end{equation}
Hence, for any pseudo-probability distribution that can be represented as $\operatorname{Pr}(\cdot) = \langle \boldsymbol c, \boldsymbol \varphi (\cdot)\rangle$, the RKHS norm is defined as $\| \operatorname{Pr}(\cdot)\|_s = \min_{\boldsymbol c} \left\{\| \boldsymbol c\|_2 \mid \boldsymbol c \in [-1,1]^{|\Omega|} \text { s.t. } \operatorname{Pr}(\cdot)=\langle \boldsymbol c, \boldsymbol \varphi(\cdot)\rangle\right\}$.

To be able to use known results on kernel approximation, we take the parity Boolean kernel in its integral form by defining the frequencies $\boldsymbol \omega$ with respect to the correlator indices. These frequency vectors are defined in terms of the standard orthonormal basis of $\mathbb{R}^n$, $\{\boldsymbol e_i\}_{i=1}^n$ with $(\boldsymbol e_i)_\ell = \delta_{i\ell}$, such that $\boldsymbol \omega = \sum_{i \in \boldsymbol i} \boldsymbol e_i$. This allows us to write $\sum_{i \in \boldsymbol i}\boldsymbol x_i$ as $\boldsymbol \omega\cdot \boldsymbol x$. 

Assuming a positive probability measure $\pi(\boldsymbol \omega)$  we obtain the integral representation
\begin{equation}
s(\bx,\bx') = \int \boldsymbol \varphi(\boldsymbol x;\boldsymbol \omega) \boldsymbol \varphi(\boldsymbol x';\boldsymbol \omega) d\pi(\boldsymbol \omega).
\end{equation}
Additionally, the feature function $\boldsymbol \varphi (\boldsymbol x;\boldsymbol \omega)$ is topologically continuous as it takes elements in an open set to elements in an open set.

With all previous definitions that are adapted from~\cite{sweke_potential_2025,sahebi_dequantization_2025} to the Boolean case, we are now in position to define when does an RFC sampling surrogate dequantize a quantum-deployed QCBM. 

\subsection{Excess Risk for Ideal Quantum Parameters}  \label{ssec:excess_risk_ideal_q_params}
We start by considering the case where we have an ideally trained quantum probability distribution. As we assume that we are only able to train in a classical regime, the ideal quantum parameters have to coincide with those optimized through a classical surrogate, $\btheta^*_{\mathcal{Q}} \equiv \btheta^*_{\mathcal{C}\!\ell}$, which is rarely the case given all the additional terms from the quantum model.

\begin{definition}[RFC Dequantization] Consider an unsupervised learning task with a ground state probability $\operatorname{Pr}^{*}(\bx)$
s.t. $\bx \in \mathbb{Z}^{n}_2$. Given a QCBM, with hypothesis class $\mathcal{Q}$, $\operatorname{Pr}_{\boldsymbol{c}_{\mathcal{Q}}} = \operatorname{Pr_{\mathcal{Q}}(\btheta^*_{\mathcal{Q}})}$ with $\btheta^*_{\mathcal{Q}}$ such that $\operatorname{Pr_\mathcal{Q}}(\btheta^*_{\mathcal{Q}}) = \inf_{\operatorname{Pr} \in {\mathcal{Q}}}\, \risk [\operatorname{Pr}]$. We say the generative model is dequantizable if the coefficient distribution $\boldsymbol{c}_{\mathcal{C}\!\ell}:\Omega\to\R$, such that for some $\varepsilon > 0$, with $M,D \in \mathcal{O}(\poly(n, \varepsilon^{-1}))$ the following upper bound of the excess risk is true with high probability
\begin{equation}\label{eq:excess_ideal}
0\leq\risk [\operatorname{Pr}^*_{D,\boldsymbol{c}_{\mathcal{C}\!\ell}}] - \risk [\operatorname{Pr}^*_{\boldsymbol{c}_{\mathcal{Q}}}] \leq \varepsilon.
\end{equation}
$\operatorname{Pr}^*_{D,\boldsymbol{c}_{\mathcal{C}\!\ell}} = \operatorname{Pr}_{\mathcal{C}\!\ell}(\btheta^*_{\mathcal{C}\!\ell})$ is the surrogate probability distribution trained with $D$ correlators sampled from $\boldsymbol{c}_{\mathcal{C}\!\ell}$, while $\risk$ denotes the true risk defined in Eq.~\ref{eq:risk}.
\end{definition}

It is possible then to bound the excess risk through a ridge regression approach~\cite{rudi_generalization_2021}. The ridge regression solution is a perfect minimizer within our Hilbert space and should coincide with that of the ideal parameters. Next, by using reweighting of the features $\boldsymbol \varphi$ it is possible to derive the dequantization conditions for Fourier distributions.

\begin{theorem}[Fourier Distribution Dequantization for Shift-Invariant Kernels~\cite{sweke_potential_2025, sahebi_dequantization_2025}]\label{thm:ideal_deq}
Consider a coefficient distribution $\boldsymbol{c}_{\mathcal{C}\!\ell}$ defined over a frequency support set of size $D$ related to the Fourier coefficients $\boldsymbol c^*_{\mathcal{Q}} = \boldsymbol{c}_{\mathcal{Q}}(\boldsymbol \theta_{\mathcal Q}^*)$ of the optimal quantum model $\operatorname{Pr}^*_{\boldsymbol{c}_{\mathcal{Q}}}$ described by Eq.~\ref{eq:fourier_prod}. The following conditions are required:
\begin{enumerate}
    \item \textbf{Efficient Sampling:} It is computationally tractable to draw samples from the distribution $\boldsymbol c_{\mathcal{C}\!\ell}$.
    \item \textbf{Polynomial Concentration throughout Frequency Support:} The maximum probability of the distribution scales is suppressed as an inverse polynomial with the dimensionality $n$ of the bit-string, specifically, $(\boldsymbol c_{\mathcal{C}\!\ell})_{\max}^{-1} \in \mathcal{O}(\operatorname{poly}(n))$.
    \item \textbf{Coefficient Alignment with Quantum Model:} The distribution $\boldsymbol c_{\mathcal{C}\!\ell}$ aligns proportionally with the magnitude of the Fourier coefficients from the optimal quantum model, such that $\boldsymbol c_{\mathcal{C}\!\ell} \propto |\boldsymbol c^*_{\mathcal{Q}}|$.
\end{enumerate}
Under these assumptions, the distribution $\boldsymbol c_{\mathcal{C}\!\ell}$ enables efficient classical dequantization for shift-invariant kernel methods.
\end{theorem}
We term a task RFC‑dequantized if an RFC surrogate, constrained to polynomial‑scale computational resources, achieves performance comparable to that of the deployed QCBM. Although, $\operatorname{Pr}^*_{\boldsymbol{c}_{\mathcal{Q}}}$ is in practice difficult to achieve for large scale problems, Theorem~\ref{thm:ideal_deq} is already a sufficient condition to show dequantization even when we have non-ideal quantum parameters. Nevertheless, we can assess where discrepancies for the non-ideal case happen. Theorem~\ref{thm:ideal_deq} is depicted in Fig.~\figref[b)]{fig:main_fig_deployment}.

\subsection{Excess Risk for Non-Ideal Quantum Parameters.} \label{ssec:excess_risk_nonideal_q_params}
Realistically, as the deployed parameters we take are the classically trained ones, the optimal quantum parameters are different from them, $\boldsymbol{\theta}_{\mathcal Q} =\boldsymbol{\theta}^*_{\mathcal{C}\!\ell} \neq \boldsymbol{\theta}^*_{\mathcal Q} $. Now, the excess risk we are interested in is given by the risk difference between the perfect minimizer of the classical deployment $\operatorname{Pr}^*_{D,\boldsymbol{c}_{\mathcal{C}\!\ell}} = \operatorname{Pr}^*_{D,\boldsymbol{c}_{\mathcal{C}\!\ell}}(\btheta^*_{\cl})$ and the non-ideal (quantum deployment with classical parameters) quantum risk $\operatorname{Pr}_{\mathcal{Q}} = \operatorname{Pr}_{\mathcal{Q}}(\boldsymbol \theta^*_{\mathcal{C}\!\ell})$:
\begin{equation}
    \mathcal{R}[\operatorname{Pr^*_{D,\ccl}}]-\mathcal{R}[\operatorname{Pr}_{\mathcal{Q}}].
\end{equation}
Contrary to the excess risk of having an ideal quantum deployed model in Eq.~\ref{eq:excess_ideal}; this more realistic excess risk can take negative values if the classical model (with classical parameters) performs better. This is possible because now $\mathcal{R}[\operatorname{Pr}_{\mathcal{Q}}]$ is not the infimum within the Hilbert space defined by the quantum model.
Nevertheless, to derive an understanding on where discrepancies happen we can bound its absolute value.
\begin{theorem}[Discrepancy Sources between Classical and Quantum Deployed Distributions]\label{thm:discrepancy}
Let $\operatorname{Pr}_{D,\boldsymbol{c}_{\mathcal{C}\!\ell}}$ denote the approximated distribution using $D$ Fourier frequencies with the optimally trained classical coefficients $\boldsymbol{c}_{\mathcal{C}\!\ell}(\btheta_{\mathcal{C}\!\ell})$, and let $\operatorname{Pr}_{\mathcal{Q}}(\boldsymbol{\theta}^*_{\mathcal{C}\!\ell})$ be the quantum‐deployed distribution evaluated at the same classical parameters with coefficients $\boldsymbol{c}_{\mathcal{Q}}(\boldsymbol \theta^*_{\mathcal{C}\!\ell})$. If $\boldsymbol{c}_{\mathcal{Q}}(\boldsymbol \theta_{\mathcal{Q}})$ are the ideal coefficients that would result from training in the full quantum feature space and $\mathcal{R}$ is the risk defined in Eq.~\ref{eq:risk}, then we have
\begin{equation}
    \begin{aligned}
        |\mathcal{R}[\operatorname{Pr^*_{D,\boldsymbol{c}_{\mathcal{C}\!\ell}}}]-\mathcal{R}[\operatorname{Pr}_{\mathcal{Q}}]|\leq C (&\|\boldsymbol {c}_{\mathcal{C}\!\ell}(\boldsymbol \theta^*_{\mathcal{C}\!\ell})-\boldsymbol c_{\mathcal{Q}}(\boldsymbol\theta^*_{\mathcal{Q}})\|\\&+\|\boldsymbol c_{\mathcal{Q}}(\boldsymbol\theta^*_{\mathcal{Q}}) - \boldsymbol c_{\mathcal{Q}}(\boldsymbol\theta^*_{\mathcal{C}\!\ell} )\|),
    \end{aligned}
\end{equation}
where $C> 0$ a positive constant depending on the performance of each model with respect to the best possible distribution.
\end{theorem}
In the next section, we will briefly discuss the relationship between the quantities, $c_{\mathcal{Q}}(\boldsymbol\theta^*_{\mathcal{Q}} ), c_{\mathcal{Q}}(\boldsymbol\theta^*_{\mathcal{C}\!\ell})$ and $ c_{\mathcal{C}\!\ell}(\boldsymbol\theta^*_{\mathcal{C}\!\ell})$. Studying this relationship is the fundamental target of our work, which we elaborate on in the following sections, both through theory and numerical experiments. 

\subsection{Discussion on the Deployment-Dequantization Theorems}
Theorem~\ref{thm:ideal_deq} assumes the knowledge of the perfect minimizer for the quantum distribution. Only if the classical distribution meets the conditions of being easy to sample from and matches the best quantum distribution up to polynomial concentration throughout the frequency support, and if it aligns well enough with the underlying case (see Fig.~\ref{fig:main_fig_deployment}). Finding these classical distributions is a whole field of study and the development of different methods could shrink in practice the potential advantage. However, if we train on classical we are not typically going to have this best scenario case.

In Theorem~\ref{thm:discrepancy} we derived a bound on the empirical risk between a classically trained surrogate and the quantum-deployed model. The decomposition explicitly separates two fundamental and intuitive contributions to the difference between classical and quantum coefficients. The first contribution $\|\boldsymbol {c}_{\mathcal{C}\!\ell}(\boldsymbol \theta^*_{\mathcal{C}\!\ell})-\boldsymbol c_{\mathcal{Q}}(\boldsymbol\theta^*_{\mathcal{Q}})\|$, arises solely from the fact that the classical model uses fewer features compared to the richer quantum model. This so-called \emph{feature-cardinality gap} quantifies precisely how limited the classical representation is compared to the ideal quantum model trained directly in the quantum feature space. The larger this gap, the more we expect the classical predictor to be inferior, as it inherently lacks the expressive power of the full quantum feature set.

The second contribution $\|\boldsymbol c_{\mathcal{Q}}(\boldsymbol\theta^*_{\mathcal{Q}}) - \boldsymbol c_{\mathcal{Q}}(\boldsymbol\theta^*_{\mathcal{C}\!\ell} )\|$, captures the error introduced purely by surrogate training, namely the mismatch between parameters trained classically and those optimally obtained if one were to train directly on the quantum features. This surrogate mismatch reflects the cost of deploying classical-trained parameters into a quantum-enhanced space without re-optimization. When this mismatch is large, the quantum predictor's performance could degrade significantly, potentially offsetting any benefits offered by the richer quantum feature representation.

Informally, Theorems~\ref{thm:ideal_deq} and \ref{thm:discrepancy} can be understood as:
\begin{enumerate}
    \item Theorems~\ref{thm:ideal_deq}: If the ideal quantum distribution is sufficiently close to the underlying distribution while remaining too complex for the classical case, then dequantization is not possible. 
    \item Theorem~\ref{thm:discrepancy}: Training with classical constraints may diminish such an advantage even more.
\end{enumerate}

Once more, the ideal coefficients $\boldsymbol{c}_{\mathcal{Q}}(\boldsymbol{\theta}^*_{\mathcal{Q}})$ are difficult to achieve in practice for large problem sizes due to trainability issues. The results of this section provide the correct intuition for understanding the simulability of the QCBM, with emphasis on obtaining $\ccl(\boldsymbol{\theta}_{\cl})$ and deploying $\cq(\boldsymbol{\theta}^*_{\cl})$. In the remainder of this work, we complement this analysis by theoretically analyzing some QCBM ansatz\"{e} and classical surrogates, and numerically exploring the effects of approximation across different values of $D$, alternative surrogate methods for the coefficients, and various circuit architectures, thereby gaining insight into how to best exploit the capabilities of the QCBM.

\section{Vanishing of Correlators and Inaccessibility of Optimal Quantum Parameters}\label{sec:concentration}

\subsection{Vanishing gradients} \label{sec:vanishing_gradients}

The curse of dimensionality is an initialization problem that happens when the ansatz $U$ is too expressive. Specifically, it refers to the issue of having an exponentially large parameter search space in $n$ to be able to attain the global minimum. When this occurs, the quantity of interest to minimize vanishes (goes to zero) as so do its gradients~\cite{arrasmith_equivalence_2022}. This leaves little room for any advantage when training with the output of a quantum computer. This concentration effect will leave us unable to achieve the ideal quantum-trained parameters, $\boldsymbol{\theta}^*_{\mathcal{Q}}$ in reality.

\begin{figure*}[htbp]
    \centering
    \resizebox{\textwidth}{!}{\input{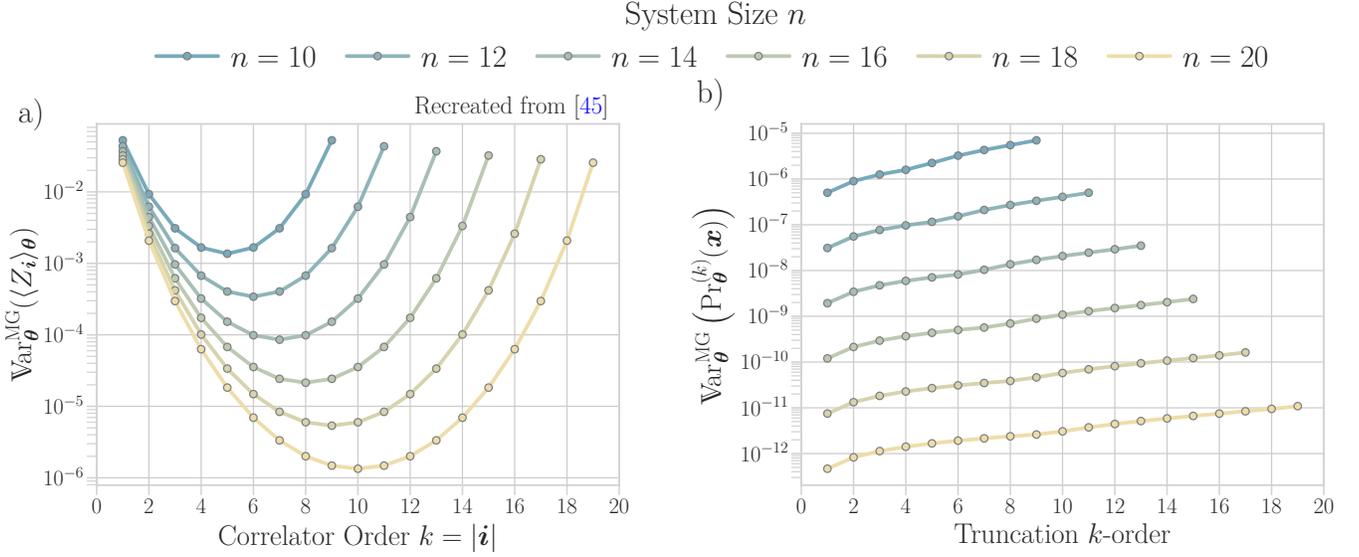}}\\
    \caption{\justifying Variances for the matchcircuit of a) individual correlators and, b) truncated probabilities. High bodied individual correlators do not necessarily vanish and the contribution of multiple correlators may be greater than that of the sum of them individually. When accumulated, the non-vanishing variance may be supported predominantly by high variance terms.}
    \label{fig:variancealgebra}
\end{figure*}

In this heuristic setting, training to minimize the loss function leads different correlators to contribute with varying influence to the parameterized probability distribution $\operatorname{Pr}_{\boldsymbol{\theta}}$. Even if the overall distribution does not vanish as the problem size increases, meaning that it remains trainable for large $n$, the chosen ansatz may still give preference to non-vanishing correlators. In short, a set of the correlators can vanish, which in turn may affect the correlations that the model is able to capture. Such phenomenon is studied for the supervised case in~\cite{mhiri_constrained_2025} where the Fourier coefficients play the role of the correlators  and the frequencies play the role of the correlation indices $\boldsymbol{i}$.

In practice, the issue of having a quantity to minimize that vanishes exponentially is that it requires an exponential number of samples to obtain a desired precision. As all such correlators are diagonal in the same basis (cf. Eq.~\ref{eq:exp_from_probs}), and therefore commute,  they can be estimated simultaneously with a single circuit instance~\cite{yen_measuring_2020}. Hence, assuming the same number of polynomial shots per correlator makes some correlators have more informative gradients than others. The variance of the actual quantity characterizes how informative the gradient of the quantity $Q$ is through Chebysev’s inequality given any  $\delta>0$,

\begin{equation}
\operatorname{Pr}_{\boldsymbol\theta}\Bigl(\bigl|Q_{\boldsymbol\theta}- \mathbb{E}_{\boldsymbol\theta}\bigl[Q_{\boldsymbol\theta}\bigr]\bigr| \ge \delta\Bigr)\le\frac{\Var_{\boldsymbol\theta}\bigl[Q_{\boldsymbol\theta}\bigr]}{\delta^2},
\end{equation}
with
$$
\Var_{\boldsymbol\theta}\bigl[Q_{\boldsymbol\theta}(\boldsymbol x)\bigr]
=
\mathbb{E}_{\boldsymbol\theta}\!\bigl[Q_{\boldsymbol\theta}^2\bigr]
- \bigl(\mathbb{E}_{\boldsymbol\theta}[Q_{\boldsymbol\theta}(\boldsymbol x)]\bigr)^2.
$$

\subsection{Correlator Vanishing with Truncated Distributions} \label{sec:vanishing_gradients_trunc}

Formally, in the Fourier development of the Born rule in Eq.~\ref{eq:decomposition_pr} we can think of having the vanishing of the overall probability distribution $\Var(\operatorname{Pr}_{\boldsymbol{\theta}})\sim\mathcal{O}(\exp(-n))$ which necessarily means that all correlators vanish $\Var(\langle Z_{\boldsymbol{i}}\rangle_{\boldsymbol{\theta}})\sim\mathcal{O}(\exp(-n))$ $\forall \boldsymbol{i}\in 2^{[n]}$. We can also think of the case where the overall probability does not vanish  $\Var(\operatorname{Pr}_{\boldsymbol{\theta}})\sim\mathcal{O}(\poly(n)^{-1})$  but some of the correlators do,  $\Var(\langle Z_{\boldsymbol{i}}\rangle_{\boldsymbol{\theta}})\sim\mathcal{O}(\exp(-n))$, for just a subset of indices $\{\boldsymbol{i}\}\subset 2^{[n]}$.

It is important to note here, that when there is \emph{provable} absence of the vanishing phenomenon, one can actually simulate classically the minimization of the quantity at hand efficiently~\cite{cerezo_does_2025} as in practice the search-space is now polynomial instead of exponential in $n$. This motivates the introduction of surrogates of the correlators in the next section. 

In the below, we analyze how our various approximations can affect the variances of the resulting distributions. First, we focus on the truncation of the distributions themselves, using a specific circuit ansatz - specifically matchgate circuits~\cite{valiant_quantum_2012,terhal_classical_2002,jozsa_matchgates_2008}. Then, we combine the distribution truncation with an approximate estimation of the correlators within the truncated distribution - specifically with tensor networks.

Both of these examples have consequences for the $\ccl, \cq, \btheta^{*}_{\mathcal Q}, \btheta^{*}_{\cl}$ relationship as follows:
\begin{enumerate}
    \item Matchcircuits with truncated distributions: classical training should be done via e.g. a classical surrogate, $\ccl$ as $\btheta^{*}_{\mathcal Q}$ is not achievable due to trainability issues (vanishing gradients).
    \item Tensor network correlator surrogation: Obtaining optimal classical parameters, $\btheta^{*}_{\cl}$ from the classical surrogate, $\ccl$. The higher the bond dimension (which means we have a closer description to the quantum case), the higher the chances are of having vanishing correlators.
\end{enumerate}

\subsection{Truncated Distribution Variance of Matchcircuits}\label{subsection:lie_algebra}

We begin with the matchcircuit ansatz\"{e} as introduced in Sec.~\ref{sssec:matchcircuits_defn}. This algebraic structure exhibits a polynomial scaling of its dimension with respect to the number of qubits. In Díaz et al.~\cite{diaz_showcasing_2023}, the authors derive an analytic expression for the variance of the expectation values when the initial state and the observable are not necessarily supported by the matchgate algebra.

First, we derive an expression for the variance of  individual correlators in the QCBM. For matchgates, the variance is given by:
\begin{equation}\label{eq:exp_mc}
    \Var^{\rm MG}_{\btheta}[\langle Z_{\bi}\rangle] = \binom{n}{k}\binom{2n}{2k}^{-1} \text{ with } k=|\bi|.
\end{equation}
However, in our case, this variance (depicted in Fig.~\figref[a]{fig:variancealgebra}) alone is insufficient as it does not take into account the dependencies with other correlators in the final (truncated) distribution. For the case of $k$-order probability truncation (Eq.~\ref{eq:truncated_prob}), we next derive the full variance of this truncated distribution in Lemma~\ref{thm:var_mg}. This illustrates how the dependencies between correlators arises. For example, when we have a pair of correlators of size $k$ and $n-k$ respectively, the contribution to the final expression increases by a factor of two.

\begin{lemma}[Variance of Truncated Probability for Matchcircuits]\label{thm:var_mg}
For a matchcircuit, with set of gates as in Eq.~\ref{eq:MG_set} forming the correspondent Lie algebra Eq.~\ref{eq:MG_algebra},
\begin{equation}\label{eq:var_prob}
\Var_{\btheta}^{\rm MG}\left(\operatorname{Pr}_{\btheta}^{(k)}(\bx)\right) = \frac{1}{{2^{2n}}}\sum_{p=1}^{n-1}\mathbb{V}_p^{(k)}
\end{equation},
where,
\begin{equation}\label{eq:individual_contributions}
\operatorname{\mathbb{V}}_p^{(k)} =
    \begin{cases}
        \binom{n}{p}^2 \binom{2n}{2p}^{-1} & \text{if } p \leq k, \\[1em]
        2\binom{n}{p}^2 \binom{2n}{2p}^{-1} & \text{if } p \leq k \text{ and } n - p \leq k, \\[1em]
        0 & \text{otherwise.}
    \end{cases}
\end{equation}
\end{lemma}

We give the proof in Appendix~\ref{app:lie_modules}. The increase in variance seen in Eq.~\ref{eq:individual_contributions}, caused by the paired dependencies between correlators, is specific to the matchcircuit and highlights the importance of the inductive bias. Figure~\figref[b]{fig:variancealgebra} shows the variance obtained through Lemma~\ref{thm:var_mg} for different problem sizes. The variance exhibits an anti-concentration phenomenon as the 
$k$-order increases for a fixed problem size 
$n$. However, from Eq.~\ref{eq:individual_contributions}, we know that this variance is cumulative, and that both the individual contributions and their dependencies behave as $\binom{n}{p}\binom{2 n}{2 p}^{-1}$. Note that the square in $\binom{n}{p}$ appearing in $\mathbb{V}_p^{(k)}$ arises from the fact that there are $\binom{n}{p}$ terms of order $p$. Consequently, the contributions scale as $ \mathcal{O}\left(\tfrac{1}{\operatorname{poly}(n)}\right)$ when $p=\mathcal{O}(1)$ and as $\mathcal{O}\left(\tfrac{1}{\exp (n)}\right)$ when $p \propto n$.

Taken together, these results show that when sampling from a parameterized matchcircuit, not all correlators remain trainable, and these are not necessarily the higher-order ones. This observation may extend to more general circuit ansätze. The correlators that do not vanish can be efficiently simulated~\cite{cerezo_does_2025}, whereas those that do vanish have not been shown to provide any scaling advantage through quantum training compared to their classical surrogates. This further supports a train-on-classical, deploy-on-quantum strategy~\cite{recio-armengol_train_2025} for the QCBM, where the surrogate captures the relevant trainable structure, and the quantum circuit is ultimately used for sampling. In other words, we argue that the entire training process should be done via the $\ccl$ as $\btheta^{*}_{\mathcal Q}$ is not achievable due to trainability issues.

\subsection{Truncated Distribution Variance for Tensor Network Correlator Surrogates} \label{subsec:var_tensor_networks}

\begin{figure*}[htbp]
    \centering
    \resizebox{\linewidth}{!}{\input{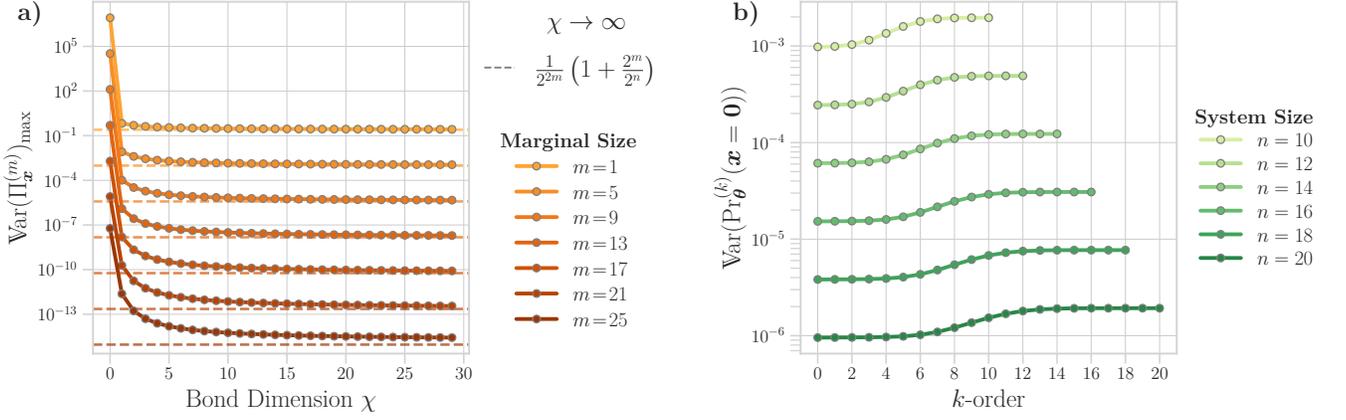}}
    \caption{\justifying Variances for the Random Matrix Product State regarding a) Marginal of size $m$ with maximum correlation, the variance of the marginal vanishes with $m$ as the bond dimension $\chi$ increases. b) Variance of the probability of measuring bitstring $\bzero$ as a function of the $k$-order truncation at a fixed bond dimension of $\chi=1000$. Analysis for other bitstrings can be found in Appendix~\ref{app:tn_proofs}.}
    \label{fig:TN_results}
\end{figure*}

The exponentially decaying variance of truncated distributions generated by matchcircuits indicates the difficulty of direct training on a quantum computed to get optimal parameters, $\btheta^{*}_{\mathcal Q}$. In this section, we compute the same variance when approximating correlators via classical surrogates, specifically tensor networks as introduced in Sec.~\ref{subsec:tensor_networks}. The resulting variance will depend on the bond dimension of the network, hence we can scale it accordingly to a regime where the variance does not vanish exponentially, and hence classical training is possible.

To compute the variance of the tensor network correlator, we will use the framework of Random Matrix Product States (MPS)~\cite{haferkamp_emergent_2021}. This framework is useful as it allows the calculation of correlation lengths between two subsystems \cite{haag_typical_2023} and considers the magic within the circuit~\cite{chen_magic_2024}. We will mainly use it to obtain an exact formula for the variance of the correlators and related quantities by enhancing the theory developed in ~\cite{haferkamp_emergent_2021}. Moreover, using this theorem  we derive other quantities of interest for PQCs with arbitrary number of qubits $n$ in Appendix~\ref{app:tn_proofs}, for example the Renyi-2 entropy. An RMPS is formally defined as follows.

\begin{definition}[Random Matrix Product State (RMPS)~\cite{haferkamp_emergent_2021}]\label{defn:random_mps}
    Given an $n$-qubit state $\ket{\psi}$ described as in Eq.~\ref{eq:mps} with local dimension $\ell$ and bond dimension $\chi$, an RMPS is such that the embedded unitary matrices $U_1, U_2, \dots\, ,U_n \in U(\ell\chi)$ are drawn i.i.d. randomly from the Haar measure, $\mu_{\ell,\chi}$.
\end{definition}

Definition~\ref{defn:random_mps} then allows one to obtain an expression for the variance of the individual correlators:
\begin{equation}
    \Var^{\rm  RMPS}_{\btheta} [\langle Z_{\bi}\rangle] = \left(
    \begin{array}{ll}
    1 & 1
    \end{array}
    \right) T_{\bj(\bi)}\binom{1}{1}.
\end{equation}
Where $T_{\bj(\bi)} = \prod_{r=1}^n T_{j_r}$ with $j_r = \1$ if $r\notin \bi$ and  $j_r = Z$ if $r\in\bi$. The matrices are then defined as $T_{\mathds{1}}=\left(\begin{array}{ll} 1 & \eta_\chi \\ 0 & \eta_\ell \end{array}\right)$ and $T_Z = \frac{1}{\ell(\ell^2\chi^2-1)}\left(\begin{array}{ll} -1 & -\chi \\ \ell\chi & \ell\chi^2 \end{array}\right)$. As seen, the final value of the variance is subject to the correlator index ${\bi}$. Favorably, this result can be used to calculate a closed equation for the variance of a similar informative quantity, the marginal of size $m$ and $k$-order equal to $m$ $\Pi^{(m)}_{\bx}$, 

\begin{equation} \label{eqn:tensor_network_marginal_distribution}
\Pi^{(m)}_{{\bx}} = \frac{1}{2^m} \sum^{m}_{p=0} \sum_{\substack{{\bi}\subseteq\{1,...,m\}\\|{\bi}|=p}}\!\!\!\! (-1)^{\sum_{i\in {\bi}} x_i}\left\langle Z_{\bi}\right\rangle_{\btheta}.
\end{equation}

These marginals represent the reduced distributions obtained by ignoring some bits. In our framework, they correspond to truncating the Fourier expansion to subsets fully contained in the chosen coordinates. Note, these are distinct  Hence, marginals are projections of the global distribution onto smaller subsets of variables, and in the Fourier picture. Formally, we can relate Eq.~\ref{eqn:tensor_network_marginal_distribution} to the truncated distribution as keeping only the correlators on the chosen variables. The variance of the marginals reads as follows,

\begin{equation}
\begin{aligned}\label{eq:tn_var}
\Var^{\rm RMPS}_{\btheta}(\Pi^{(m)}_{{\bx}} )_{\max} &= 
\frac{1}{1+\chi^2} \left(\frac{(2 \chi-1)(1+\chi)}{(2(1-4 \chi))^2}\right)^m \\
&\quad \times \Bigg[2^m(1+\chi)(1+2 \chi) \\ &
+ 2^n(\chi-1)(2 \chi-1) \left(\frac{\chi^2-1}{4 \chi^2-1}\right)^{n-m}\Bigg],
\end{aligned}
\end{equation}
and when considering a large amount of computational resources, which translates into a high bond dimension we find the limit
\begin{equation}
\lim_{\chi\rightarrow \infty} \Var^{\rm RMPS}(\Pi^{(m)}_{{\bx}} )_{\max} = \frac{1}{2^{2m}}\left(1+ \frac{2^m}{2^n} \right).
\end{equation}
This result shows that when a MPS is constructed by random matrices, as the bond dimension increases, the vanishing of the marginals is dominated by the size of the \emph{marginal} $m$ and not that of the system size $n$. This behavior is depicted in Fig.~\figref[a)]{fig:TN_results}.

For completeness, we obtain the contribution of each $k$-order truncation of the probability of measuring bitstring $\bx = \bzero \equiv 00\dots0$ for different numbers of qubits within the RMPS,
\begin{equation}
\Var\left(\operatorname{Pr} ^{(k)}_{\btheta}(\bx = \bzero)\right) = \frac{1}{2^n} \sum_{p=0}^k \sum_{\bi \in \{\1,Z\}^{\otimes n}}\!\!\!\!\!\! \left(\begin{array}{ll}
1 & 1
\end{array}\right) T_{\bi}\binom{1}{1}.
\end{equation} 
As seen in Fig.~\figref[b)]{fig:TN_results} this probability does not vanish as the order increases because it aggregates contributions from all previous terms. Despite this, the same does not apply for the individual contributions, as seen in the marginals case where vanishing occurs. Moreover, for values of $k$ close to $n/2$, the accumulation is more pronounced due to the combinatorial scaling of the number of contribution terms as $\binom{n}{k}$

\section{Pauli Propagating Surrogates of IQP Circuits}\label{sec:PPS}
We have discussed above the use of tensor network surrogates, which motivates the optimal classical parameters $\btheta^*_{\cl}$ by using the coefficients of the classical surrogate $\ccl$. Here, we give a second example using Pauli Propagation to surrogate IQP circuits, again training to find optimal classical parameters $\btheta^*_{\cl}$. However, in doing so, we present a closed form for the Pauli propagation for the IQP circuit where we leverage the symmetries within the circuit structure, which may be of independent interest. One could also apply Pauli propagation more generally to generic circuit ansatz\"{e}.

As introduced above, the expectation value of a general observable $O$ in an IQP circuit is given by
\begin{equation}
    \langle O \rangle_{\text{IQP}}=\Tr\Bigl( H^{\otimes n} U_c^\dagger H^{\otimes n}  O H^{\otimes n} U_c H^{\otimes n}\ket{\bzero}\!\!\bra{\bzero}\Bigr).
\end{equation}
After using the relations
$
H\ket{0}=\ket{+},\quad HZH=X,\quad HXH=Z,
$
the truncated probability $\Pr^{(k)}_{\btheta}$
can be rewritten as
\begin{equation}
\operatorname{Pr}_{\boldsymbol{\theta}}^{(k)}\!(\bx)
\!=\!\!\frac{1}{2^n}\!\! \sum^{k}_{p=0} \sum_{\substack{{\bi}\subseteq\{1,...,n\}\\|{\bi}|=p}}\!\!\!\!\!\!\! (-1)^{\sum\limits_{i\in {\bi}} x_i}\!\!\Tr\Bigl( U_c^\dagger\!\sum_{\bi} \!X_{\bi}U_c\,\ket{+}\!\!\bra{+}\Bigr).
\end{equation}
Here the sum $\sum_{\bi}X_{\bi}$ is again taken over all Pauli strings (or correlators) of weight up to $k$. The circuit $H^{\otimes n} U_c H^{\otimes n}$ is an IQP circuit characterized by a unitary of $g$ individual gates, $ U_c = U_g \cdots U_2 U_1,$ where we have $[U_i,U_j]=0, \forall i, j \in \{1, \dots g\}$. This complete commutativity fixes the propagation of any Pauli observable regardless of the order of the gates. Consequently, thanks to Pauli propagation it is clear that the truncated probability in Eq.~\ref{eq:empirical_corr} appears from the interactions among qubits in the circuit as trigonometric functions. These weights are determined by the rotation parameters associated with the entangling and single–qubit gates and can be interpreted as the contributions of the corresponding Pauli correlators.

The surrogate approach relies on propagating an observable $X_{\boldsymbol{i}}$, which acts as $X$ on a given subset $\boldsymbol{i}\subset\{1,\dots,n\}$ (with $|\boldsymbol{i}|=k$) and as the identity elsewhere, through the circuit. The adjoint action of a rotation gate with generator $P_{\boldsymbol{j}}$ is given by
\begin{equation} R_{P_{\boldsymbol{j}}}^\dagger(\theta)\,X_{\boldsymbol{i}}\,R_{P_{\boldsymbol{j}}}(\theta)
= \cos\theta\,X_{\boldsymbol{i}} + i\sin\theta\,(P_{\boldsymbol{j}}X_{\boldsymbol{i}}).
\end{equation}
Thus, if $X_{\boldsymbol{i}}$ anticommutes with $P_{\boldsymbol{j}}$, the gate “flips” the observable; if they commute, it remains unchanged. Propagating $X_{\boldsymbol{i}}$ through the entire IQP circuit yields
\begin{equation}
U_c^\dagger\,X_{\boldsymbol{i}}\,U_c = \prod_{j\in S(\boldsymbol{i})}\Bigl[\cos\theta_j + i\sin\theta_j\,P_{\boldsymbol{j}}\Bigr]\,X_{\boldsymbol{i}},
\end{equation}
where $ S(\boldsymbol{i}) = \{\, j \,:\, \{X_{\boldsymbol{i}},P_{\boldsymbol{j}}\}=0 \}
$ is the set of gate indices for which the generator anticommutes with $X_{\boldsymbol{i}}$. Expanding the product over the $M=|S(\boldsymbol{i})|$ gates, we obtain
\begin{equation}
U_c^\dagger\,X_{\boldsymbol{i}}\,U_c = \sum_{\br\in\{0,1\}^{M}} \alpha_{\boldsymbol{i}}(\br)\,\Bigl(\prod_{j\in S(\boldsymbol{i})}P_{\boldsymbol{j}}^{\,r_j}\Bigr)X_{\boldsymbol{i}},
\end{equation}
with
\begin{equation}
\alpha_{\boldsymbol{i}}(\br)= \prod_{j\in S(\boldsymbol{i})}\Bigl[i\sin\theta_j\Bigr]^{r_j}\Bigl[\cos\theta_j\Bigr]^{\,1-r_j},
\end{equation}
carrying the trigonometric information of the development.

In many practical IQP circuits the generators are chosen to be $Z$--like. In that case each $P_{\boldsymbol{j}}$ is replaced by its corresponding $Z$--operator. Since on any qubit $q$ we have $Z\,X = iY,$ a single sine factor converts $X$ into $Y$. However, if several gates act on the same qubit the sine factors may appear in pairs; when an even number of flips occur on a qubit the net effect is to leave $X$ unchanged (recalling that $Z^2 = \1$). Since the support of $X_{\boldsymbol{i}}$ is the set of qubits on which it acts non-trivially ($\boldsymbol{i}$), for each qubit $q\in \boldsymbol{i}$ we define a local parity function
\begin{equation}
\ell(q,\br) = \sum_{\substack{j\in S(\boldsymbol{i}) \\ q\in \mathrm{supp}(P_{\boldsymbol{j}})}} r_j \quad (\mathrm{mod}\,2).
\end{equation}
We then introduce the indicator function
\begin{equation}
f_{\boldsymbol{i}}(\br) = \prod_{q\in \boldsymbol{i}} \frac{1+(-1)^{\ell(q,\br)}}{2},
\end{equation}
which is equal to one if and only if the total number of sine contributions ($Z$--flips) on each qubit in $\boldsymbol{i}$ is even. Thus, the expectation value measured in the $X$ basis is given by
\begin{equation}\label{eq:surr_iqp}
\langle +|^{\otimes n}U_c^\dagger\,X_{\boldsymbol{i}}\,U_c|+\rangle^{\otimes n} = \sum_{\br\in\{0,1\}^{M}} f_{\boldsymbol{i}}(\br)\, c_{\boldsymbol{i}}(\br).
\end{equation}

 Although additional gates add extra cosine or sine factors, the structure of the expansion, together with the parity condition imposed by $f_{\boldsymbol{i}}(\br)$, guarantees that only those terms that preserve the original $k$--order of $X_{\boldsymbol{i}}$ survive. In other words, while extra gates contribute additional parameters, their sine contributions always appear in even numbers on each qubit, so that the final observable retains its original support.

Eq.~\ref{eq:surr_iqp} gives the recipe for building a surrogate. The computational cost of simulating the surrogate increases exponentially with $M$, which is the number of gates that anticommute with the observable. This is because the surrogate expansion is over all binary strings of length $M$, so that there are $2^M$ terms to compute in the worst case. In practice, if many gates contribute; that is, if many of the gates in the circuit anticommute with the observable, the number of terms in the sum grows very rapidly, causing the simulation to be intractable.

\begin{figure*}[htbp]
    \centering
    \includegraphics[width=\textwidth]{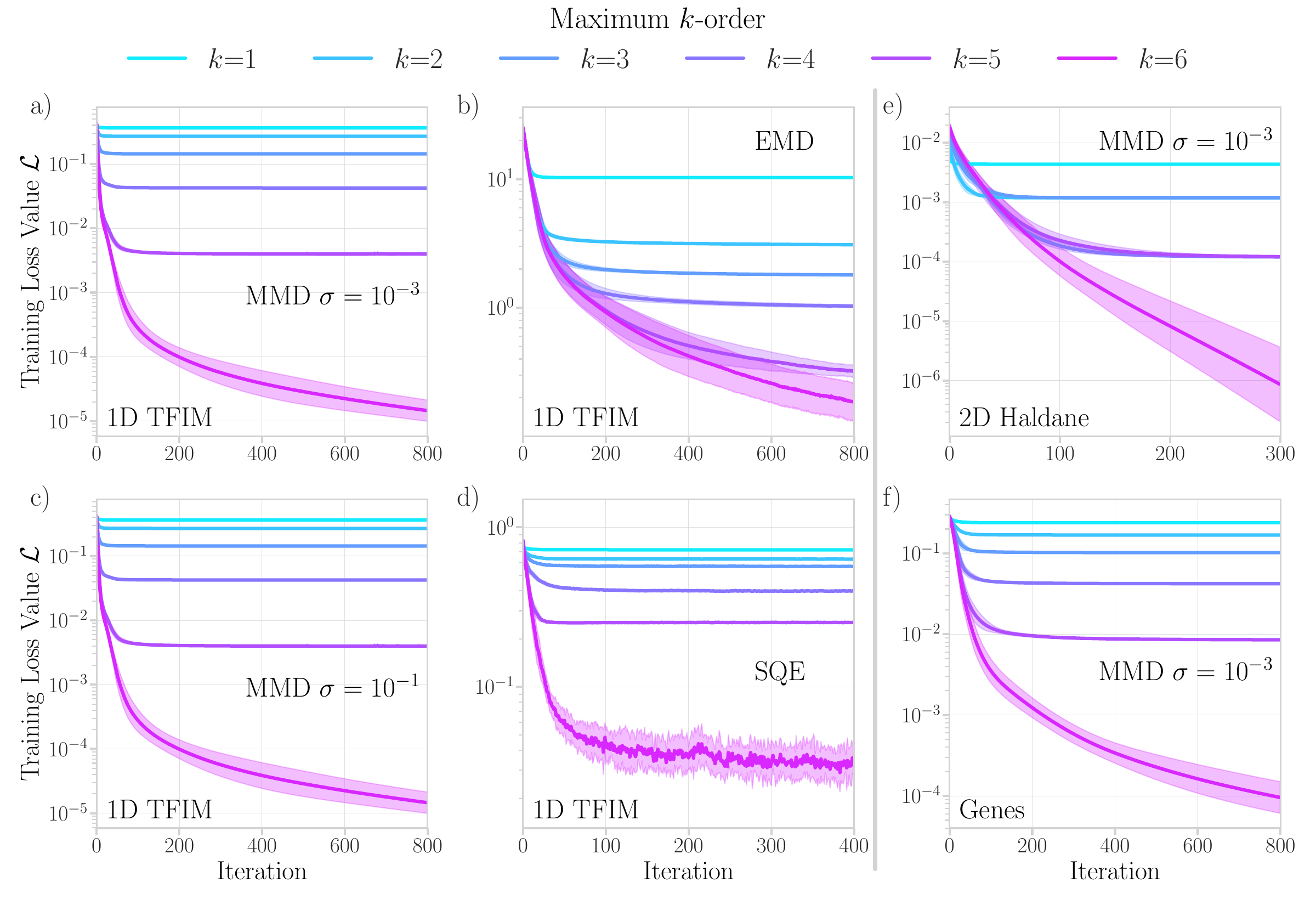}
    \caption{\justifying \textbf{Loss function discrepancy during training for different $k$-order correlation truncations with $n=6$.} The circuit ansatz is that of a strongly entangling layered circuit~\cite{schuld_circuit-centric_2020}, with six layers for full connectivity among qubits and a sufficient number of parameters. Panels a)-d) use the one-dimensional Transverse Field Ising Model ground state probability dataset; panel e)  uses the two-dimensional Haldane ground state probability dataset; panel f) uses the gene activation probability dataset. The Maximum Mean Discrepancy Loss (MMD) with bandwidth hyperparameter $\sigma = 10^{-3}$ is used for experiments a), e) and f). The MMD with $\sigma=10^{-1}$ is used in c). Panel b) uses the Earth Movers Distance (EMD) loss and panel d) the Squared Euclidean Loss (SQE). the shaded region represents the standard deviation error between ten different random initializations.} 
    \label{fig:truncation}
\end{figure*}

The lower the $k$-order in the truncated probability $\Pr^{(k)}_{\btheta}$, the fewer gates in the circuit anticommute with the observable. For instance, if the observable acts on a small number of qubits, it is less likely to anticommute with many gates. An approach to approximating this probability is to simulate only those terms in the surrogate expansion whose corresponding binary vector $\br$ has a Hamming weight $|\br|$ up to a fixed threshold $h_{\rm max}$ in number of flips. This number of flips plays the equivalent to the weight truncation in the vanilla PPS and is more convenient in this IQP architecture. In the full expansion, one would need to sum over $2^M$ binary strings. By restricting the sum to only those $\br\in\{0,1\}^M$ with Hamming weight at most $h$, the number of terms reduces to $\sum_{w=0}^{h}\binom{M}{w}$.
This is a significant reduction when $h$ is small compared to $M$, as the number of terms grows polynomially rather than exponentially. In effect, this truncation limits the number of gate-induced flips considered in the propagation and thus reduces the computational complexity.

The approximated truncated probability $\widetilde{\mathrm{Pr}}^{(k)}_{\btheta}$ for a $Z$-like only IQP circuit is then given by:
\begin{equation} \label{eqn:iqp_pps_main_result}
    \widetilde{\mathrm{Pr}}^{(k)}_{\btheta}(\bx) = \frac{1}{2^n}\sum_{p=0}^k \sum_{\substack{{\bi}\\|{\bi}|=p}} \sum_{w=0}^h \sum_{\substack{{\br}\\|{\br}|=w}} f_{\boldsymbol{i}}(\br)\, \alpha_{\boldsymbol{i}}(\br).
\end{equation}

\section{Numerical Results}\label{sec:numerics}
\begin{figure*}[htp]
\resizebox{\linewidth}{!}{\input{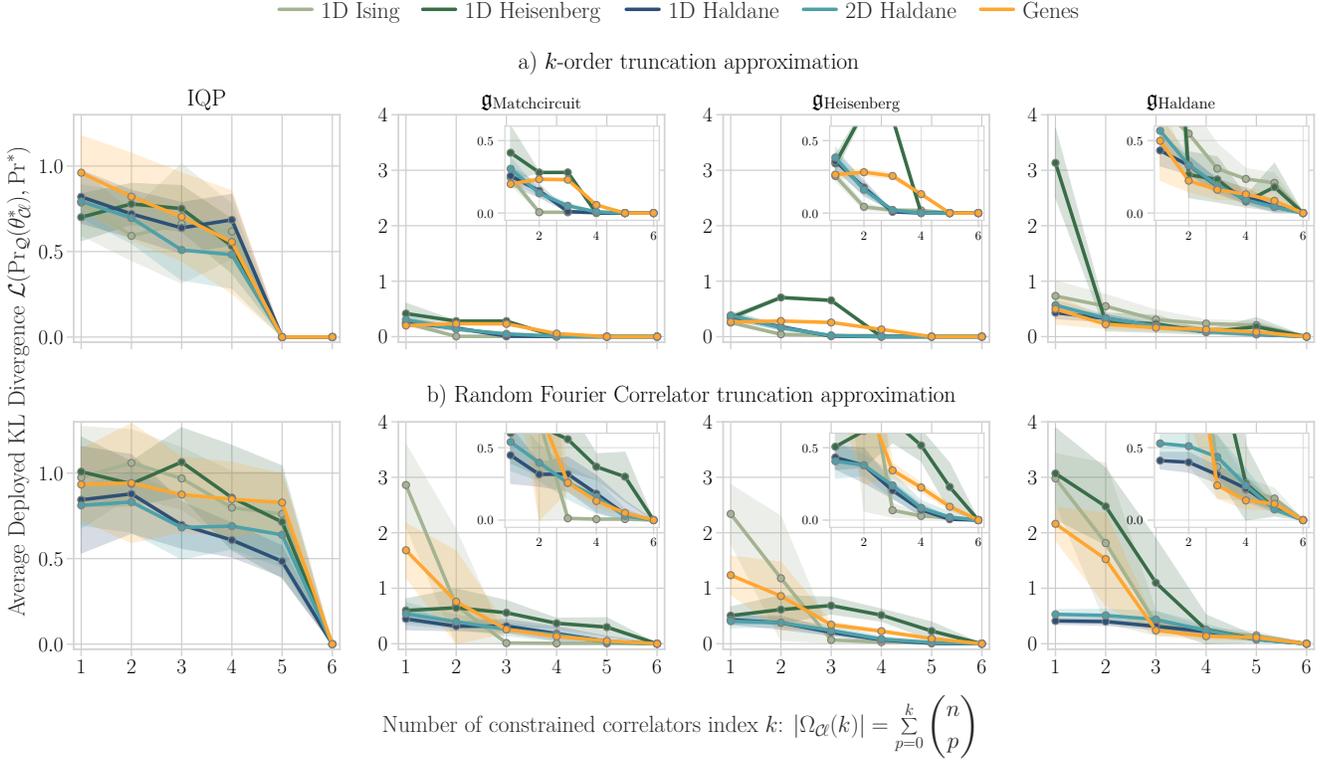}}
\caption{\label{fig:k_order_generalization}\justifying\textbf{KL-divergence between deployed circuit and dataset}  The parameters are transferred from surrogate trained models which constrains parameters up to the $k$-order truncation approximation. Training is done for 100 iterations and 100 different experiments to produce the average point and standard deviation shaded area. The PQC ansatz draws 40 gates from a) the matchcircuit algebra, and b) the Haldane one-dimensional chain.}
\end{figure*}

\subsection{Loss functions}
When training with the approximated truncated probability distribution, only the losses that have a positive defined domain are possible. This means that losses which use for example logarithms . Among this valid set we have for two distributions $p$ and $q$,
\begin{itemize}
\item Maximum mean discrepancy (MMD) $\mathrm{MMD}^2(p,q) \;=\;
        \mathbb{E}_{\boldsymbol{x},\,\boldsymbol{x}' \sim p}\bigl[K(\boldsymbol{x}, \boldsymbol{x}')\bigr] 
    -\; 2\,\mathbb{E}_{\boldsymbol{x} \sim p,\,\boldsymbol{y} \sim q}\bigl[K(\boldsymbol{x}, \boldsymbol{y})\bigr]
    +\; \mathbb{E}_{\boldsymbol{y},\,\boldsymbol{y}' \sim q}\bigl[K(\boldsymbol{y}, \boldsymbol{y}')\bigr].$ with $K \in \mathbb{R}^{2^n \times 2^n}$ a kernel matrix with entries $K({\boldsymbol{x},\boldsymbol{y}})$ which is typically taken to be a Gaussian kernel which controls the proximity between bitstrings with the bandwidth $\sigma$.
\item Squared Euclidean loss (SQE) $
\ell_2^2(p, q) = \sum_{\boldsymbol{x}} \big(p(\boldsymbol{x}) - q(\boldsymbol{x})\big)^2$.
\item Earth movers distance (EMD) $\operatorname{EMD}(p, q) = \sum_{\boldsymbol{x}} \big|\mathrm{CDF}_p(\boldsymbol{x}) - \mathrm{CDF}_q(\boldsymbol{x})\big|$, with CDF the cumulative distribution function.
\end{itemize}
When deploying, we can use any valid function, as we do with the Kullbach-Leibler divergence,  ${\mathrm{KL}}(\operatorname{Pr}_{\mathcal H} \,\|\, \operatorname{Pr}_{\btheta})
    \;=\;
    \sum_{\boldsymbol{x}}\operatorname{Pr}_{\mathcal H}(\boldsymbol{x})
    \,\log\!\left[\frac{\operatorname{Pr}_{\mathcal H}(\boldsymbol{x})}{\operatorname{Pr}_{\btheta}(\boldsymbol{x})}\right]$.
Further details are available in Appendix~\ref{app:losses}. In Appendix~\ref{app:anova} we explore the possibility of having MMD losses with pre-designed kernels.

\subsection{Training with $k$-order truncation}\label{subsec:general_results}
Given our two truncation approximation methods, we begin with $k$-order truncation approximation, as this enables a structured comparison with the datasets in Sec.~\ref{sec:data}. Fig.~\ref{fig:truncation} shows how the truncation in Eq.~\ref{eq:truncated_prob} impacts the training of the QCBM. We primarily use the MMD and the EMD to train, though our methods are applicable to any such loss function. . These loss functions admit negative function domains for the datasets considered. Across all numerical experiments, the learning rate is fixed and identical to ensure a fair comparison. In the left panels \figref[a)]{fig:truncation}-\figref[d)]{fig:truncation}, we use the 1D TFIM described in Eq.~\ref{eq:ising_1d} which has correlation values constant in order of magnitude throughout the whole support. Between the MMD loss with bandwidth $\sigma = 10^{-3}$ Fig.~\figref[a)]{fig:truncation}, and with bandwidth  $\sigma = 10^{-1}$ Fig.~\figref[c)]{fig:truncation}, there is no clearly manifested difference when it comes to the actual values the loss takes given the same initializations, in both cases one can distinguish between the loss histories through iterations for the different $k$-order truncations. This distinguishability is also present for the EMD and SQE losses, although for the EMD loss one needs more iterations to be able to distinguish between orders, for $k=5$ and $k=6$ this is more obvious at approximately 700 iterations. 

Panels \figref[e)]{fig:truncation} and \figref[f)]{fig:truncation} now use different datasets but the same MMD loss with bandwidth $\sigma = 10^{-3}$. As seen in Fig.~\figref[c)]{fig:correlations_hamiltonians} the 2D Haldane dataset has only $k=2$ and $k=4$ orders; this explains why in \figref[e)]{fig:truncation}, the loss history for $k=3$ merges with that one of $k=2$, the same applies for $k=5$ and $k=4$. Nevertheless, when $k=6$ the loss decreases considerably. Here we have only added the sixth order term to convey a symmetry condition that allows the model to fit correctly. Panel \figref[f)]{fig:truncation}, shows the case for the gene activation target dataset, where the correlations, similarly to the 1D TFIM, decrease polynomially with the $k$-order. This explains the similarity between panels \figref[a)]{fig:truncation} and \figref[f)]{fig:truncation}.
\subsection{Generalization}\label{subsec:generalization}
Generalization typically refers to the performance of a model on unseen data. In our setting, this notion is adapted to quantify performance with respect to unconstrained correlators. It describes how an under-determined parameterized distribution becomes restricted by the inductive bias introduced by the chosen model. In essence, we analyze $\cq(\theta^*_{\cl})$, which formalizes the transfer of parameters from the classical model to the quantum model.

This corresponds to a realistic scenario in which the full probability distribution is evaluated using the angles learned from a truncated model. It mimics the practical usual setting where training is performed with a polynomial size set of correlators and sampling is later carried with the full exponential spectrum.

In an ideal scenario the unconstrained correlators have useful information~\cite{xu_theory_2020,ethayarajh_understanding_2022} such that when the model is deployed there is an advantage. In this way the randomness is leveraged in favor of practicality. This generalization is achieved with a correct choice of the inductive bias.

We denote by $\operatorname{Pr_{\mathcal{Q}}}$ the probability distribution where all correlators are considered. From Theorem~\ref{thm:ideal_deq} we know that the advantage mainly relies on the underlying distribution complexity and the generative capacity of the unconstrained QCBM. 

\subsubsection{Inductive bias comparison}
\begin{figure*}[htp]
\resizebox{\linewidth}{!}{\input{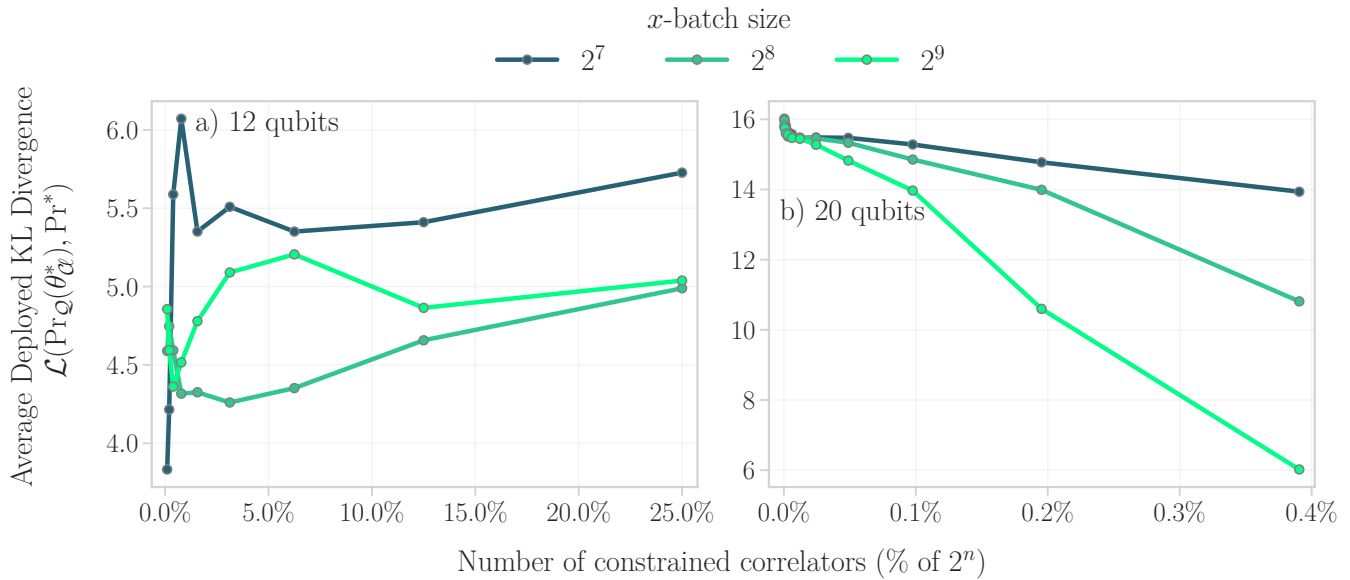}}
\caption{\label{fig:batches}\justifying \textbf{Average deployed KL divergence} representing evaluation at inference time, as a function of the bitstring batch size for different fraction of correlators fixed at the beginning of training. Parameters are initialized with the technique in~\cite{recio-armengol_train_2025}}
\end{figure*}
In Fig.~\ref{fig:k_order_generalization} we show the combination of the different ansatzë and datasets when trained up to certain $k$-order truncation approximation (Def.~\ref{defn:k-order_truncation}) in the top row, and with the RFC sampling truncation approximation (Def.~\ref{defn:random-sampling_truncation}) in the bottom row. This illustrative study is done for 6 qubits to show the impact that the inductive bias has in the unconstrained correlators set.

For the RFC truncation approximation we choose a constrained correlator set size $|\Omega_{\cl}|$ equal to that of the $k$-order truncation approximation. This means that the size increases as $\sum_{p=0}^k\binom{n}{k}$ for both cases. 

When comparing along the columns, which implies same ansatz, both approximation methods behave similarly. In the IQP case, it is only when we consider order 5 or 6 that the deployed loss reduces substantially. Furthermore, there is no difference between datasets throughout the different $k$ values as averages and standard deviation curves collide with each other. The KL divergence value remains below 1 and hence comparable to the other cases,  but as we are selecting the commuting gates at random there could be a further refinement for better performance.

In the case of specific algebras, we observe an overall improved performance when using the $k$-order truncation approximation compared to the RFC truncation approximation. This indicates that lower-order correlators in the data encode information relevant to higher orders, which the circuit models can effectively exploit. For both the Matchcircuit and Heisenberg ansätze, the TFIM dataset yields a KL divergence notably closer to zero than other datasets, specifically for $k=2$ in the $k$-order truncation and $k=3$ in the RFC truncation approximation. This behavior is consistent with the relative simplicity of the one-dimensional TFIM, which is known to be classically learnable~\cite{lewis_improved_2024}. Notably, for the Haldane algebra the one-dimensional and two-dimensional Haldane datasets have a lower KL divergence value with respect to the other datasets for low values of $|\Omega_{\cl}|$ in the RFC truncation aproximation.

For the genes dataset, the performance throughout the different ansatzë and truncation approximation is similar. Nevertheless, when compared with the Heisenberg data (which is physically motivated) the KL divergence is lower for the genes dataset, even when using a Heisenberg chain circuit.

\subsubsection{Scaling analysis}
Extending beyond the previous example, we assess the deployment on 12 and 20 qubit IQP systems (Fig.~\ref{fig:batches}) to examine how performance scales with system size. In this case we do use an $\boldsymbol{x}$-batch, which lowers the number of bitstrings considered for each training step for the ground state of the TFIM in the same number of sites as qubits. In these large systems case we use the Pauli Propagation surrogate with low number of flips $h_{\rm max} = 2$. As done in Ref.~\cite{recio-armengol_train_2025} the initial parameters are given by the existing correlations in the data, for example if a  $Z$-multiqubit gate applies in qubits 1,3,6 then we calculate the correlator from data and make $\theta_{1,3,6}^{\rm init} \equiv \langle S^*_{1,3,6}\rangle$. This initialization ensures convergence in the training curves, in difference to randomly initialization where there is no convergence.

\begin{figure*}[htbp]
    \centering
    \resizebox{\linewidth}{!}{\input{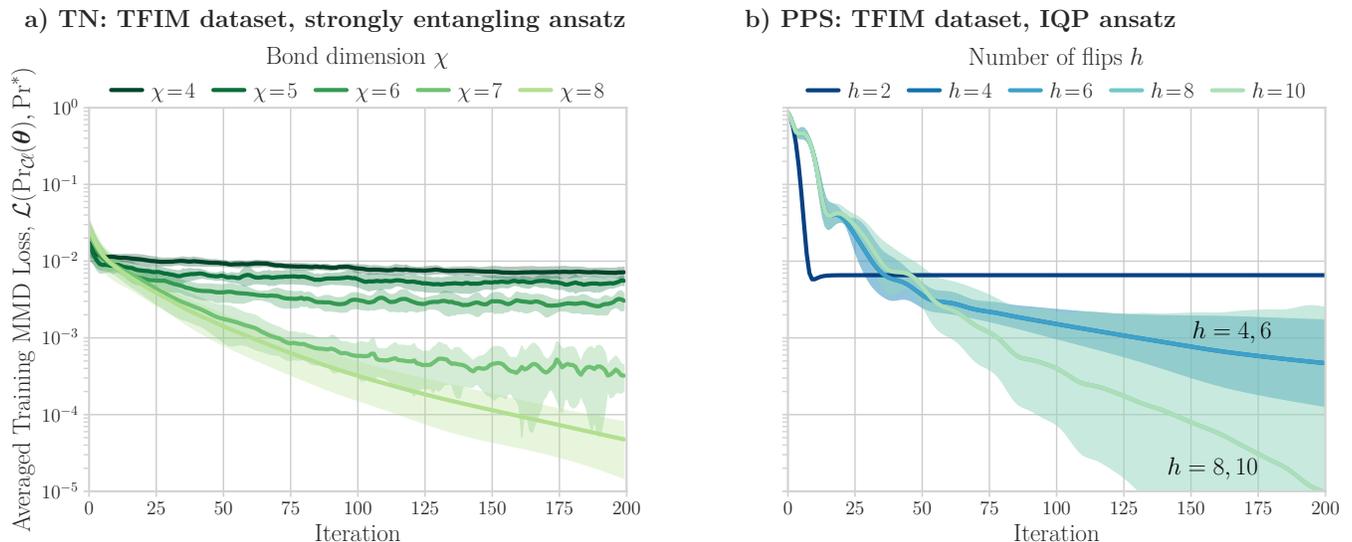}}
    \caption{\justifying \textbf{Surrogate training curves} averaged over 10 different parameter initializations.}
    \label{fig:surrogate_training_curves}
\end{figure*}

In this scenario we assume we have only access to low polynomial number of constrained correlators, therefore we limit the number of correlators to $25\%$ in the 12 qubits case, and to $0.4\%$ in the 20 qubits case. For the 12 qubits system there is no apparent improvement when increasing the constraint correlator size. For the 20 qubits system, even if the percentage is low there is a clear decrease meaning an improvement in performance at inference time. Nonetheless, the KL divergence for the 12 qubits is lower than that of the 20 qubits. One would expect that with an increased $h_{\rm max}$ the discrepancies decrease, implying a potential improved tendency.

\subsection{Surrogating Correlators}\label{sec:train_surr}
When surrogating the QCBM instance, the new surrogated analytical expression for the overall loss and the individual correlators are different from the quantum deployed case, see Fig.~\ref{fig:approximations}. To isolate the correlator surrogate from the truncation approximation we consider the full set of correlators in the simple $n=6$ case. As there are no limits in this difference, the classical training has the possibility of being better than that of the quantum machine as the implicit bias is also different. 

\subsubsection{Overall surrogated loss training}
Fig.~\ref{fig:surrogate_training_curves} shows the training for the two different surrogates with the same data and different ansatzë. In the strongly entangling ansatz case with TN surrogate the training varies in a structured manner as the bond dimension increases. Since the ground state data is obtained via the DMRG method (see Appendix~\ref{app:data_obtention}), the necessary bond dimension of the ground state with six sites is equal to 8. Coincidentally, the training turns smooth and converges consistently when for the surrogate we also have $\chi=8.$

In the Pauli Propagation case, the structured improvement is not that evident as for different number of flips there is overlap. This is because the trigonometric equations for each correlator are the same. Also, even if the number of flips is not the same as what the quantum model would need to have an exact description, the training for $h=8,10$ does converge. In this case, although the analytical expressions of the surrogate is not exact, the surrogate model is expressive enough to be able to decrease the training KL divergence significantly.

\subsubsection{Individual surrogated correlator analysis}
\begin{figure*}[t]
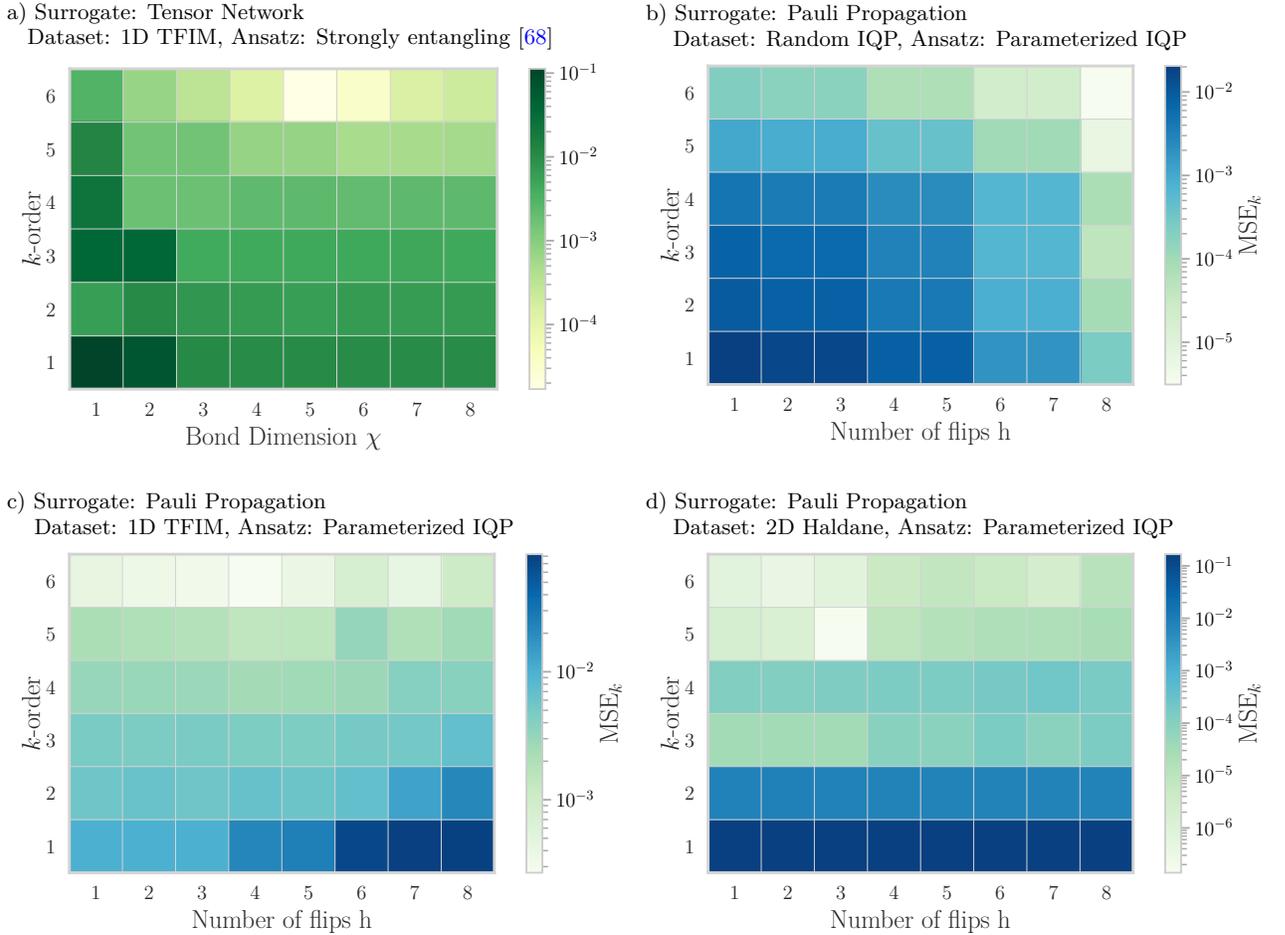

    \centering
    \begin{adjustbox}{width=\textwidth}
          \newlength{\panelW}
  \newlength{\panelV}
  \newlength{\colseptop}
  \newlength{\colsepbot}
  \newlength{\rowsep}
  \setlength{\panelW}{0.48\linewidth} 
  \setlength{\panelV}{0.5\linewidth} 
  \setlength{\colseptop}{0.03\linewidth} 
  \setlength{\colsepbot}{0.01\linewidth} 
  \setlength{\rowsep}{3.7em}         

  \begin{tikzpicture}[every node/.style={inner sep=0, outer sep=0}]

    \node[anchor=north west] (A) at (0,0)
      {\resizebox{\panelW}{!}{\input{figures/fig12a_tn_training_heatmap.pgf}}};

    \node[anchor=south west, yshift=0.6ex] at (A.north west)
      {\parbox[t]{\panelW}{\raggedright
        a) Surrogate: Tensor Network\\
        \hspace{.9em}Dataset: 1D TFIM, Ansatz: Strongly entangling~\cite{schuld_circuit-centric_2020}\hspace{0.3em}
      }};

    \node[anchor=north west] (B) at ($(A.north east)+(\colseptop,0)$)
      {\resizebox{\panelV}{!}{\input{figures/fig12b_mse_h_p_untrained.pgf}}};

    \node[anchor=south west, yshift=0.6ex] at (B.north west)
      {\parbox[t]{\panelV}{\raggedright
        b) Surrogate: Pauli Propagation\\
        \hspace{1.2em}Dataset: Random IQP, Ansatz: Parameterized IQP\hspace{0.3em}
      }};

    \node[anchor=north west] (C) at ($(A.south west)+(0,-\rowsep)$)
      {\resizebox{\panelV}{!}{\input{figures/fig12c_mse_h_p_trained_1d_ising.pgf}}};

    \node[anchor=south west, yshift=0.6ex] at (C.north west)
      {\parbox[t]{\panelV}{\raggedright
        c) Surrogate: Pauli Propagation\\
        \hspace{1.2em}Dataset: 1D TFIM, Ansatz: Parameterized IQP\hspace{0.3em}
      }};

    \node[anchor=north west] (D) at ($(C.north east)+(\colsepbot,0)$)
      {\resizebox{\panelV}{!}{\input{figures/fig12d_mse_h_p_trained_2d_haldane.pgf}}};

    \node[anchor=south west, yshift=0.6ex] at (D.north west)
      {\parbox[t]{\panelV}{\raggedright
        d) Surrogate: Pauli Propagation\\
        \hspace{1.2em}Dataset: 2D Haldane, Ansatz: Parameterized IQP\hspace{0.3em}
      }};

  \end{tikzpicture}
    \end{adjustbox}
    \caption{\justifying Training with the TN surrogate for different bond dimensions. a) Averaged MMD loss over 10 runs with the standard deviation shown in the shaded region. b) MSE$_k$ from Eq.~\ref{eq:MSE_order}. \textbf{MSE$_k$ for parameterized IQP circuit with difficult to learn random IQP circuit output as a target}. Both IQP circuits, target and trained, have a gate depth of $g=20$ . Each data point represents the average MSE$_k$ over different circuit initializations and different set of gates where we perform 20 iterations of training such that convergence is assured. The number of flips $h$ are with respect to the IQP-specific Pauli propagation scheme, and the $k$-order is the maximum truncation order in Eq.~\ref{eq:truncated_prob}. \justifying\textbf{MSE$_k$ for trained circuits} $g=20$ and averaged over 20 runs. a) $n=6$ Ising chain b) 2D Haldane Lattice $n_x=2$ and $n_y=3$.}
    \label{fig:combined_grid}
\end{figure*}

To now compare the effect of the truncation in the truncation level of individual $k$-order sets, we introduce the Mean Squared Error (MSE) of order $k$

\begin{equation}\label{eq:MSE_order}
\operatorname{MSE}_k = \frac{1}{\binom{n}{k}} \sum_{|\bi|=k} E_{\bi}^2\text{ where } E_{\bi} = \langle X_{\boldsymbol i}\rangle -\langle\widetilde{  X_{\boldsymbol i}}\rangle. 
\end{equation}
This serves as a metric to measure wether if each set of $k$-order correlators, trains better or worse in average when a higher  is considered within the surrogate. Each value in Fig.~\ref{fig:combined_grid}
shows the final MSE$_k$ when the truncation approximation is just taken up to the fixed $k$ order such that no influence of higher orders impacts the order of interest.

The MSE$_k$ for the TN with 1D TFIM data and strongly ansatz is shown in ~\figref[a)]{fig:combined_grid}. As the correlator order of the truncation increases, the individual MSE$_k$ of the last order within the truncation does not necessarily improve with the bond dimension. This has a strong dependence on the data.

The problem shown in Fig.~\figref[b)]{fig:combined_grid} is a coarse grained version of the identity problem~\cite{hinsche_learnability_2021} as the objective is for the IQP circuit to learn itself, this means that the target probability distribution is given by random instances of a same constructed parameterized IQP circuit. Here the truncation is taken cumulatively up to the $k$-order as in Eq.~\ref{eq:truncated_prob} to isolate the effect of higher orders. Systematically for each order we see that as the number of flips increases, the ${\rm MSE}_k$ improves. This is interpreted as the fact that the IQP target circuit is complex enough for the correlators to benefit from a larger trigonometric development in the parameters.

On the contrary, Fig.~\figref[c)]{fig:combined_grid} shows the same analysis of MSE$_k$ for the target ground state of the TFIM one-dimensional chain Eq.~\ref{eq:ising_1d}, and Fig.~\figref[d)]{fig:combined_grid} for the Haldane lattice Eq.~\ref{eq:haldane_2d} 

For the TFIM in ~\figref[a)]{eq:ising_1d}, the increase along the number of flips axis shows that the final correlator does not necessarily have to improve when we increase the number of terms per correlator. This means that a simpler model suffices. This is in accordance to the behavior of the TFIM in~\figref[a)]{fig:combined_grid}.

For the two-dimensional Haldane lattice in~\figref[b)]{eq:ising_1d} we have constant values along the $h$ axis for the most significant correlator orders which are the second and fourth orders (see Fig.~\figref[c)]{fig:correlations_hamiltonians}). Again, higher complexity within the correlators does not improve the MSE$_k$

We can conclude that the complexity of the target distribution dictates the number of flips or bond dimension there should be for a low  classical deployed MSE$_k$. However, these results do not inform us of how the MSE$_k$ is when transferring the parameters to the complete quantum circuit.

\section{Conclusion}
\label{sec:conclusion}
\subsubsection{Overview of Main Results}
In this work, we point out that the QCBM is a quantum Fourier model to which the dequantization conditions from the supervised case apply as given in Theorem~\ref{thm:ideal_deq}. However, we evidence that training all correlators is not always possible, exemplifying it via the matchcircuit case which has a polynomial growing algebra in the number of qubits. 

Due to this, if we decide to train on classical hardware the chances of having a quantum advantage reduces due to the discrepancy sources appearing in Theorem~\ref{thm:discrepancy}.

We underscore the importance of selecting the most significant correlators and numerically analyze two different techniques from the point of view of inductive bias. However, there is also much potential for future work in optimizing these methods to select the strategy that suits best an advantageous quantum deployment. Nevertheless, our results point towards the direction of having a strong dependence on the dataset and the ansatz.

With the objective of reducing the discrepancies between classically surrogated correlators and the quantumly deployed correlators, we assess how different surrogate techniques and circuit structures may present different training patterns. In the case of tensor networks the training was gradually improving, for the PPS case the model did learn, however in a less structured manner, and is much more sensitive to hyperparameters.

Lastly, we point out that given these individual discrepancies, we were not able to observe a convergence in all cases between the deployed model to the fully un-truncated, un-approximated case. This indicates that the convergence may be difficult, perhaps impossible, to achieve in general without large computational resources. This is especially true as the number of constrained correlators and amount of data introduced increases.

\subsubsection{Further Results in Appendix} \label{sssec:further_appendix_results}
Finally, we argue that the Fourier transform of the Born rule Eq.~\ref{eq:decomposition_pr} allows for further analysis and supplementary results, outside of what is strictly necessary for this subject of this work. This gives further evidence for the usefulness of our framework.

In particular, in the Appendices, we expand on:
\begin{enumerate}
    \item How the Gaussian kernel in the MMD loss may be used to selectively emphasize specific correlators during training. Numerical experiments reveal that neglecting relevant correlators in the kernel may lead to the misleading conclusion that the QCBM is effectively learning the target distribution, when in fact it is not due to the incomplete correlator set.
    \item For a scrambling unitary QCBM and theoretical assumptions, we show how more iterations are needed for higher order correlators.
    \item In Haar random cases, averaged over the full landscape, the dependencies between correlators are zero (Appendix.~\ref{app:general}). However, we exemplify that this is not necessarily the case in practice, (Appendix.~\ref{app:dependencies}).
    \item We derive further quantities of interest for the case of Random Matrix Product States~in Appendix~\ref{app:tn_proofs}.
\end{enumerate}

\section{Discussion}\label{sec:discussion}

We have emphasized in this work the delicate interplay between classically trained and quantumly deployed models, particularly in the generative setting with quantum circuit Born machines we focused on. Such translations lead to inherent inductive biases which can be helpful or harmful to any potential quantum advantage. As such, there are several potential directions for future work. A natural step forward is understanding theoretically how the inductive bias can be utilized to positively, rather than negatively, impact any quantum advantage. For example, leveraging symmetries of the problem at hand is a well known topic of study~\cite{meyer_exploiting_2023}. These symmetry-aware models could be tailored to the context of train on classical and deploy on quantum. For example, problem informed ansatz\"{e} which take into account underlying graphs for graph-native problems~\cite{bako_problem-informed_2025,makarski_circuit_2025}. Linked to the question of symmetry, is the output characterization of the model. Specifically, we have only considered the computational basis in this work, but choosing the correct basis to sample from can give better explainability~\cite{gil-fuster_opportunities_2024} when choosing specific underlying distributions and can lead to multidimensional distributions as there exist for the supervised case~\cite{casas_multidimensional_2023}. Indeed, basis-enhancement has already been proposed as a tool to enhance QCBMs~\cite{rudolph_generation_2022}.

Moreover, different resources can be at play when having a separation between a quantum and a classical computation~\cite{thomas_role_2024}. Therefore, developing knowledge on how these affect the discrepancies here described is a promising idea. 

Next, relating to some of the specific circuit families we study here, IQP circuits do not achieve universality in the sense of being able to attain any choice of distribution. However, in~\cite{kurkin_note_2025, kurkin_universality_2025} universality is theoretically obtained through the inclusion of additional qubits. Studying how these ancillary qubits modify the correlators, discrepancy sources and surrogates is a logical next step. A second direction is to apply these techniques to other scenarios where there is sampling advantage. For example, in boson sampling-like circuits or continuous variable cases~\cite{cepaite_continuous_2022, barthe_parameterized_2025, romero_variational_2021}. Photonic QCBMs have already been put in practice~\cite{sedrakyan_photonic_2024, salavrakos_error-mitigated_2025} and one could also adapt the correlator surrogates to these situations. 

Finally, it remains an open question whether a QCBM trained up to a given precision can benefit from a warm-start procedure~\cite{puig_variational_2025, mhiri_unifying_2025}. The approximations introduced here could provide a pathway to such a warm start and help identify the conditions under which subsequent gradient evaluation would be more suitably performed on a quantum device.

In summary, the cooperation and tension between quantum and classical resources for quantum machine learning is a fruitful area for study. 

\section*{Code Availability}
The code associated with the results of this paper is available at \url{https://github.com/quantumsoftwarelab/QCBM_correlator_surrogates.git}.

\section*{Acknowledgments}
The authors thank Abel Jansma and Phalgun Lolur for their input in the initial exploration phase of this work. MHG thanks Chirag Wadhwa, Adithya Sireesh, Raul Garcia-Patron Sanchez and Mina Doosti for fruitful discussions; Oliver Thomson Brown for clarifying questions regarding tensor networks; Marco Cerezo for highlighting potential issues with the MMD kernel; and Hugo Thomas for constructive comments on the manuscript. Konstantinos Georgopoulos provided line management to KM. MHG, BC, RG, EK acknowledge support from the EPSRC Quantum Advantage Pathfinder research program within the UK’s National Quantum Computing Center. AK was supported by a Langmuir Talent Development Fellowship from the Institute of Genetics and Cancer, and a philanthropic donation from Hugh and Josseline Langmuir.

\newpage

\bibliographystyle{ieeetr}
\bibliography{refs}

\clearpage
\appendix

\makeatletter
\let\addcontentsline\orig@addcontentsline
\makeatother

\onecolumngrid
{\Large \textbf{Appendix}}
\begingroup
\setcounter{tocdepth}{2}
\renewcommand{\contentsname}{Appendix contents}
\tableofcontents
\endgroup

\counterwithin*{equation}{section}
\renewcommand\theequation{\thesection\arabic{equation}}    
\newpage

\section{Sampling and Quantum Advantage}\label{subsec:sampling_advantage}
In this work, we have shown a unified framework where training quantum generative models can be reduced to computing expectation values, despite the inherent apparent reliance on sampling in the model itself. There is a delicate interplay between these two mindsets as it has been shown in many cases that calculating expectation values can be approximately classically simulable on average~\cite{angrisani_classically_2024}, but yet the same does not necessarily hold for particular instances of the corresponding sampling (`inference') problem. Under widely accepted complexity-theoretic assumptions, it has been demonstrated that classically sampling from the probability distribution produced by Instantaneous Quantum Polynomial (IQP) circuits~\cite{shepherd_instantaneous_2009}, which are characterized by states of the form $H^{\otimes n} U_c H^{\otimes n}|\bzero\rangle$ and $U_c$ composed of commuting gates, is \#P-hard. Here, \#P refers to the class of counting problems associated with the number of solutions to a nondeterministic polynomial-time problem, which are believed to be intractable for classical computation. Furthermore, efficiently being able to sample from IQP instances would imply the collapse of the polynomial hierarchy under certain assumptions~\cite{marshall_improved_2024}. This average hardness persists even in the approximate setting~\cite{bremner_average-case_2016} and in the presence of noise~\cite{bremner_achieving_2017}. Although a generic QCBM contains non-commuting gates and thus falls outside the strictly commuting-circuit class, its structure could still allow advantageous differences between quantum and classical distributions. Therefore, in our analysis we do not only limit ourselves to the study of IQP instances.

Returning to the learning problem, a fundamental aspect of classical simulability concerns the complexity of the dataset. If the distribution can be efficiently learned using a classical algorithm, then employing a quantum computer, at least for training, becomes less desirable. However, the quantum framework’s sampling capabilities can still be leveraged to generate samples more efficiently. 

To harness these advantage results, the IQP architecture is employed for the QCBM in~\cite{coyle_born_2020}. The study demonstrates that, under gradient-based training, the circuits encountered cannot be efficiently sampled by classical methods within a multiplicative error bound in the worst-case scenario.

From a theoretical perspective, Hinsche et al.~\cite{hinsche_learnability_2021} explored the problem of learning the output distributions of quantum circuits using identity testing. Their results demonstrate that, within the probably approximately correct (PAC) framework, learning the distribution of a two-qubit gate brickwork random quantum circuit is, on average, a computationally hard task. Specifically, they show that super-logarithmic depth Clifford circuits cannot be learned sample-efficiently within the statistical query model. Conversely, they prove that local Clifford circuits, when provided with access to samples, are computationally efficiently PAC-learnable by a classical learner.

Beyond IQP instancies, there exists work establishing the computational power of shallow random circuits ~\cite{haferkamp_closing_2020,bermejo-vega_architectures_2018}. Under standard conjectures, there are shallow circuits that can generate distribution no poly-time classical algorithm can. To consider these we use the algebraic ansatz\"{e} so that we can have a controlled evaluation.

In our analysis, we assess how well general PQCs capture correlations through their inductive bias. The use of correlations allows for a more refined assessment of the quantum learner's performance. In practical scenarios, when both quantum and classical learners fail to approximate a given distribution, complexity arguments alone provide no clear indication of which learner fails to a greater or lesser extent. We now return to the correlation decomposition which allows us to first truncate the probability distribution and pinpoint the classical limits.

\section{Loss Functions for Truncated Probability Distributions}
\label{app:losses}
In this work, we consider three distinct loss functions to train QCBMs by quantifying discrepancies between approximated probability distribution vector $p =\operatorname{Pr}^{\rm approx}_{\btheta}$ in Eq.~\ref{eq:truncated_prob} or Eq.~\ref{eq:random_corr_prob} and a given target distribution vector over discrete binary bitstrings $q=\operatorname{Pr}_{\mathcal H}$.

\paragraph{The Earth Mover’s Distance~\cite{rubner_metric_1998},} or first Wasserstein distance, measures the minimum cost of transforming one probability distribution into another. For discrete distributions, it can be expressed as
\begin{equation}
\operatorname{EMD}(p, q) = \sum_{\boldsymbol{x}} \big|\mathrm{CDF}_p(\boldsymbol{x}) - \mathrm{CDF}_q(\boldsymbol{x})\big|,
\end{equation}
where $\operatorname{CDF}_p(\boldsymbol{x})=\sum_{\boldsymbol{y} \leq \boldsymbol{x}} p(\boldsymbol{y})$ and $\operatorname{CDF}_q(\boldsymbol{x})=\sum_{\boldsymbol{y} \leq \boldsymbol{x}} q(\boldsymbol{y})$ are cumulative distribution functions of the respective distributions $p$ and $q$. Here, the ordering relation $\boldsymbol y \leq \boldsymbol x$ refers to a fixed lexicographic ordering of bitstrings. The EMD, rooted in optimal transport theory, captures cumulative discrepancies between the two distributions.

\paragraph{Squared Euclidean Loss (SQE).}
The squared Euclidean loss measures the elementwise squared difference between two distributions,
\begin{equation}
\ell_2^2(p, q) = \sum_{\boldsymbol{x}} \big(p(\boldsymbol{x}) - q(\boldsymbol{x})\big)^2.
\end{equation}
This measure quantifies the deviation of the distribution $p$ from $q$ in a straightforward Euclidean space. As it captures discrepancies in individual probabilities symmetrically without considering lexicographic ordering, it emphasizes larger deviations.

\paragraph{Maximum Mean Discrepancy (MMD).}
The Maximum Mean Discrepancy (MMD)~\cite{gretton_kernel_2012} is a kernel-based metric used to distinguish between two probability distributions based on their embeddings in a reproducing kernel Hilbert space (RKHS). Formally, for distributions $p$ and $q$, MMD is defined as:
\begin{equation}
\begin{aligned}
    \mathrm{MMD}^2(p,q) \;=\;&
        \mathbb{E}_{\boldsymbol{x},\,\boldsymbol{x}' \sim p}\bigl[K(\boldsymbol{x}, \boldsymbol{x}')\bigr] \\
    &-\; 2\,\mathbb{E}_{\boldsymbol{x} \sim p,\,\boldsymbol{y} \sim q}\bigl[K(\boldsymbol{x}, \boldsymbol{y})\bigr] \\
    &+\; \mathbb{E}_{\boldsymbol{y},\,\boldsymbol{y}' \sim q}\bigl[K(\boldsymbol{y}, \boldsymbol{y}')\bigr].
\end{aligned}
\label{eq:mmd_def}
\end{equation}
where $K(\cdot, \cdot)$ is often a Gaussian kernel function, $K(\boldsymbol{x}, \boldsymbol{y}) = \exp\bigl(-\|\boldsymbol{x}-\boldsymbol{y}\|^2/(2\sigma^2)\bigr)$ being $\sigma$ the bandwidth hyperparameter. The larger the bandwidth is, the less samples per bitstrings are necessary to have a substantial overlap between distributions.

An alternative form of the MMD loss can be obtained explicitly in terms of the probability vectors of the two distributions, see Appendix~\ref{app:anova} for the derivation from Eq.~\ref{eq:mmd_def},
\begin{equation}
\mathrm{MMD}^2(p,q) = (p-q)^{\mathsf T}\,K\,(p-q),
\label{eq:mmd_matrix}
\end{equation}
with $K \in \mathbb{R}^{2^n \times 2^n}$ the kernel matrix with entries $K({\boldsymbol{x},\boldsymbol{y}})$.

Different works have suggested and discussed the potential use of a quantum kernel~\cite{coyle_quantum_2021,rudolph_trainability_2023}. We argue that as generative tasks use polynomial-sized batches to estimate the loss, we can mainly focus on classical-type kernels and analyze them from the perspective of $k$-order correlations. Further details of this method are in Appendix~\ref{app:anova}, where we show how it is also possible to truncate the correlations that appear within the kernel when comparing the two probability distributions.

\paragraph{Evaluation Metric.} When considering non-truncated probabilities for evaluation purposes we can use distances that not need to have positive function domain as we are assured that the non-truncated probability will always be non-negative. Specifically, the Kullback-Leibler (KL) divergence, which is a $f$-divergence, is defined by the logarithmic function that has positive domain,
\begin{equation}
    {\mathrm{KL}}(\operatorname{Pr}_{\mathcal H} \,\|\, \operatorname{Pr}_{\btheta})
    \;=\;
    \sum_{\boldsymbol{x}}\operatorname{Pr}_{\mathcal H}(\boldsymbol{x})
    \,\log\!\left[\frac{\operatorname{Pr}_{\mathcal H}(\boldsymbol{x})}{\operatorname{Pr}_{\btheta}(\boldsymbol{x})}\right].
\end{equation}
A more thorough analysis on how to compute local terms of $f$-divergences whose domain space is restricted can be found in Subsec. III.B of~\cite{leadbeater_f-divergences_2021}.

\paragraph{Note on adversarial losses.} In adversarial learning scenarios as in QGANs~\cite{wang_qgan_2019}, one often trains a \emph{discriminator} $D(\boldsymbol{x})$ to separate samples by classification from $\operatorname{Pr}_{\mathcal H}$ versus samples from $\operatorname{Pr}_\btheta$. Intuitively, if the discriminator itself is limited to features or tests that involve \emph{up to $k$-order combinations} of the input variables, then it can only detect statistical differences manifest at those first $k$-order correlations. Hence, any mismatch in higher-order correlations remains invisible to such a restricted discriminator, mirroring the truncation in the Born distribution. If the discriminator is more expressive, and hence able to test higher-order dependencies it
can in principle detect discrepancies beyond the $k$-order level. We leave the quantitative analysis of QGANs under $k$-order truncation for future work, although~\cite{letcher_tight_2024} already analyses discriminators which are based on fully-connected neural networks with rectifying linear unit activation functions to prove that for first order correlators within the discriminator there is no concentration.

Overall, limiting the Born machine to specific correlations confines the primary measurable differences between $\operatorname{Pr}_\theta^{\rm approx}$ and $\operatorname{Pr}_{\mathcal H}$ (under MMD, $f$-divergences, or a discriminator) to those same correlation statistics, leaving correlators effectively unconstrained. 

\section{Obtaining Reference Data with DMRG}
\label{app:data_obtention}
\begin{figure}[htbp]
    \resizebox{\linewidth}{!}{\input{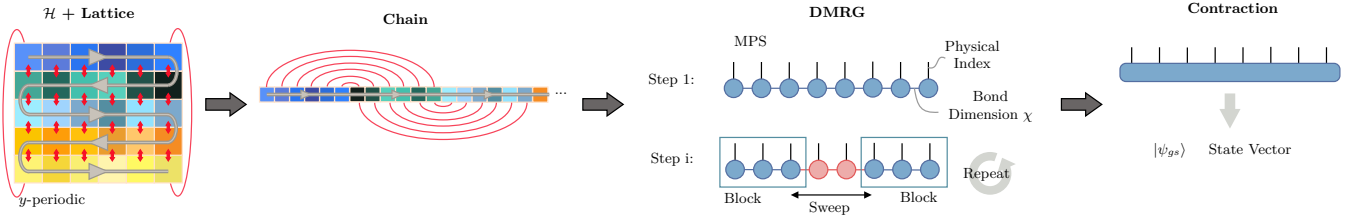}}
    \caption{\justifying\textbf{Extraction of ground states for Hamiltonians defined in $y$-periodic square lattices.} Obtention of the target data from ground state whose underlying Hamiltonian is described by long-range terms. The Hamiltonian terms are defined over a lattice and then unraveled to be fed in to the DMRG algorithm. After full contraction the probabilities of the ground state are obtained. Adapted from \cite{thomson_unravelling_2024} and \cite{verstraete_density_2023}.
    }    \label{fig:data_lattice}
\end{figure}

Low-energy states of a Hamiltonian form a linear subspace in the total Hilbert space whose dimension scales polynomially in the system size. In addition, the states in this subspace can be described using tensor networks (TN)~\cite{banuls_tensor_2023}. The computational resources required for TN schemes, such as Matrix Product States (MPS) and Projected Entangled Pair States (PEPS), can scale linearly in the system size, providing an efficient method of computing properties such as correlation functions, entanglement properties and phase transitions among others.
Ground states of the Hamiltonians are obtained via the Density Matrix Renormalization Group (DMRG) algorithm~\cite{white_density_1992}, for different lattice geometries, namely, one-dimensional chains and $y$-periodic rectangular lattices.
While DMRG is mainly suited for one-dimensional problems, one can nevertheless enforce two-dimensional lattice correlations for small problems.
DMRG takes the representation of a many-body wavefunction as an MPS with a chosen physical index bond dimension, then iteratively updates local blocks in a block sweep to compress the wavefunction by discarding states of minimal entanglement. 
By repeating this sweeping process and adjusting the bond dimensions as necessary, DMRG refines the MPS representation to find a highly accurate ground-state approximation. The full process is illustrated in Fig.~\ref{fig:data_lattice}.

\section{Proofs on Truncation}\label{app:general}
In this section, we give the explicit proofs relating to probability distribution truncation, and Fourier representations of the Born rule.

\subsection{Proof of Main Background Lemmas}
We begin with the Born rule from the main text:
\begin{equation}\label{eq:born_rule_app}
\operatorname{Pr}_{\boldsymbol{\theta}}(\boldsymbol{x}) =\Tr(\ketbra{\bx}{\bx} \rho_{\boldsymbol{\theta}}) = |\langle \boldsymbol{x} | \psi(\boldsymbol{\theta}) \rangle|^2.
\end{equation}

Now, it is possible to write this probability in terms of expectation values (correlators):
\begin{lemma}[Lemma~\ref{lemma:decomposition_full} repeated.]
The probability of any bit-string $\bx = (x_1,x_2,...,x_n)\in\{0,1\}^n$ can be expressed in terms of $Z$-basis correlators, namely
\begin{equation} \label{applemma:decomposition_pr}
\pr(\bx) = \frac{1}{2^n} \sum^{n}_{p=0} \sum_{\substack{{\bi}\subseteq\{1,...,n\}\\|{\bi}|=p}} (-1)^{\sum_{i\in {\bi}} x_i}\left\langle Z_{\bi}\right\rangle_{\btheta},
\end{equation}
where
\begin{equation}
Z_{\bi} =  \left(\bigotimes_{i\in {\bi}} Z_i\right) \otimes \1_{\boldsymbol{\noi}}
\end{equation}
with $\boldsymbol{\noi} = \{ i \notin {\bi}\}$.
\end{lemma}

\begin{proof}
Given that the projector is $\prod_{x_i}\!= \ket{x_i}\!\!\bra{x_i}\!=\!\frac{1}{2}(\1_i+(-1)^{x_i}Z_i)$ and $\operatorname{Pr}(\bx)= \left\langle \psi \left| \prod_{\bx}\right| \psi \right\rangle = \left\langle  \prod_{\bx}\right\rangle$ it suffices to expand $\prod_{\bx} = \bigotimes^n_{i=1} \ket{x_i}\!\!\bra{x_i}$ and regroup the sums in $p$ and $|\bi|$.
\end{proof}

Equivalently, we can rewrite the correlators in terms of bit string probabilities.
\begin{lemma}[Lemma~\ref{lem:exp_from_probs} repeated.]\label{lemma:app
_exp_from_probs}
A $Z$-basis $k$-order correlator acting on a subset of ${\bi}\subseteq \{1,2,...,n\}$ qubits can be written in terms of the bit-string probabilities,  
\begin{equation}\label{eq:app_exp_from_probs}
\left\langle Z_{\bi} \right\rangle = \sum_{\bx} (-1)^{\sum_{i \in{\bi}}x_i}\operatorname{Pr}(\bx).
\end{equation}
\end{lemma}

\begin{proof}
Expanding the definition of the correlator,
\begin{equation*}
\begin{aligned}
\left\langle Z_{\bi}\right\rangle & =\bra{\bx^{\prime}} \sum_{\bx^{\prime}} \alpha_{\bx^{\prime}}^*(\btheta) Z_{{\bi}}\sum_{\bx} \alpha_{\bx}(\btheta)\ket{\bx}\\
& =\left\langle \bx^{\prime}\right| \sum_{\bx^{\prime}} \alpha_{\bx^{\prime}}(\btheta) \sum_{\bx} \alpha_{\bx}(\btheta)(-1)^{\sum_{i \in {\bi}} x_i}|\bx\rangle \\
& =\sum_{\bx}(-1)^{\sum_{i \in {\bi}} x_i}\left|\alpha_{\bx}(\btheta)\right|^2 \\
& =\sum_{\bx}(-1)^{\sum_{i \in {\bi}} x_i} \operatorname{Pr}_\btheta(\bx).
\end{aligned}
\end{equation*}
This shows the claim.
\end{proof}

\subsection{A Toy Example - Three Qubits} \label{app_ssec:toy_example_three_qbs}
To clarify the above, and the intuition for our work, we give a toy example of a correlator decomposition for a three qubit distribution (three qubit QCBM). For $n=3$, the general decomposition given by
\begin{equation}
\operatorname{Pr}(\boldsymbol{x})=\frac{1}{2^3}\sum_{p=0}^3 \sum_{\substack{\boldsymbol{i}\subseteq{1,2,3}\\|\boldsymbol{i}|=p}}
(-1)^{\sum_{i\in\boldsymbol{i}}x_i},\langle Z_{\boldsymbol{i}}\rangle
\end{equation}
takes the explicit form
\begin{equation}
\operatorname{Pr}(\boldsymbol{x})=\frac{1}{8}\left[1+\sum_{i=1}^{3}(-1)^{x_i}\langle Z_i\rangle + \sum_{i<j}(-1)^{x_i+x_j}\langle Z_i Z_j\rangle + (-1)^{x_1+x_2+x_3}\langle Z_1 Z_2 Z_3\rangle\right].
\end{equation}

This decomposition can be expressed compactly using the ordered Hadamard matrix $H_8$ as:

\begin{equation}\label{eq:matrix_form}
\left(\begin{array}{c}
\operatorname{Pr}(000) \\
\operatorname{Pr}(001) \\
\operatorname{Pr}(010) \\
\operatorname{Pr}(100) \\
\operatorname{Pr}(011) \\
\operatorname{Pr}(101) \\
\operatorname{Pr}(110) \\
\operatorname{Pr}(111)
\end{array}\right)= \frac{1}{2^3}
\underbrace{\left(
\begin{array}{c|ccc|ccc|c}
1 & 1 & 1 & 1 & 1 & 1 & 1 & 1 \\
1 & -1 & 1 & 1 & -1 & -1 & 1 & -1 \\
1 & 1 & -1 & 1 & -1 & 1 & -1 & -1 \\
1 & 1 & 1 & -1 & 1 & -1 & -1 & -1 \\ 
1 & -1 & -1 & 1 & 1 & -1 & -1 & 1 \\
1 & -1 & 1 & -1 & -1 & 1 & -1 & 1 \\
1 & 1 & -1 & -1 & -1 & -1 & 1 & 1 \\ 
1 & -1 & -1 & -1 & 1 & 1 & 1 & -1
\end{array}
\right)}_{H_8}
\underbrace{\left(
\begin{array}{c}
\mathds{1} \\\hline
\langle Z_1 \rangle \\
\langle Z_2 \rangle \\
\langle Z_3 \rangle \\\hline
\langle Z_1 Z_2 \rangle \\
\langle Z_1 Z_3 \rangle \\
\langle Z_2 Z_3 \rangle \\\hline
\langle Z_1 Z_2 Z_3 \rangle
\end{array}
\right)}_{\text{correlators}}
\end{equation}
where vertical lines separate the matrix columns into blocks corresponding to $0^{\rm th}$, $1^{\rm st}$, $2^{\rm nd}$, and $3^{\rm rd}$-order correlators, while horizontal lines group the correlators by order. Here, it is clear to see that for any truncation, 

Explicitly, the $k$-order truncations are:
\begin{align*}
\operatorname{Pr}^{(0)}(\boldsymbol{x})&=\frac{1}{8}, \\[6pt]
\operatorname{Pr}^{(1)}(\boldsymbol{x})&=\frac{1}{8}\left[1+\sum_{i=1}^{3}(-1)^{x_i}\langle Z_i\rangle\right], \\[6pt]
\operatorname{Pr}^{(2)}(\boldsymbol{x})&=\frac{1}{8}\left[1+\sum_{i=1}^{3}(-1)^{x_i}\langle Z_i\rangle + \sum_{i<j}(-1)^{x_i+x_j}\langle Z_i Z_j\rangle\right], \\[6pt]
\operatorname{Pr}^{(3)}(\boldsymbol{x})&=\operatorname{Pr}(\boldsymbol{x}).
\end{align*}

To illustrate, consider the GHZ state $(\ket{000}+\ket{111})/\sqrt{2}$, for which:
\begin{equation}
\langle Z_i\rangle=0, \quad \langle Z_i Z_j\rangle=1, \quad \langle Z_1 Z_2 Z_3\rangle=0.
\end{equation}
Then, we have:
\begin{equation}
\operatorname{Pr}^{(0)}(\boldsymbol{x})=\frac{1}{8},\quad\operatorname{Pr}^{(1)}(\boldsymbol{x})=\frac{1}{8},\quad \operatorname{Pr}^{(2)}(\boldsymbol{x})=\begin{cases}
\frac{1}{2}, & \boldsymbol{x}=000,111 \\
0, & \text{otherwise},
\end{cases}
\end{equation}
showing explicitly that second-order correlators fully characterize the GHZ state's distribution for three qubits.

\subsection{Theoretical analysis}
We now show how to theoretically bound the truncation error when considering unitaries from the maximally expressive uniform Haar distribution~\cite{holmes_connecting_2022}.

\begin{theorem}[Truncation Error] 
We define the distance between the exact and truncated probabilities as
\begin{equation}
D_k\equiv D^{(k)}(\bx) := \operatorname{Pr}_{\btheta}(\bx)-\operatorname{Pr}^{(k)}_{\btheta}(\bx).
\end{equation}
When using unitaries distributed Haar-randomly over the whole unitary space we obtain that on average
\begin{equation}
\mathbb{E}[D_k^2] = \frac{2^n - N_k}{2^{3n}} \text{ with } N_k = \sum_{p=0}^{k}\binom{n}{p}.
\end{equation}
Alternatively, without the assumption of randomness, we obtain the deterministic upper bound through the triangle inequality as
\begin{equation}
\left|D_k^2\right| \leq  \left( 1 - \frac{N_k}{2^n} \right)^2,
\end{equation}
due to the fact that $\left\langle Z_{\bi} \right\rangle \in [-1,1]$.
\end{theorem}
This distance is only meaningful from the perspective of expectation values, since the distance between two probabilities cannot exceed unity. Consequently, $D_k^2$
is naively bounded by 4, reflecting potential negative contributions to the expectation values. While the Haar-random evaluation yields a value below the deterministic bound, it fails to account for second-order correlators, as integrating over the full space of unitaries enforces $\Tr \left(\langle Z_{\bi}\rangle \langle Z_{\bj}\rangle \right) = 0,\, \forall \bi \neq \bj$. Therefore, barely calculating the variance through Haar unitaries is non-realistic and we proceed by adding specific structure to the analysis of the variance for the cases of Tensor Networks (subsection~\ref{subsec:tensor_networks}) and Lie algebraic approaches (subsection~\ref{subsection:lie_algebra}).

\subsubsection{Deterministic upper bound}
We consider the discrepancy measure
\begin{equation}
D_k = \frac{1}{2^{n}} \sum_{p=0}^{n} \sum_{{\bi} \subseteq \{1, \dots, n\}, |{\bi}| = p} \langle Z_{{\bi}} \rangle 
- \frac{1}{2^{n}} \sum_{p=0}^{k} \sum_{{\bi} \subseteq \{1, \dots, n\}, |{\bi}| = p} \langle Z_{{\bi}} \rangle.
\end{equation}

Hence,
\begin{equation}
D_k^2 = \left[ \frac{1}{2^n} \sum_{p=k+1}^{n} \sum_{|{\bi}|=p} \langle Z_{{\bi}} \rangle \right]^2.
\end{equation}

Using the fact that $ |\langle Z_{{\bi}} \rangle| \leq 1 $, 

\begin{equation}
D_k^2
\leq \left(1 - \frac{  N_k}{2^n} \right)^2.
\end{equation}
where $N_k = \sum_{p=0}^k \binom{n}{p}$ is the number of all possible Pauli strings up to order $k$.

\subsubsection{Haar-averaged bound}
Alternatively, we can calculate the average value of the error $D_k$ over the full unitary group using the Haar randomness properties:
\begin{equation}
    \mathbb{E}[\langle Z_{{\bi}} \rangle] = 0 \text{ for } {\bi} \neq \emptyset, \quad \mathbb{E}[\langle Z_{{\bi}} \rangle^2] = \frac{1}{2^n}, \quad \mathbb{E}[\langle Z_{{\bi}} \rangle \langle Z_{{\bj}} \rangle] = 0 \\\ \forall \bi\neq\bj, \quad \E[\langle Z_{\varnothing}\rangle] = \E[\langle \1 \rangle] = \E[\Tr(\rho_0)]=1.
\end{equation}
Expanding the squared difference,
\begin{equation}
    \mathbb{E}[D_k^2] = \mathbb{E} \left[ \left( \frac{1}{2^n}\sum_{p=k+1}^{n} \sum_{|{\bi}|=p} \langle Z_{{\bi}} \rangle \right)^2  \right] 
    = \sum_{p=k+1}^{n} \sum_{|{\bi}|=p} \left( \frac{1}{2^n} \right)^2 \mathbb{E}\left[\langle Z_{{\bi}} \rangle^2\right],
\end{equation}
and substituting,
\begin{equation}
    \mathbb{E}[D_k^2] = (2^n - N_k) \left( \frac{1}{2^n} \right)^2 \frac{1}{2^n}.
\end{equation}
The final expression is,
\begin{equation}
    \mathbb{E}[D_k^2] = \frac{2^n - N_k}{2^{3n}}.
\end{equation}

\section{Main Theorems}

In this appendix, we provide the theoretical background and proofs for the main theorems we use as the foundation of the work, specifically Theorem~\ref{thm:ideal_deq} and Theorem~\ref{thm:discrepancy}.

\subsection{Bochner's theorem for Theorem~\ref{thm:ideal_deq}}
In the supervised scenario the datapoints $x\in \mathbb{R}$ are uploaded to the and hence the output function of the circuit has the dependence $f(x, \boldsymbol \theta)$. The Fourier description arises from the data reuploading model~\cite{perez-salinas_data_2020} as $f( x, \boldsymbol \theta) = \sum_{\omega \in \Omega} c_{\omega}(\boldsymbol \theta) e^{-i\omega x}$~\cite{schuld_effect_2021} where here $\Omega$ is the frequency set.
Bochner's theorem implies that any continuous shift-invariant kernel $K(x, y)=k(x-y)$ admits the integral representation

$$
K(x, y)=\int_{\mathbb{R}^d} e^{i \omega \cdot(x-y)} \mathrm{d} \mu(\omega),
$$

where $\mu$ is a nonnegative spectral measure that can be normalized to a probability distribution when $k(0)=1$. On the Boolean cube $G=\mathbb{Z}_2^n$, an analogous discrete integral holds:

$$
K(x, y)=\sum_{S \subseteq[n]} \mu_S \chi_S(x) \chi_S(y)=\int_{\widehat{G}} \chi(x) \chi(y) \mathrm{d} \mu(\chi),
$$

where $\widehat{G}$ is the dual group of parity characters $\chi_S(x)=(-1)^{S \cdot x} \quad$. These spectral representations justify Random Fourier Features (RFF) by interpreting the kernel as an expectation over frequencies or characters and approximating it via Monte Carlo sampling from $\mu$ to obtain an explicit low-dimensional feature map $\phi$ such that
\begin{equation}
K(x,y)\;\approx\;\frac1m\sum_{i=1}^m\phi(x;\omega_i)\,\overline{\phi(y;\omega_i)}. \end{equation}

Integral representation in Euclidean Space. A continuous shift-invariant kernel \(K(x,y)=k(x-y)\) on \(\mathbb{R}^d\) is positive-definite if and only if 
$k(\delta) =\int_{\mathbb{R}^d}e^{\,i\omega\cdot\delta}\,\mathrm{d}\mu(\omega)$,
for some finite nonnegative measure $\mu$ on $\mathbb{R}^d$. When $\mu$ admits a density $p(\omega)$, this becomes 
\begin{equation}
k(\delta)=\int_{\mathbb{R}^d} p(\omega) e^{i \omega \cdot \delta} \mathrm{~d} \omega=\mathbb{E}_{\omega \sim p}\left[e^{i \omega \cdot(x-y)}\right]
\end{equation}
where $\delta = x-y$.
Now as with the main text, we analyze the above in the discrete scenario.Viewing $\{0,1\}^n$ as the abelian group $G=\mathbb{Z}_2^n$ under bitwise XOR, its Pontryagin dual $\widehat{G}$ is again $\mathbb{Z}_2^n$, with characters
$$
\chi_S(x)=(-1)^{\sum_{i \in S} x_i}, \quad S \subseteq[n],
$$
which form an orthonormal basis for functions on the cube. Any positive-definite, XOR-invariant kernel $K(x, y)=f(x \oplus y)$ admits the spectral (integral) decomposition
$$
f(g)=\int_{\widehat{G}} \chi(g) \mathrm{d} \mu(\chi)=\sum_{S \subseteq[n]} \mu_S(-1)^{S \cdot g}, \quad \mu_S \geq 0,
$$

so that
\begin{equation}
K(x,y) =\sum_{S\subseteq[n]}\mu_S\,\chi_S(x)\,\chi_S(y), \quad\mu_S\ge0.
\end{equation}

Interpreting the kernel as an expectation 
\begin{equation}
K(x,y) =\mathbb{E}_{\omega\sim p}\bigl[e^{\,i\omega\cdot(x-y)}\bigr] \quad\text{or}\quad K(x,y) =\mathbb{E}_{S\sim\mu}\bigl[\chi_S(x)\,\chi_S(y)\bigr]
\end{equation}
leads to explicit feature maps
$\phi(x ; \omega)=e^{i \omega \cdot x}$ (continuous) or $\phi(x ; S)=\chi_s(x)$ (discrete), satisfying $K(x,y)=\mathbb{E}\bigl[\phi(x)\,\overline{\phi(y)}\bigr].$ By drawing \(m\) i.i.d. samples \(\{\omega_i\}\) or \(\{S_i\}\) from the spectral measure and forming $ \hat K_m(x,y) =\frac{1}{m}\sum_{i=1}^m\phi(x;\omega_i)\,\overline{\phi(y;\omega_i)}$, one obtains the RFF approximation~\cite{rahimi_random_2007}, which converges to $K(x, y)$ at a rate $O\left(m^{-1 / 2}\right)$ and enables scalable kernel methods in high dimensions

Thanks to this discretization of the kernel and the technicalities presented in Sec.~\ref{ssec:deployment_technical_term}, we can directly translate the conditions presented in ~\cite{sweke_potential_2025,sahebi_dequantization_2025}

\subsection{Discrepancy Sources Derivation} \label{app_ssec:derivation_main_theorem_2_discrepancy}

We use the following adapted result to prove Theorem~\ref{thm:discrepancy}, relating to sources of discrepancy between classically trained/deployed and quantumly deployed models.
\begin{theorem}
(Adapted from Remark~6 in~\cite{cucker_mathematical_2002,de_vito_learning_2005,rudi_generalization_2021}). 
Assume that $\int y^2\, d\rho$ is finite, so that $\opr^*_\rho \in L^2(X,\rho_X)$ and it minimizes the risk functional $\risk$ over all measurable functions. 
When $\int s(\bx,\bx')\, d\rho_X$ is finite, the range of $\Pi$ (and of the corresponding integral operator $\mathcal{L}$) is the closure of $\mathcal{H}$ in $L^2(X,\rho_X)$. 
If both conditions hold, then for any $\opr \in L^2(X,\rho_X)$ the following equality holds:
\begin{equation}
\label{eq:proj_risk}
\risk[\opr]-\min_{g\in\mathcal{H}}\risk[g]
= \|\opr-\Pi\opr^*_\rho\|_{\rho_X}^2
+2\langle \opr,(\1-\Pi)\opr^*_\rho\rangle_{\rho_X}.
\end{equation}
The second term vanishes for all $\opr\in\mathcal{H}$, and also for functions defined by $M$ random features. 
Moreover, if there exists $\opr_{\mathcal{H}}\in\mathcal{H}$ minimizing $\risk$, then Assumption~6 is equivalent to the existence of $r\geq\frac12$ and $g\in L^2(X,\rho_X)$ such that
\begin{equation}
\Pi \opr^*_\rho = \mathcal{L}^r g,
\qquad 
R := \|g\|_{L^2(X,\rho_X)}.
\end{equation}
Here, $\Pi : L^2(X,\rho_X)\to \overline{\mathcal{H}}$ denotes the orthogonal projector onto the closure of the reproducing kernel Hilbert space $\mathcal{H}$ in $L^2(X,\rho_X)$, i.e.
\[
\Pi f = \arg\min_{h\in\mathcal{H}} \|f-h\|_{\rho_X}.
\]
\end{theorem}
By reverting to our notation where $\operatorname{Pr}^*_{D,\boldsymbol{c}_{\mathcal{C}\!\ell}} = \operatorname{Pr}^*_{D,\boldsymbol{c}_{\mathcal{C}\!\ell}}(\btheta^*_{\cl})$ and $\operatorname{Pr}_{\mathcal{Q}} = \operatorname{Pr}_{\mathcal{Q}}(\boldsymbol \theta^*_{\mathcal{C}\!\ell})$,we can proof the discrepancy sources result.
\begin{equation}
\begin{aligned}
\risk[\operatorname{Pr}_{D,\mathcal{C}\!\ell}]-\risk[\operatorname{Pr}_{\mathcal{Q}}] &= (\risk[\operatorname{Pr}_{D,\mathcal{C}\!\ell}]-\risk[\operatorname{Pr}^*_{\mathcal{Q}}])-(\risk[\operatorname{Pr}_{\mathcal{Q}}] - \risk[\operatorname{Pr}^*_{\mathcal{Q}}] ) \\ 
&= \|\opr^*_{D,\cl}-\Pi_\Q\opr^*\|^2 - \|\opr_{\Q}-\Pi_\Q\opr^*\|^2 \\
&= \langle \opr^*_{D,\cl} - \opr_{\Q} , (\opr^*_{D,\cl}-\Pi_\Q\opr^*)+(\opr_{\Q}-\Pi_\Q\opr^*)\rangle.
\end{aligned}
\end{equation}
where we have used the equality $\|u\|^2-\|v\|^2=\langle u, u\rangle-\langle v, v\rangle=\langle u-v, u+v\rangle$
Applying Cauchy-Schwarz $| \langle u,v\rangle |\leq \|u\|\|v\|$,
\begin{equation}
\begin{aligned}
    |\risk[\operatorname{Pr}_{D,\mathcal{C}\!\ell}]-\risk[\operatorname{Pr}_{\mathcal{Q}}]| &\leq \| \opr^*_{D,\cl} - \opr_\Q\| \| (\opr^*_{D,\cl}-\Pi_{\Q}\opr^*)+(\opr_{\Q}-\Pi_\Q\opr^*)\|  \\
    &= C \| \opr^*_{D,\cl} - \opr_\Q \| \\
    &= C \| \langle \ccl(\btheta^*_{\cl}) - \cq(\btheta^*_{\cl}), \boldsymbol{\varphi}_{\cl}+ \boldsymbol{\varphi}_{\Q} \rangle \| \\
    &\leq C \|(\ccl(\btheta^*_{\cl}) - \cq(\btheta^*_\Q)) + (\cq(\btheta^*_\Q)-\cq(\btheta^*_{\cl}))\|\\
    &\leq C (\|(\ccl(\btheta^*_{\cl}) - \cq(\btheta^*_\Q))\| + \|(\cq(\btheta^*_\Q)-\cq(\btheta^*_{\cl}))\|)
\end{aligned}
\end{equation}
with $C = \| (\opr^*_{D,\cl}-\Pi_{\Q}\opr^*)+(\opr_{\Q}-\Pi_\Q\opr^*)\| > 0$, and used the fact that we can embed the classical features in the quantum space with zero padding, $\|\boldsymbol{\varphi}_{\cl}+ \boldsymbol{\varphi}_{\Q}\| = 1$.

This gives the discrepancy sources for Theorem~\ref{thm:discrepancy}.

\section{Matchcircuits}\label{app:lie_modules}
In this section, we provide the derivation for Eqs.~\ref{eq:exp_mc} and \ref{eq:var_prob} in the main text, computing the variance of truncated distributions generated by matchcircuits.

When working with matchgate or matchcircuits, it is convenient to introduce a Majorana representation of Pauli operators acting on $n$ qubits.  
The $2n$ Majorana operators are defined as
\begin{equation}
\begin{aligned}
& c_1 = X_1 \mathds{1} \cdots \mathds{1}, \quad 
  c_3 = Z_1 X_2 \mathds{1} \cdots \mathds{1}, \quad \ldots, \quad 
  c_{2n-1} = Z_1 Z_2 \cdots Z_{n-1} X_n, \\
& c_2 = Y_1 \mathds{1} \cdots \mathds{1}, \quad 
  c_4 = Z_1 Y_2 \mathds{1} \cdots \mathds{1}, \quad \ldots, \quad 
  c_{2n} = Z_1 Z_2 \cdots Z_{n-1} Y_n,
\end{aligned}
\end{equation}
and satisfy the canonical anti-commutation relations
\begin{equation}
\{ c_\mu, c_\nu \} = 2\delta_{\mu\nu}, \qquad \mu,\nu \in \{1,\ldots,2n\}.
\end{equation}

\begin{lemma}
Let $\mathcal{B}$ denote the space of linear operators acting on an $n$-qubit Hilbert space. Then $\mathcal{B}$ admits the orthogonal decomposition
\[
\mathcal{B} = \bigoplus_{\kappa=0}^{2n} \mathcal{B}_\kappa,
\]
where each subspace $\mathcal{B}_\kappa$ is spanned by all possible products of $\kappa$ distinct Majorana operators and has dimension $\binom{2n}{\kappa}$.
\end{lemma}

\begin{proof}[Proof sketch]
The $2n$ Majoranas generate a basis for the Clifford algebra $\mathcal{C}\ell_{2n}$, whose elements are linear combinations of products of Majoranas.  
By grouping terms according to the number $\kappa$ of distinct Majoranas appearing in each monomial, one obtains the stated orthogonal decomposition of $\mathcal{B}$.  
\end{proof}

This decomposition allows any operator $M \in \mathcal{B}$ to be written as
\begin{equation}
M = \sum_{\kappa=0}^{2n} M_\kappa, \qquad M_\kappa \in \mathcal{B}_\kappa,
\end{equation}
where each $M_\kappa$ is a homogeneous polynomial of degree $\kappa$ in the Majoranas.  
The component $M_\kappa$ can be obtained through orthogonal projection,
\begin{equation}
M_\kappa = \sum_{j=1}^{\dim(\mathcal{B}_\kappa)} \mathrm{Tr}[B_j M]\, B_j,
\end{equation}
with $\{B_j\}$ a Hermitian orthonormal basis of $\mathcal{B}_\kappa$ under the Hilbert--Schmidt inner product.  
In particular, $\mathcal{B}_0 = \mathrm{span}_{\mathbb{C}}\{\mathds{1}\}$, and $\mathcal{B}_{2n} = \mathrm{span}_{\mathbb{C}}\{Z^{\otimes n}\}$ corresponds to the fermionic parity operator.  
Furthermore, basis elements in $\mathcal{B}_\kappa$ are related to those in $\mathcal{B}_{2n-\kappa}$ by multiplication with the parity operator $P$.

Because each Pauli-$Z$ operator corresponds to the product of two Majoranas, a correlator involving $m$ Pauli-$Z$ terms corresponds to $\kappa = 2m$ Majorana operators,
\begin{equation}
Z_{\boldsymbol i} = \prod_{j\in\boldsymbol i} c_{2j-1} c_{2j} \qquad\text{so that}\qquad Z_{\boldsymbol i} \in \mathcal B_{2|\boldsymbol i|} 
\end{equation}
Now, having introduced the required terminology and background, we are in a position to prove the main Lemma~\ref{thm:var_mg} of this section, .
\begin{theorem}[Variance of matchcircuits~\cite{diaz_showcasing_2023}]
Consider a parameterized matchcircuit
\(
U_{\mathrm{MG}} = U_{\mathrm{MG}}(\boldsymbol{\theta})
\)
acting on an initial state $\rho$ and measured with observable $O$.  
The variance of the expectation value over the parameters $\boldsymbol{\theta}$ is given by
\begin{equation}\label{eq:matchcircuit}
\Var^{\mathrm{MG}}_{\boldsymbol{\theta}}(O)
 = \Var_{\boldsymbol{\theta}}\!\left[ 
   \mathrm{Tr}\!\left(U_{\mathrm{MG}}\, \rho\, U_{\mathrm{MG}}^\dagger O\right)
 \right]
 = \sum_{\kappa=1}^{2n-1}
   \frac{
     \mathcal{P}_\kappa(\rho)\, \mathcal{P}_\kappa(O)
     + \mathcal{C}_\kappa(\rho)\, \mathcal{C}_\kappa(O)
   }{
     \dim(\mathcal{B}_\kappa)
   }.
\end{equation}
Here, the \emph{\(\kappa\)-purity} of an operator $M\in\mathcal{B}$ is defined as  
\(
\mathcal{P}_\kappa(M)
   = \langle M_\kappa, M_\kappa \rangle_{\mathds{1}},
\)
and its \emph{\(\kappa\)-coherence} as  
\(
\mathcal{C}_\kappa(M)
   = i^{\kappa \bmod 2}
     \langle M_\kappa, M_{2n-\kappa} \rangle_P.
\)
The inner product is given by
\(
\langle M_1, M_2 \rangle_\Gamma
   = \mathrm{Tr}[\Gamma\, M_1^\dagger M_2],
\)
for $M_1, M_2, \Gamma \in \mathcal{B}$. Finally, $dim(\mathcal{B}_\kappa)=\binom{2n}{\kappa}$.
\end{theorem}

The initial state is typically taken as
$
\rho = \ket{\boldsymbol{0}}\!\bra{\boldsymbol{0}} = \frac{1}{2^{n}}\sum_{\boldsymbol{i}\in 2^{[n]}}\langle Z_{\boldsymbol i} \rangle,
$ with 
\begin{equation}\label{eq:state_k_proj}
\rho_{\kappa}= \begin{cases} \frac{1}{2^n} \sum_{\boldsymbol j: |\boldsymbol j| = \kappa/2} \langle Z_{\bj} \rangle Z_{\bj} &\text{ if } \kappa \text{ is even,} \\  0 &\text{ if } \kappa \text{ is odd.}
\end{cases}
\end{equation}
This renders the following values for the even-$\kappa$ purity and coherence
\begin{equation}
\mathcal{P}_{\kappa=2k}(\rho) = \frac{1}{2^{2n}}\binom{n}{k}\operatorname{Tr}(\1) = \frac{1}{2^n}\binom{n}{k}.
\end{equation}
As $Z^{\otimes n} \ket{\bzero}\!\!\bra{\bzero} =  +\ket{\bzero}\!\!\bra{\bzero}$ then $\mathcal{C}_{\kappa}(\rho) = \mathcal{P}_{\kappa}(\rho)$.

A general correlator has the form
\(
O = \langle Z_{\boldsymbol{i}} \rangle
\)
such that,
\begin{equation}
\langle Z_{\bi} \rangle_{\kappa} = \begin{cases} Z_{\bi} &\text{if } |\bi|=\kappa/2 \\ 0 &\text{o.w.}\end{cases} \implies \mathcal{P}_\kappa(\langle Z_{\bi})\rangle) = \begin{cases} 2^n &\text{ if } |\bi| = \kappa/2 \\ 0 &\text{o.w.} \end{cases} \qquad \text{and} \qquad \mathcal{C}_\kappa(\langle Z_{\bi}\rangle) = 0
\end{equation}
Putting everything together with Eq.~\ref{eq:matchcircuit}, we just have one term contributing to the overall sum making the variance for a $k$-order correlator,
\begin{equation}
\Var^{\rm MG}_{\btheta}[\langle Z_{\bi}\rangle] = \binom{n}{k}\binom{2n}{2k}^{-1} \text{ with } k=|\bi|
\end{equation}
In the context of a truncated expansion of the probability distribution,
the observable associated with the \(k\)-order truncation reads
\begin{equation}
O = \operatorname{Pr}^{(k)}_{\boldsymbol{\theta}}(\boldsymbol x) = \frac{1}{2^n}
\sum_{p=0}^{k}
\sum_{\substack{\boldsymbol{i} \subseteq [n]\\ |\boldsymbol{i}|=p}}
(-1)^{\sum_{i\in\boldsymbol{i}} x_i}
\big\langle Z_{\boldsymbol{i}} \big\rangle_{\boldsymbol{\theta}},
\end{equation}
corresponding to correlators supported on $\kappa = 0,2,4,...,2k$ Majoranas. As $-Z\ket{0} = XZ\ket{1}$, the signs do not contribute to the purity or coherence (see~\cite{diaz_showcasing_2023}) we can drop the signed terms for each individual bitstring.
\begin{equation}
\left[\operatorname{Pr}^{(k)}_{\boldsymbol{\theta}}(\boldsymbol x)\right]_{\kappa} = \begin{cases} \frac{1}{2^{n}}\sum_{|\bj|=\kappa/2}Z_{\bj}& \text{ if } \frac{\kappa}{2}\leq k \\ 0 & \text{o.w.}\end{cases}
\end{equation}

\begin{equation}
\mathcal{P}_{\kappa}\left(\operatorname{Pr}^{(k)}_{\btheta}(\boldsymbol x)\right) =
\begin{cases}
\frac{1}{2^{2n}}\operatorname{Tr}[(\sum_{|\bj|=\kappa/2}Z_{\bj})^2]= \frac{1}{2^n}\binom{n}{\kappa/2}& \text{if } \frac{\kappa}{2} \leq k\\ 0
& \text{o.w.}
\end{cases}
\end{equation}

\begin{equation}
\mathcal{C}_{\kappa}\left(\operatorname{Pr}^{(k)}_{\btheta}(\boldsymbol x)\right) =
\begin{cases}
\frac{1}{2^{2n}}\operatorname{Tr}[Z^{\otimes n}(\sum_{|\bj|=\kappa/2}Z_{\bj})(\sum_{|\bj|=n-\kappa/2}Z_{\bj})]= \frac{1}{2^n}\binom{n}{\kappa/2}& \text{if } \frac{\kappa}{2} \leq k \text{ and } n-\frac{\kappa}{2}\leq k\\ 0
& \text{o.w.}
\end{cases}
\end{equation}
Combining these results with Eq~\ref{eq:matchcircuit} and Eq.~\ref{eq:state_k_proj}, the variance is 
\begin{equation}\label{eq_app:var_prob}
\Var_{\btheta}^{\rm MG}\left(\operatorname{Pr}_{\btheta}^{(k)}(\bx)\right) = \frac{1}{{2^{2n}}}\sum_{p=1}^{n-1}\mathbb{V}_p^{(k)}
\end{equation}
where,
\begin{equation}
\operatorname{\mathbb{V}}_p^{(k)} =
\begin{cases}
\binom{n}{p}^2 \binom{2n}{2p}^{-1} & \text{if } p \leq k, \\[1em]
2\binom{n}{p}^2 \binom{2n}{2p}^{-1} & \text{if } p \leq k \text{ and } n - p \leq k, \\[1em]
0 & \text{otherwise.}
\end{cases}
\end{equation}

which is the expression for Eq.~\ref{eq:var_prob} in the main text.

\section{Proofs for Tensor Networks}\label{app:tn_proofs}
Next, we provide the relevant proofs for Section~\ref{subsec:var_tensor_networks} in the main text, specifically computing variances of the truncated distributions from surrogate tensor networks. 

As we are interested in integrating polynomials over the unitary groups with respect to the Haar measure, we make use of the Weingarten calculus \cite{mele_introduction_2024,haferkamp_emergent_2021}

\begin{definition}[Weingarten Calculus]
The $t$-th moment operator with respect to the probability measure is obtained via the Weingarten coefficients $\operatorname{Wg}\left(\pi^{-1} \sigma, \ell \chi \right)$ for dimension $\ell \chi$ as
\begin{equation}
    \underset{U \sim \mu_H}{\mathbb{E}}\left[U^{\otimes t} \otimes U^{\dagger \otimes t}\right]=\sum_{\pi, \sigma \in S_t} \operatorname{Wg}\left(\pi^{-1} \sigma, \ell \chi\right) \ket{\sigma}\!\!\bra{\pi}.
\end{equation}
The permutation tensor $\ket{\sigma}$ is defined in terms of the representation $r$ in $S_t$ as $|\sigma\rangle=(\1 \otimes r(\sigma))|\Omega\rangle$, where $|\Omega\rangle=\sum_{j=1}^{q^t}|j, j\rangle$ is the maximally  entangled state vector.
\end{definition}
For $t=2$, $S_2 = \{\1,\F\}$ and the Weingarten coefficients are 
\begin{equation}
\begin{aligned}
    \operatorname{Wg}(\1^{-1} \1,\ell \chi) &= \operatorname{Wg}(\F^{-1} \F,\ell \chi) =  \frac{1}{(\ell \chi)^2-1},\\
    \operatorname{Wg}(\1^{-1} \F,\ell \chi) &= \operatorname{Wg}(\F^{-1} \1,\ell \chi) =  \frac{-\ell \chi}{(\ell \chi)^2-1}.
\end{aligned}
\end{equation}
Along with Eq.~\ref{eq:mps}, it is then possible to evaluate each individual $\E_{U_i} U_i^{\otimes 2}\otimes  U_i^{\dagger\otimes 2} \equiv \E_{U_i} U_i^{\otimes2,2} $ and obtain that for an initial state $\rho=\ket{\psi}\!\!\bra{\psi}\equiv\ket{\bzero}\!\!\bra{\bzero}$ that
\begin{equation}
\E_{U_i} \!\left[\Tr(U \rho U^{\dagger})^2\right]\! =\!\! \sum_{\bullet \in\{\1,\F\}^n} \raisebox{-2em}{\resizebox{.19\textwidth}{!}{\tikzset{every picture/.style={line width=0.75pt}} 

\begin{tikzpicture}[x=0.75pt,y=0.75pt,yscale=-1,xscale=1]

\draw    (31.88,45.44) -- (31.88,32.51) ;
\draw  [color={rgb, 255:red, 155; green, 155; blue, 155 }  ,draw opacity=1 ][fill={rgb, 255:red, 204; green, 204; blue, 204 }  ,fill opacity=1 ] (8.57,6.95) -- (53.6,6.95) -- (53.6,31.94) -- (8.57,31.94) -- cycle ;
\draw  [dash pattern={on 4.5pt off 4.5pt}]  (31.88,45.44) -- (7.12,57.36) ;
\draw  [fill={rgb, 255:red, 0; green, 0; blue, 0 }  ,fill opacity=1 ] (28.49,45.44) .. controls (28.49,43.57) and (30.01,42.05) .. (31.88,42.05) .. controls (33.76,42.05) and (35.27,43.57) .. (35.27,45.44) .. controls (35.27,47.31) and (33.76,48.83) .. (31.88,48.83) .. controls (30.01,48.83) and (28.49,47.31) .. (28.49,45.44) -- cycle ;
\draw   (32.52,48.94) .. controls (31.39,49.76) and (30.32,50.54) .. (30.32,51.44) .. controls (30.32,52.35) and (31.39,53.13) .. (32.52,53.94) .. controls (33.65,54.76) and (34.72,55.54) .. (34.72,56.44) .. controls (34.72,57.35) and (33.65,58.13) .. (32.52,58.94) .. controls (31.39,59.76) and (30.32,60.54) .. (30.32,61.44) .. controls (30.32,62.35) and (31.39,63.13) .. (32.52,63.94) .. controls (33.65,64.76) and (34.72,65.54) .. (34.72,66.44) .. controls (34.72,67.35) and (33.65,68.13) .. (32.52,68.94) .. controls (31.39,69.76) and (30.32,70.54) .. (30.32,71.44) .. controls (30.32,72.35) and (31.39,73.13) .. (32.52,73.94) .. controls (33.21,74.44) and (33.88,74.93) .. (34.3,75.44) ;
\draw  [fill={rgb, 255:red, 0; green, 0; blue, 0 }  ,fill opacity=1 ] (28.49,75.44) .. controls (28.49,73.57) and (30.01,72.05) .. (31.88,72.05) .. controls (33.76,72.05) and (35.27,73.57) .. (35.27,75.44) .. controls (35.27,77.31) and (33.76,78.83) .. (31.88,78.83) .. controls (30.01,78.83) and (28.49,77.31) .. (28.49,75.44) -- cycle ;
\draw    (31.88,88.37) -- (31.88,75.44) ;
\draw  [color={rgb, 255:red, 155; green, 155; blue, 155 }  ,draw opacity=1 ][fill={rgb, 255:red, 204; green, 204; blue, 204 }  ,fill opacity=1 ] (5.77,88.25) -- (50.8,88.25) -- (50.8,113.24) -- (5.77,113.24) -- cycle ;
\draw    (108.28,88.37) -- (108.28,75.44) ;
\draw  [color={rgb, 255:red, 155; green, 155; blue, 155 }  ,draw opacity=1 ][fill={rgb, 255:red, 204; green, 204; blue, 204 }  ,fill opacity=1 ] (84.97,6.95) -- (130,6.95) -- (130,31.94) -- (84.97,31.94) -- cycle ;
\draw  [fill={rgb, 255:red, 0; green, 0; blue, 0 }  ,fill opacity=1 ] (104.89,45.44) .. controls (104.89,43.57) and (106.41,42.05) .. (108.28,42.05) .. controls (110.16,42.05) and (111.67,43.57) .. (111.67,45.44) .. controls (111.67,47.31) and (110.16,48.83) .. (108.28,48.83) .. controls (106.41,48.83) and (104.89,47.31) .. (104.89,45.44) -- cycle ;
\draw   (108.92,48.94) .. controls (107.79,49.76) and (106.72,50.54) .. (106.72,51.44) .. controls (106.72,52.35) and (107.79,53.13) .. (108.92,53.94) .. controls (110.05,54.76) and (111.12,55.54) .. (111.12,56.44) .. controls (111.12,57.35) and (110.05,58.13) .. (108.92,58.94) .. controls (107.79,59.76) and (106.72,60.54) .. (106.72,61.44) .. controls (106.72,62.35) and (107.79,63.13) .. (108.92,63.94) .. controls (110.05,64.76) and (111.12,65.54) .. (111.12,66.44) .. controls (111.12,67.35) and (110.05,68.13) .. (108.92,68.94) .. controls (107.79,69.76) and (106.72,70.54) .. (106.72,71.44) .. controls (106.72,72.35) and (107.79,73.13) .. (108.92,73.94) .. controls (109.61,74.44) and (110.28,74.93) .. (110.7,75.44) ;
\draw  [fill={rgb, 255:red, 0; green, 0; blue, 0 }  ,fill opacity=1 ] (104.89,75.44) .. controls (104.89,73.57) and (106.41,72.05) .. (108.28,72.05) .. controls (110.16,72.05) and (111.67,73.57) .. (111.67,75.44) .. controls (111.67,77.31) and (110.16,78.83) .. (108.28,78.83) .. controls (106.41,78.83) and (104.89,77.31) .. (104.89,75.44) -- cycle ;
\draw  [color={rgb, 255:red, 155; green, 155; blue, 155 }  ,draw opacity=1 ][fill={rgb, 255:red, 204; green, 204; blue, 204 }  ,fill opacity=1 ] (82.17,88.25) -- (127.2,88.25) -- (127.2,113.24) -- (82.17,113.24) -- cycle ;
\draw    (108.28,45.44) -- (108.28,32.51) ;

\draw  [dash pattern={on 4.5pt off 4.5pt}]  (146.32,60.16) -- (190.72,60.16) ;
\draw    (31.88,75.44) -- (108.28,45.44) ;
\draw    (226.28,88.37) -- (226.28,75.44) ;
\draw  [color={rgb, 255:red, 155; green, 155; blue, 155 }  ,draw opacity=1 ][fill={rgb, 255:red, 204; green, 204; blue, 204 }  ,fill opacity=1 ] (202.97,6.95) -- (248,6.95) -- (248,31.94) -- (202.97,31.94) -- cycle ;
\draw  [fill={rgb, 255:red, 0; green, 0; blue, 0 }  ,fill opacity=1 ] (222.89,45.44) .. controls (222.89,43.57) and (224.41,42.05) .. (226.28,42.05) .. controls (228.16,42.05) and (229.67,43.57) .. (229.67,45.44) .. controls (229.67,47.31) and (228.16,48.83) .. (226.28,48.83) .. controls (224.41,48.83) and (222.89,47.31) .. (222.89,45.44) -- cycle ;
\draw   (226.92,48.94) .. controls (225.79,49.76) and (224.72,50.54) .. (224.72,51.44) .. controls (224.72,52.35) and (225.79,53.13) .. (226.92,53.94) .. controls (228.05,54.76) and (229.12,55.54) .. (229.12,56.44) .. controls (229.12,57.35) and (228.05,58.13) .. (226.92,58.94) .. controls (225.79,59.76) and (224.72,60.54) .. (224.72,61.44) .. controls (224.72,62.35) and (225.79,63.13) .. (226.92,63.94) .. controls (228.05,64.76) and (229.12,65.54) .. (229.12,66.44) .. controls (229.12,67.35) and (228.05,68.13) .. (226.92,68.94) .. controls (225.79,69.76) and (224.72,70.54) .. (224.72,71.44) .. controls (224.72,72.35) and (225.79,73.13) .. (226.92,73.94) .. controls (227.61,74.44) and (228.28,74.93) .. (228.7,75.44) ;
\draw  [fill={rgb, 255:red, 0; green, 0; blue, 0 }  ,fill opacity=1 ] (222.89,75.44) .. controls (222.89,73.57) and (224.41,72.05) .. (226.28,72.05) .. controls (228.16,72.05) and (229.67,73.57) .. (229.67,75.44) .. controls (229.67,77.31) and (228.16,78.83) .. (226.28,78.83) .. controls (224.41,78.83) and (222.89,77.31) .. (222.89,75.44) -- cycle ;
\draw  [color={rgb, 255:red, 155; green, 155; blue, 155 }  ,draw opacity=1 ][fill={rgb, 255:red, 204; green, 204; blue, 204 }  ,fill opacity=1 ] (200.17,88.25) -- (245.2,88.25) -- (245.2,113.24) -- (200.17,113.24) -- cycle ;
\draw    (226.28,45.44) -- (226.28,32.51) ;

\draw    (108.28,75.44) -- (138.72,63.36) ;
\draw    (195.84,57.52) -- (226.28,45.44) ;
\draw  [dash pattern={on 4.5pt off 4.5pt}]  (251.04,63.52) -- (226.28,75.44) ;

\draw (60,41.28) node [anchor=north west][inner sep=0.75pt]    {$\chi $};
\draw (87.97,10.35) node [anchor=north west][inner sep=0.75pt]    {$| 0\rangle ^{\otimes 2,2}$};
\draw (85.17,91.65) node [anchor=north west][inner sep=0.75pt]    {$| 0\rangle ^{\otimes 2,2}$};
\draw (203.17,91.65) node [anchor=north west][inner sep=0.75pt]    {$| 0\rangle ^{\otimes 2,2}$};
\draw (205.97,10.35) node [anchor=north west][inner sep=0.75pt]    {$| 0\rangle ^{\otimes 2,2}$};
\draw (17.42,70.61) node [anchor=north west][inner sep=0.75pt]    {$l$};
\draw (11.57,10.35) node [anchor=north west][inner sep=0.75pt]    {$| 0\rangle ^{\otimes 2,2}$};
\draw (8.77,91.65) node [anchor=north west][inner sep=0.75pt]    {$| 0\rangle ^{\otimes 2,2}$};

\end{tikzpicture}}}.
\end{equation}
Again, vertical edges denote contractions over $\mathbb{C}^\ell$ and now diagonals are over $\mathbb{C}^\chi$. The contraction over any permutation $\1$ or $\F$ with the zero state is the unity, $\langle\pi|0\rangle^{\otimes 2,2} = |\langle0|0\rangle|^2=1$ and hence we omit the $\ket{0}^{\otimes 2,2}$ in future diagrams. Moreover, contractions over $\mathbb{C}^q$ with $q$ being $\chi$ or $\ell$ for the permutations render $\langle \1 | \1 \rangle = \langle \F | \F \rangle = q^2$ and $\langle \1 | \F \rangle = \langle \F | \1 \rangle = q$. The wiggly line (\!{\resizebox{.035\textwidth}{!}{\tikzset{every picture/.style={line width=0.75pt}} 

\begin{tikzpicture}[x=0.75pt,y=0.75pt,yscale=-1,xscale=1]

\draw   (23.08,34.3) .. controls (24.71,38.04) and (26.27,41.6) .. (28.08,41.6) .. controls (29.89,41.6) and (31.45,38.04) .. (33.08,34.3) .. controls (34.71,30.56) and (36.27,27) .. (38.08,27) .. controls (39.89,27) and (41.45,30.56) .. (43.08,34.3) .. controls (44.71,38.04) and (46.27,41.6) .. (48.08,41.6) .. controls (49.89,41.6) and (51.45,38.04) .. (53.08,34.3) .. controls (54.71,30.56) and (56.27,27) .. (58.08,27) .. controls (59.89,27) and (61.45,30.56) .. (63.08,34.3) .. controls (63.21,34.61) and (63.35,34.91) .. (63.48,35.21) ;

\end{tikzpicture}}}\!) accounts for the Weingarten coefficients.

We will now focus on the calculation of the Renyi-2 entropy $S_2(\rho_{k,n})$ which depends on $\F_{\bi}$. This entropy serves as a proxy for vanishing loss function detection~\cite{sack_avoiding_2022}. The truncated probability depending on observables $Z_{\bi}$ and exact marginals $\Pi_{\bx}^{(m)}$ of size $m$ are the projection to $\ket{\bx}\!\!\bra{\bx}$. Let $Q_{\bi}$ denote the variable that the quantity of interest depends on, then, to calculate the individual elements of $\E\left(\Tr[( Q_{\bi} U \rho U^{\dagger}) ^2]\right)$ we use 
\begin{equation}
 \raisebox{-.4em}{\resizebox{.08\textwidth}{!}{\tikzset{every picture/.style={line width=0.75pt}} 

\begin{tikzpicture}[x=0.75pt,y=0.75pt,yscale=-1,xscale=1]

\draw    (19.1,35.5) -- (35.6,35.5) ;
\draw  [color={rgb, 255:red, 61; green, 97; blue, 63 }  ,draw opacity=1 ][fill={rgb, 255:red, 190; green, 194; blue, 159 }  ,fill opacity=1 ] (36.09,23.01) -- (71.28,23.01) -- (71.28,47.99) -- (36.09,47.99) -- cycle ;
\draw    (71.28,35.5) -- (87.51,35.5) ;

\draw (43.19,26.4) node [anchor=north west][inner sep=0.75pt]  [font=\large]  {$Q_{i}$};

\end{tikzpicture}}}= \sum_{\bullet \in\1,\F} \!\!\raisebox{-1.5em}{\resizebox{.08\textwidth}{!}{\tikzset{every picture/.style={line width=0.75pt}} 

\begin{tikzpicture}[x=0.75pt,y=0.75pt,yscale=-1,xscale=1]

\draw    (48.06,16.94) -- (48.06,36.24) ;
\draw   (7.08,16.3) .. controls (8.71,20.04) and (10.27,23.6) .. (12.08,23.6) .. controls (13.89,23.6) and (15.45,20.04) .. (17.08,16.3) .. controls (18.71,12.56) and (20.27,9) .. (22.08,9) .. controls (23.89,9) and (25.45,12.56) .. (27.08,16.3) .. controls (28.71,20.04) and (30.27,23.6) .. (32.08,23.6) .. controls (33.89,23.6) and (35.45,20.04) .. (37.08,16.3) .. controls (38.71,12.56) and (40.27,9) .. (42.08,9) .. controls (43.89,9) and (45.45,12.56) .. (47.08,16.3) .. controls (47.21,16.61) and (47.35,16.91) .. (47.48,17.21) ;
\draw  [fill={rgb, 255:red, 0; green, 0; blue, 0 }  ,fill opacity=1 ] (44.76,16.94) .. controls (44.76,15.12) and (46.24,13.64) .. (48.06,13.64) .. controls (49.88,13.64) and (51.36,15.12) .. (51.36,16.94) .. controls (51.36,18.76) and (49.88,20.24) .. (48.06,20.24) .. controls (46.24,20.24) and (44.76,18.76) .. (44.76,16.94) -- cycle ;
\draw    (80.28,16.94) -- (48.06,16.94) ;
\draw  [color={rgb, 255:red, 61; green, 97; blue, 63 }  ,draw opacity=1 ][fill={rgb, 255:red, 190; green, 194; blue, 159 }  ,fill opacity=1 ] (30.69,36.01) -- (65.88,36.01) -- (65.88,60.99) -- (30.69,60.99) -- cycle ;

\draw (29.69,38.41) node [anchor=north west][inner sep=0.75pt]  [font=\Large]  {$\ket{Q_{i}}$};

\end{tikzpicture}}} \quad \text{ yielding } \quad \E\left(\Tr[( Q_{\bi} U \rho U^{\dagger}) ^2]\right) = \sum_{\bullet \in\{\1,\F\}^n} \!\!\raisebox{-.1em}{\resizebox{.2\textwidth}{!}{\tikzset{every picture/.style={line width=0.75pt}} 

\begin{tikzpicture}[x=0.75pt,y=0.75pt,yscale=-1,xscale=1]

\draw  [color={rgb, 255:red, 61; green, 97; blue, 63 }  ,draw opacity=1 ][fill={rgb, 255:red, 190; green, 194; blue, 159 }  ,fill opacity=1 ] (29.81,14.01) -- (65,14.01) -- (65,38.99) -- (29.81,38.99) -- cycle ;
\draw    (65.1,26.5) -- (81.6,26.5) ;
\draw  [color={rgb, 255:red, 61; green, 97; blue, 63 }  ,draw opacity=1 ][fill={rgb, 255:red, 190; green, 194; blue, 159 }  ,fill opacity=1 ] (82.09,14.01) -- (117.28,14.01) -- (117.28,38.99) -- (82.09,38.99) -- cycle ;
\draw    (117.28,26.5) -- (133.51,26.5) ;
\draw  [dash pattern={on 4.5pt off 4.5pt}]  (133.51,26.5) -- (184.93,26.5) ;
\draw    (184.93,26.5) -- (201.17,26.5) ;
\draw  [color={rgb, 255:red, 61; green, 97; blue, 63 }  ,draw opacity=1 ][fill={rgb, 255:red, 190; green, 194; blue, 159 }  ,fill opacity=1 ] (201.42,13.76) -- (236.6,13.76) -- (236.6,38.74) -- (201.42,38.74) -- cycle ;
\draw  [dash pattern={on 4.5pt off 4.5pt}]  (29.32,26.5) -- (13.7,26.5) ;
\draw  [fill={rgb, 255:red, 0; green, 0; blue, 0 }  ,fill opacity=1 ] (72,26.5) .. controls (72,25.78) and (72.58,25.2) .. (73.3,25.2) .. controls (74.02,25.2) and (74.6,25.78) .. (74.6,26.5) .. controls (74.6,27.22) and (74.02,27.8) .. (73.3,27.8) .. controls (72.58,27.8) and (72,27.22) .. (72,26.5) -- cycle ;
\draw  [fill={rgb, 255:red, 0; green, 0; blue, 0 }  ,fill opacity=1 ] (21,26.5) .. controls (21,25.78) and (21.58,25.2) .. (22.3,25.2) .. controls (23.02,25.2) and (23.6,25.78) .. (23.6,26.5) .. controls (23.6,27.22) and (23.02,27.8) .. (22.3,27.8) .. controls (21.58,27.8) and (21,27.22) .. (21,26.5) -- cycle ;
\draw  [fill={rgb, 255:red, 0; green, 0; blue, 0 }  ,fill opacity=1 ] (126.25,26.5) .. controls (126.25,25.78) and (126.83,25.2) .. (127.55,25.2) .. controls (128.27,25.2) and (128.85,25.78) .. (128.85,26.5) .. controls (128.85,27.22) and (128.27,27.8) .. (127.55,27.8) .. controls (126.83,27.8) and (126.25,27.22) .. (126.25,26.5) -- cycle ;
\draw  [fill={rgb, 255:red, 0; green, 0; blue, 0 }  ,fill opacity=1 ] (190.25,26.5) .. controls (190.25,25.78) and (190.83,25.2) .. (191.55,25.2) .. controls (192.27,25.2) and (192.85,25.78) .. (192.85,26.5) .. controls (192.85,27.22) and (192.27,27.8) .. (191.55,27.8) .. controls (190.83,27.8) and (190.25,27.22) .. (190.25,26.5) -- cycle ;
\draw  [dash pattern={on 4.5pt off 4.5pt}]  (236.84,26.5) -- (252.75,26.5) ;
\draw  [fill={rgb, 255:red, 0; green, 0; blue, 0 }  ,fill opacity=1 ] (242.9,26.5) .. controls (242.9,25.78) and (243.48,25.2) .. (244.2,25.2) .. controls (244.92,25.2) and (245.5,25.78) .. (245.5,26.5) .. controls (245.5,27.22) and (244.92,27.8) .. (244.2,27.8) .. controls (243.48,27.8) and (242.9,27.22) .. (242.9,26.5) -- cycle ;
\draw  [fill={rgb, 255:red, 0; green, 0; blue, 0 }  ,fill opacity=1 ] (19,26.5) .. controls (19,24.68) and (20.48,23.2) .. (22.3,23.2) .. controls (24.12,23.2) and (25.6,24.68) .. (25.6,26.5) .. controls (25.6,28.32) and (24.12,29.8) .. (22.3,29.8) .. controls (20.48,29.8) and (19,28.32) .. (19,26.5) -- cycle ;
\draw  [fill={rgb, 255:red, 0; green, 0; blue, 0 }  ,fill opacity=1 ] (70,26.5) .. controls (70,24.68) and (71.48,23.2) .. (73.3,23.2) .. controls (75.12,23.2) and (76.6,24.68) .. (76.6,26.5) .. controls (76.6,28.32) and (75.12,29.8) .. (73.3,29.8) .. controls (71.48,29.8) and (70,28.32) .. (70,26.5) -- cycle ;
\draw  [fill={rgb, 255:red, 0; green, 0; blue, 0 }  ,fill opacity=1 ] (124.25,26.5) .. controls (124.25,24.68) and (125.73,23.2) .. (127.55,23.2) .. controls (129.37,23.2) and (130.85,24.68) .. (130.85,26.5) .. controls (130.85,28.32) and (129.37,29.8) .. (127.55,29.8) .. controls (125.73,29.8) and (124.25,28.32) .. (124.25,26.5) -- cycle ;
\draw  [fill={rgb, 255:red, 0; green, 0; blue, 0 }  ,fill opacity=1 ] (188.25,26.5) .. controls (188.25,24.68) and (189.73,23.2) .. (191.55,23.2) .. controls (193.37,23.2) and (194.85,24.68) .. (194.85,26.5) .. controls (194.85,28.32) and (193.37,29.8) .. (191.55,29.8) .. controls (189.73,29.8) and (188.25,28.32) .. (188.25,26.5) -- cycle ;
\draw  [fill={rgb, 255:red, 0; green, 0; blue, 0 }  ,fill opacity=1 ] (242.2,26.5) .. controls (242.2,24.68) and (243.68,23.2) .. (245.5,23.2) .. controls (247.32,23.2) and (248.8,24.68) .. (248.8,26.5) .. controls (248.8,28.32) and (247.32,29.8) .. (245.5,29.8) .. controls (243.68,29.8) and (242.2,28.32) .. (242.2,26.5) -- cycle ;

\draw (38.19,17.4) node [anchor=north west][inner sep=0.75pt]  [font=\large]  {$Q_{1}$};
\draw (91.19,17.4) node [anchor=north west][inner sep=0.75pt]  [font=\large]  {$Q_{2}$};
\draw (211.19,17.4) node [anchor=north west][inner sep=0.75pt]  [font=\large]  {$Q_{n}$};

\end{tikzpicture}}}.
\end{equation}

\subsection{Entropy}
\begin{figure*}[htbp]
    \centering
    \resizebox{.8\linewidth}{!}{\input{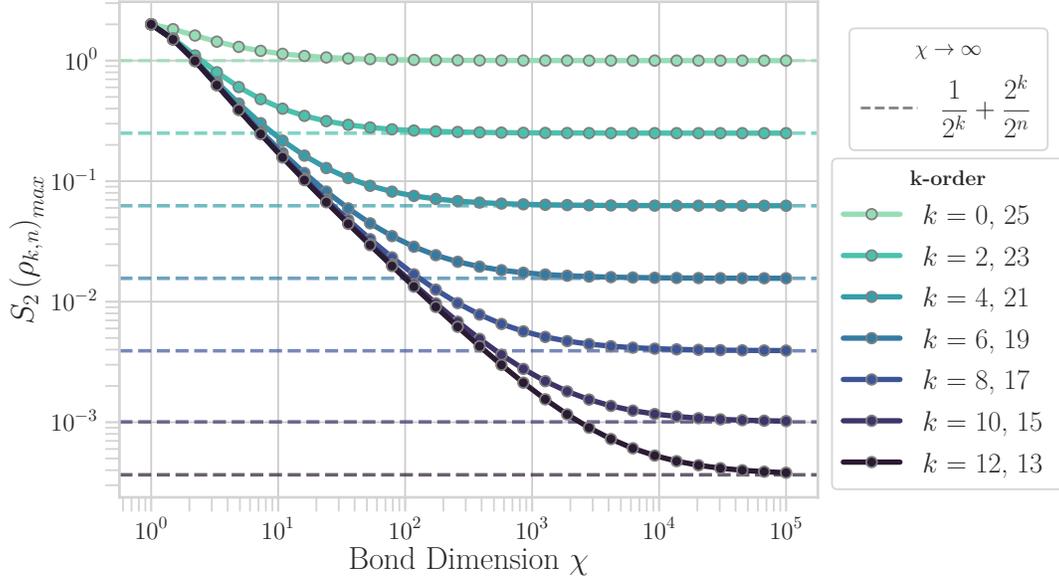}}
    \caption{\justifying Renyi-2 entropy as a function of the bond dimension for a subsystem of $k$ qubits in a system of 25 qubits and RMPS assumption.}
    \label{fig:entropy_TN_results}
\end{figure*}
Our contribution to this theory is the use of transition matrices $T$ that enable the obtention of exact values for the quantities of interest or at worst an efficient numerical implementation to obtain the exact value. This is possible because the matrices allow for the exact combination of each different string of $\bullet \in\{\1,\F\}^n$.

For \raisebox{-.1em}{\resizebox{.05\textwidth}{!}{\tikzset{every picture/.style={line width=0.75pt}} 

\begin{tikzpicture}[x=0.75pt,y=0.75pt,yscale=-1,xscale=1]

\draw  [color={rgb, 255:red, 84; green, 142; blue, 161 }  ,draw opacity=1 ][fill={rgb, 255:red, 186; green, 235; blue, 252 }  ,fill opacity=1 ] (56.81,22.01) -- (92,22.01) -- (92,46.99) -- (56.81,46.99) -- cycle ;
\draw    (92,34.5) -- (108.23,34.5) ;
\draw    (39.93,34.5) -- (56.17,34.5) ;

\draw (67.91,26.4) node [anchor=north west][inner sep=0.75pt]  [font=\Large]  {$\mathds{1}$};

\end{tikzpicture}}} = \raisebox{-.6em}{\resizebox{.05\textwidth}{!}{\tikzset{every picture/.style={line width=0.75pt}} 

\begin{tikzpicture}[x=0.75pt,y=0.75pt,yscale=-1,xscale=1]

\draw    (89.06,21.53) -- (89.06,40.83) ;
\draw   (48.08,20.89) .. controls (49.71,24.63) and (51.27,28.19) .. (53.08,28.19) .. controls (54.89,28.19) and (56.45,24.63) .. (58.08,20.89) .. controls (59.71,17.15) and (61.27,13.59) .. (63.08,13.59) .. controls (64.89,13.59) and (66.45,17.15) .. (68.08,20.89) .. controls (69.71,24.63) and (71.27,28.19) .. (73.08,28.19) .. controls (74.89,28.19) and (76.45,24.63) .. (78.08,20.89) .. controls (79.71,17.15) and (81.27,13.59) .. (83.08,13.59) .. controls (84.89,13.59) and (86.45,17.15) .. (88.08,20.89) .. controls (88.21,21.2) and (88.35,21.51) .. (88.48,21.81) ;
\draw  [fill={rgb, 255:red, 0; green, 0; blue, 0 }  ,fill opacity=1 ] (85.76,21.53) .. controls (85.76,19.71) and (87.24,18.23) .. (89.06,18.23) .. controls (90.88,18.23) and (92.36,19.71) .. (92.36,21.53) .. controls (92.36,23.36) and (90.88,24.83) .. (89.06,24.83) .. controls (87.24,24.83) and (85.76,23.36) .. (85.76,21.53) -- cycle ;
\draw    (121.28,21.53) -- (89.06,21.53) ;
\draw  [color={rgb, 255:red, 84; green, 142; blue, 161 }  ,draw opacity=1 ][fill={rgb, 255:red, 186; green, 235; blue, 252 }  ,fill opacity=1 ] (71.69,40.6) -- (106.88,40.6) -- (106.88,65.59) -- (71.69,65.59) -- cycle ;

\draw (76.69,42) node [anchor=north west][inner sep=0.75pt]  [font=\Large]  {$\ket{\mathbb{1}}$};

\end{tikzpicture}}}:
\newlength{\defaultarraycolsep}
\setlength{\defaultarraycolsep}{\arraycolsep}
\setlength{\arraycolsep}{1.5em}
\begin{equation}\label{eq:array_I}
\begin{array}{ll}
\1\raisebox{-.15em}{\resizebox{.06\textwidth}{!}{\tikzset{every picture/.style={line width=0.75pt}} 

\begin{tikzpicture}[x=0.75pt,y=0.75pt,yscale=-1,xscale=1]

\draw  [color={rgb, 255:red, 84; green, 142; blue, 161 }  ,draw opacity=1 ][fill={rgb, 255:red, 186; green, 235; blue, 252 }  ,fill opacity=1 ] (56.81,22.01) -- (92,22.01) -- (92,46.99) -- (56.81,46.99) -- cycle ;
\draw    (92,34.5) -- (108.23,34.5) ;
\draw    (39.93,34.5) -- (56.17,34.5) ;

\draw (67.91,26.4) node [anchor=north west][inner sep=0.75pt]  [font=\Large]  {$\mathds{1}$};

\end{tikzpicture}}}\1 = 1 & \1 \raisebox{-.15em}{\resizebox{.06\textwidth}{!}{\tikzset{every picture/.style={line width=0.75pt}} 

\begin{tikzpicture}[x=0.75pt,y=0.75pt,yscale=-1,xscale=1]

\draw  [color={rgb, 255:red, 84; green, 142; blue, 161 }  ,draw opacity=1 ][fill={rgb, 255:red, 186; green, 235; blue, 252 }  ,fill opacity=1 ] (56.81,22.01) -- (92,22.01) -- (92,46.99) -- (56.81,46.99) -- cycle ;
\draw    (92,34.5) -- (108.23,34.5) ;
\draw    (39.93,34.5) -- (56.17,34.5) ;

\draw (67.91,26.4) node [anchor=north west][inner sep=0.75pt]  [font=\Large]  {$\mathds{1}$};

\end{tikzpicture}}} \F = \frac{\chi\left(\ell^2-1\right)}{\ell^2 \chi^2-1}\\[1.5em]
\F \raisebox{-.15em}{\resizebox{.06\textwidth}{!}{\tikzset{every picture/.style={line width=0.75pt}} 

\begin{tikzpicture}[x=0.75pt,y=0.75pt,yscale=-1,xscale=1]

\draw  [color={rgb, 255:red, 84; green, 142; blue, 161 }  ,draw opacity=1 ][fill={rgb, 255:red, 186; green, 235; blue, 252 }  ,fill opacity=1 ] (56.81,22.01) -- (92,22.01) -- (92,46.99) -- (56.81,46.99) -- cycle ;
\draw    (92,34.5) -- (108.23,34.5) ;
\draw    (39.93,34.5) -- (56.17,34.5) ;

\draw (67.91,26.4) node [anchor=north west][inner sep=0.75pt]  [font=\Large]  {$\mathds{1}$};

\end{tikzpicture}}} \1 = 0 & \F \raisebox{-.15em}{\resizebox{.06\textwidth}{!}{\tikzset{every picture/.style={line width=0.75pt}} 

\begin{tikzpicture}[x=0.75pt,y=0.75pt,yscale=-1,xscale=1]

\draw  [color={rgb, 255:red, 84; green, 142; blue, 161 }  ,draw opacity=1 ][fill={rgb, 255:red, 186; green, 235; blue, 252 }  ,fill opacity=1 ] (56.81,22.01) -- (92,22.01) -- (92,46.99) -- (56.81,46.99) -- cycle ;
\draw    (92,34.5) -- (108.23,34.5) ;
\draw    (39.93,34.5) -- (56.17,34.5) ;

\draw (67.91,26.4) node [anchor=north west][inner sep=0.75pt]  [font=\Large]  {$\mathds{1}$};

\end{tikzpicture}}} \F = \frac{\ell\left(\chi^2-1\right)}{\ell^2 \chi^2-1}
\end{array}
\end{equation}
\setlength{\arraycolsep}{\defaultarraycolsep}

Moreover,for \raisebox{-.1em}{\resizebox{.05\textwidth}{!}{\tikzset{every picture/.style={line width=0.75pt}} 

\begin{tikzpicture}[x=0.75pt,y=0.75pt,yscale=-1,xscale=1]

\draw    (39.93,29.7) -- (56.17,29.7) ;
\draw  [color={rgb, 255:red, 51; green, 74; blue, 82 }  ,draw opacity=1 ][fill={rgb, 255:red, 150; green, 175; blue, 184 }  ,fill opacity=1 ] (56.42,17.21) -- (91.6,17.21) -- (91.6,42.19) -- (56.42,42.19) -- cycle ;
\draw    (91.93,29.7) -- (108.17,29.7) ;

\draw (66.51,21.6) node [anchor=north west][inner sep=0.75pt]  [font=\Large]  {$\mathbb{F}$};

\end{tikzpicture}}} = \raisebox{-.6em}{\resizebox{.05\textwidth}{!}{\tikzset{every picture/.style={line width=0.75pt}} 

\begin{tikzpicture}[x=0.75pt,y=0.75pt,yscale=-1,xscale=1]

\draw    (86.06,21.53) -- (86.06,40.83) ;
\draw   (45.08,20.89) .. controls (46.71,24.63) and (48.27,28.19) .. (50.08,28.19) .. controls (51.89,28.19) and (53.45,24.63) .. (55.08,20.89) .. controls (56.71,17.15) and (58.27,13.59) .. (60.08,13.59) .. controls (61.89,13.59) and (63.45,17.15) .. (65.08,20.89) .. controls (66.71,24.63) and (68.27,28.19) .. (70.08,28.19) .. controls (71.89,28.19) and (73.45,24.63) .. (75.08,20.89) .. controls (76.71,17.15) and (78.27,13.59) .. (80.08,13.59) .. controls (81.89,13.59) and (83.45,17.15) .. (85.08,20.89) .. controls (85.21,21.2) and (85.35,21.51) .. (85.48,21.81) ;
\draw  [fill={rgb, 255:red, 0; green, 0; blue, 0 }  ,fill opacity=1 ] (82.76,21.53) .. controls (82.76,19.71) and (84.24,18.23) .. (86.06,18.23) .. controls (87.88,18.23) and (89.36,19.71) .. (89.36,21.53) .. controls (89.36,23.36) and (87.88,24.83) .. (86.06,24.83) .. controls (84.24,24.83) and (82.76,23.36) .. (82.76,21.53) -- cycle ;
\draw    (118.28,21.53) -- (86.06,21.53) ;
\draw  [color={rgb, 255:red, 51; green, 74; blue, 82 }  ,draw opacity=1 ][fill={rgb, 255:red, 150; green, 175; blue, 184 }  ,fill opacity=1 ] (68.69,40.6) -- (103.88,40.6) -- (103.88,65.59) -- (68.69,65.59) -- cycle ;

\draw (70.69,44) node [anchor=north west][inner sep=0.75pt]  [font=\Large]  {$\ket{\mathbb{F}}$};

\end{tikzpicture}}}, we have the same result as for \raisebox{-.1em}{\resizebox{.05\textwidth}{!}{\tikzset{every picture/.style={line width=0.75pt}} 

\begin{tikzpicture}[x=0.75pt,y=0.75pt,yscale=-1,xscale=1]

\draw  [color={rgb, 255:red, 84; green, 142; blue, 161 }  ,draw opacity=1 ][fill={rgb, 255:red, 186; green, 235; blue, 252 }  ,fill opacity=1 ] (56.81,22.01) -- (92,22.01) -- (92,46.99) -- (56.81,46.99) -- cycle ;
\draw    (92,34.5) -- (108.23,34.5) ;
\draw    (39.93,34.5) -- (56.17,34.5) ;

\draw (67.91,26.4) node [anchor=north west][inner sep=0.75pt]  [font=\Large]  {$\mathds{1}$};

\end{tikzpicture}}} but swapping $\1$ and $\F$,

\setlength{\arraycolsep}{1.5em}
\begin{equation}\label{eq:array_F}
\begin{array}{ll}
\1\raisebox{-.1em}{\resizebox{.05\textwidth}{!}{\tikzset{every picture/.style={line width=0.75pt}} 

\begin{tikzpicture}[x=0.75pt,y=0.75pt,yscale=-1,xscale=1]

\draw    (39.93,29.7) -- (56.17,29.7) ;
\draw  [color={rgb, 255:red, 51; green, 74; blue, 82 }  ,draw opacity=1 ][fill={rgb, 255:red, 150; green, 175; blue, 184 }  ,fill opacity=1 ] (56.42,17.21) -- (91.6,17.21) -- (91.6,42.19) -- (56.42,42.19) -- cycle ;
\draw    (91.93,29.7) -- (108.17,29.7) ;

\draw (66.51,21.6) node [anchor=north west][inner sep=0.75pt]  [font=\Large]  {$\mathbb{F}$};

\end{tikzpicture}}}\1 = \frac{\ell\left(\chi^2-1\right)}{\ell^2 \chi^2-1} & \1 \raisebox{-.1em}{\resizebox{.05\textwidth}{!}{\tikzset{every picture/.style={line width=0.75pt}} 

\begin{tikzpicture}[x=0.75pt,y=0.75pt,yscale=-1,xscale=1]

\draw    (39.93,29.7) -- (56.17,29.7) ;
\draw  [color={rgb, 255:red, 51; green, 74; blue, 82 }  ,draw opacity=1 ][fill={rgb, 255:red, 150; green, 175; blue, 184 }  ,fill opacity=1 ] (56.42,17.21) -- (91.6,17.21) -- (91.6,42.19) -- (56.42,42.19) -- cycle ;
\draw    (91.93,29.7) -- (108.17,29.7) ;

\draw (66.51,21.6) node [anchor=north west][inner sep=0.75pt]  [font=\Large]  {$\mathbb{F}$};

\end{tikzpicture}}} \F = 0\\ [1.5em]
\F \raisebox{-.1em}{\resizebox{.05\textwidth}{!}{\tikzset{every picture/.style={line width=0.75pt}} 

\begin{tikzpicture}[x=0.75pt,y=0.75pt,yscale=-1,xscale=1]

\draw    (39.93,29.7) -- (56.17,29.7) ;
\draw  [color={rgb, 255:red, 51; green, 74; blue, 82 }  ,draw opacity=1 ][fill={rgb, 255:red, 150; green, 175; blue, 184 }  ,fill opacity=1 ] (56.42,17.21) -- (91.6,17.21) -- (91.6,42.19) -- (56.42,42.19) -- cycle ;
\draw    (91.93,29.7) -- (108.17,29.7) ;

\draw (66.51,21.6) node [anchor=north west][inner sep=0.75pt]  [font=\Large]  {$\mathbb{F}$};

\end{tikzpicture}}} \1 = \frac{\chi\left(\ell^2-1\right)}{\ell^2 \chi^2-1} & \F \raisebox{-.1em}{\resizebox{.05\textwidth}{!}{\tikzset{every picture/.style={line width=0.75pt}} 

\begin{tikzpicture}[x=0.75pt,y=0.75pt,yscale=-1,xscale=1]

\draw    (39.93,29.7) -- (56.17,29.7) ;
\draw  [color={rgb, 255:red, 51; green, 74; blue, 82 }  ,draw opacity=1 ][fill={rgb, 255:red, 150; green, 175; blue, 184 }  ,fill opacity=1 ] (56.42,17.21) -- (91.6,17.21) -- (91.6,42.19) -- (56.42,42.19) -- cycle ;
\draw    (91.93,29.7) -- (108.17,29.7) ;

\draw (66.51,21.6) node [anchor=north west][inner sep=0.75pt]  [font=\Large]  {$\mathbb{F}$};

\end{tikzpicture}}} \F = 1
\end{array}
\end{equation}
\setlength{\arraycolsep}{\defaultarraycolsep}
By defining the variables,
\begin{equation}
\eta_\ell=\frac{\ell\left(\chi^2-1\right)}{\ell^2 \chi^2-1}, \qquad \eta_\chi=\frac{\chi\left(\ell^2-1\right)}{\ell^2 \chi^2-1}
\end{equation}
we can calculate the maximum Renyi-2 entropy over different subsystems given by the subsystem with contiguous qubits,
\begin{equation}
S_2\left(\rho_{k,n}\right)_{\max }=\raisebox{-.3em}{\resizebox{.4\textwidth}{!}{\tikzset{every picture/.style={line width=0.75pt}} 

\begin{tikzpicture}[x=0.75pt,y=0.75pt,yscale=-1,xscale=1]

\draw  [color={rgb, 255:red, 51; green, 74; blue, 82 }  ,draw opacity=1 ][fill={rgb, 255:red, 150; green, 175; blue, 184 }  ,fill opacity=1 ] (27.81,12.97) -- (63,12.97) -- (63,37.96) -- (27.81,37.96) -- cycle ;
\draw    (63.1,25.46) -- (79.6,25.46) ;
\draw  [color={rgb, 255:red, 51; green, 74; blue, 82 }  ,draw opacity=1 ][fill={rgb, 255:red, 150; green, 175; blue, 184 }  ,fill opacity=1 ] (80.09,12.97) -- (115.28,12.97) -- (115.28,37.96) -- (80.09,37.96) -- cycle ;
\draw    (115.28,25.46) -- (131.51,25.46) ;
\draw  [dash pattern={on 4.5pt off 4.5pt}]  (131.51,25.46) -- (182.93,25.46) ;
\draw    (182.93,25.46) -- (199.17,25.46) ;
\draw  [color={rgb, 255:red, 51; green, 74; blue, 82 }  ,draw opacity=1 ][fill={rgb, 255:red, 150; green, 175; blue, 184 }  ,fill opacity=1 ] (199.42,12.97) -- (234.6,12.97) -- (234.6,37.96) -- (199.42,37.96) -- cycle ;
\draw  [dash pattern={on 4.5pt off 4.5pt}]  (27.32,25.46) -- (11.7,25.46) ;
\draw  [color={rgb, 255:red, 84; green, 142; blue, 161 }  ,draw opacity=1 ][fill={rgb, 255:red, 186; green, 235; blue, 252 }  ,fill opacity=1 ] (251.81,12.97) -- (287,12.97) -- (287,37.96) -- (251.81,37.96) -- cycle ;
\draw    (287,25.46) -- (303.23,25.46) ;
\draw  [color={rgb, 255:red, 84; green, 142; blue, 161 }  ,draw opacity=1 ][fill={rgb, 255:red, 186; green, 235; blue, 252 }  ,fill opacity=1 ] (304.09,12.97) -- (339.28,12.97) -- (339.28,37.96) -- (304.09,37.96) -- cycle ;
\draw    (339.28,25.46) -- (355.51,25.46) ;
\draw  [dash pattern={on 4.5pt off 4.5pt}]  (355.51,25.46) -- (406.93,25.46) ;
\draw    (406.93,25.46) -- (423.17,25.46) ;
\draw  [color={rgb, 255:red, 84; green, 142; blue, 161 }  ,draw opacity=1 ][fill={rgb, 255:red, 186; green, 235; blue, 252 }  ,fill opacity=1 ] (423.42,12.97) -- (458.6,12.97) -- (458.6,37.96) -- (423.42,37.96) -- cycle ;
\draw    (234.93,25.46) -- (251.17,25.46) ;
\draw  [fill={rgb, 255:red, 0; green, 0; blue, 0 }  ,fill opacity=1 ] (70,25.46) .. controls (70,24.75) and (70.58,24.16) .. (71.3,24.16) .. controls (72.02,24.16) and (72.6,24.75) .. (72.6,25.46) .. controls (72.6,26.18) and (72.02,26.76) .. (71.3,26.76) .. controls (70.58,26.76) and (70,26.18) .. (70,25.46) -- cycle ;
\draw  [fill={rgb, 255:red, 0; green, 0; blue, 0 }  ,fill opacity=1 ] (19,25.46) .. controls (19,24.75) and (19.58,24.16) .. (20.3,24.16) .. controls (21.02,24.16) and (21.6,24.75) .. (21.6,25.46) .. controls (21.6,26.18) and (21.02,26.76) .. (20.3,26.76) .. controls (19.58,26.76) and (19,26.18) .. (19,25.46) -- cycle ;
\draw  [fill={rgb, 255:red, 0; green, 0; blue, 0 }  ,fill opacity=1 ] (124.25,25.46) .. controls (124.25,24.75) and (124.83,24.16) .. (125.55,24.16) .. controls (126.27,24.16) and (126.85,24.75) .. (126.85,25.46) .. controls (126.85,26.18) and (126.27,26.76) .. (125.55,26.76) .. controls (124.83,26.76) and (124.25,26.18) .. (124.25,25.46) -- cycle ;
\draw  [fill={rgb, 255:red, 0; green, 0; blue, 0 }  ,fill opacity=1 ] (188.25,25.46) .. controls (188.25,24.75) and (188.83,24.16) .. (189.55,24.16) .. controls (190.27,24.16) and (190.85,24.75) .. (190.85,25.46) .. controls (190.85,26.18) and (190.27,26.76) .. (189.55,26.76) .. controls (188.83,26.76) and (188.25,26.18) .. (188.25,25.46) -- cycle ;
\draw  [fill={rgb, 255:red, 0; green, 0; blue, 0 }  ,fill opacity=1 ] (241.7,25.46) .. controls (241.7,24.75) and (242.28,24.16) .. (243,24.16) .. controls (243.72,24.16) and (244.3,24.75) .. (244.3,25.46) .. controls (244.3,26.18) and (243.72,26.76) .. (243,26.76) .. controls (242.28,26.76) and (241.7,26.18) .. (241.7,25.46) -- cycle ;
\draw  [fill={rgb, 255:red, 0; green, 0; blue, 0 }  ,fill opacity=1 ] (294.65,25.46) .. controls (294.65,24.75) and (295.23,24.16) .. (295.95,24.16) .. controls (296.67,24.16) and (297.25,24.75) .. (297.25,25.46) .. controls (297.25,26.18) and (296.67,26.76) .. (295.95,26.76) .. controls (295.23,26.76) and (294.65,26.18) .. (294.65,25.46) -- cycle ;
\draw  [fill={rgb, 255:red, 0; green, 0; blue, 0 }  ,fill opacity=1 ] (349.4,25.46) .. controls (349.4,24.75) and (349.98,24.16) .. (350.7,24.16) .. controls (351.42,24.16) and (352,24.75) .. (352,25.46) .. controls (352,26.18) and (351.42,26.76) .. (350.7,26.76) .. controls (349.98,26.76) and (349.4,26.18) .. (349.4,25.46) -- cycle ;
\draw  [fill={rgb, 255:red, 0; green, 0; blue, 0 }  ,fill opacity=1 ] (411.9,25.46) .. controls (411.9,24.75) and (412.48,24.16) .. (413.2,24.16) .. controls (413.92,24.16) and (414.5,24.75) .. (414.5,25.46) .. controls (414.5,26.18) and (413.92,26.76) .. (413.2,26.76) .. controls (412.48,26.76) and (411.9,26.18) .. (411.9,25.46) -- cycle ;
\draw  [dash pattern={on 4.5pt off 4.5pt}]  (458.84,25.46) -- (474.75,25.46) ;
\draw  [fill={rgb, 255:red, 0; green, 0; blue, 0 }  ,fill opacity=1 ] (464.9,25.46) .. controls (464.9,24.75) and (465.48,24.16) .. (466.2,24.16) .. controls (466.92,24.16) and (467.5,24.75) .. (467.5,25.46) .. controls (467.5,26.18) and (466.92,26.76) .. (466.2,26.76) .. controls (465.48,26.76) and (464.9,26.18) .. (464.9,25.46) -- cycle ;

\draw (153.5,5.7) node [anchor=north west][inner sep=0.75pt]  [font=\large]  {$k$};
\draw (359.5,5.1) node [anchor=north west][inner sep=0.75pt]  [font=\large]  {$n-k$};
\draw (37.91,17.36) node [anchor=north west][inner sep=0.75pt]  [font=\large]  {$\mathbb{F}$};
\draw (90.19,17.36) node [anchor=north west][inner sep=0.75pt]  [font=\large]  {$\mathbb{F}$};
\draw (209.51,17.36) node [anchor=north west][inner sep=0.75pt]  [font=\large]  {$\mathbb{F}$};
\draw (262.91,17.36) node [anchor=north west][inner sep=0.75pt]  [font=\large]  {$\mathds{1}$};
\draw (315.19,17.36) node [anchor=north west][inner sep=0.75pt]  [font=\large]  {$\mathds{1}$};
\draw (434.51,17.36) node [anchor=north west][inner sep=0.75pt]  [font=\large]  {$\mathds{1}$};

\end{tikzpicture}}} = \left(\begin{array}{ll}
1 & 1
\end{array}\right) T_{\mathbb F}^k \cdot T_{\mathds{1}}^{n - k}\binom{1}{1}.
\end{equation}
Where we introduce the transition matrices $T_\1$ and $T_\F$, from Eq.~\ref{eq:array_I} and Eq~\ref{eq:array_F} respectively, to be able to obtain a general formula in $n$ and $k$. We proceed by diagonalizing such matrices and simplifying,
\begin{equation}
T_{\mathds{1}}=\left(\begin{array}{ll}
1 & \eta_\chi \\
0 & \eta_\ell
\end{array}\right) \implies 
P_{\mathds{1}}=\left(\begin{array}{cc}
1 & \frac{\eta_\chi}{\eta_\ell-1} \\
0 & 1
\end{array}\right),\,
D_{\mathds{1}}=\left(\begin{array}{ll}
1 & 0 \\
0 & \eta_\ell
\end{array}\right),\,
P_{\mathds{1}}^{-1}=\left(\begin{array}{cc}
1 & \frac{-\eta_\chi}{\eta_\ell-1} \\
0 & 1
\end{array}\right)
\end{equation}

\begin{equation}
T_{\mathbb F}=\left(\begin{array}{ll}
\eta_\ell & 0 \\
\eta_\chi & 1
\end{array}\right) \implies
P_\mathbb F=\left(\begin{array}{cc}
0 & \frac{\eta_\ell-1}{\eta_{\chi}} \\
1 & 1
\end{array}\right),\,
D_{\mathbb{F}}=\left(\begin{array}{ll}
1 & 0 \\
0 & \eta_\ell
\end{array}\right),\,
P_{\mathbb F}^{-1}=\left(\begin{array}{cc}
\frac{-\eta_\chi}{\eta_\ell-1} & 1 \\
 \rule{0pt}{10pt} 
\frac{\eta_\chi}{\eta_\ell-1} & 0
\end{array}\right)
\end{equation}
We continue by diagonalizing the matrices to get the closed form,
\begin{equation}
\begin{aligned}
\operatorname{S_2}\left(\rho_{k,n}\right)_{\max}&=\left(\begin{array}{ll}
1 & 1
\end{array}\right) P_{\mathbb F} D_{\mathbb{F}}^k P_{\mathbb{F}}^{-1} \, P_\mathds{1} D_{\mathds{1}}^{n-k} P_{\mathds{1}}^{-1}\binom{1}{1}
\\
&=\left(\begin{array}{ll}
1 & \frac{(\eta_\ell+\eta_\chi-1) }{\eta_\chi}\eta_\ell^k
\end{array}\right)\left(
\begin{array}{cc}
 \frac{\eta_\chi}{1-\eta_\ell} & 1-\frac{\eta_\chi^2}{(\eta_\ell-1)^2} \\
 \rule{0pt}{15pt} 
 \frac{\eta_\chi}{\eta_\ell-1} & \frac{\eta_\chi^2}{(\eta_\ell-1)^2} \\
\end{array}
\right)
\left(
\begin{array}{c}
 \frac{\eta_\chi}{1-\eta_\ell}+1 \\
  \rule{0pt}{15pt} 
 \eta_\ell^{n-k} \\
\end{array}
\right)
\\
&=\frac{\eta_\ell^{-k} \left((\eta_\ell-\eta_\chi-1) 
\eta_\ell^k \left((\eta_\ell+\eta_\chi-1) \eta_\ell^k-
\eta_\chi\right)+\eta_\ell^{n}(\eta_\ell+\eta_\chi-1)  \left(
\eta_\chi \eta_\ell^k+\eta_\ell-\eta_\chi-1\right)\right)}
{(\eta_\ell-1)^2}
\\
&=\frac{1}{(
\eta_\ell-1)^2}
\left[\eta_\ell^{n-k}\left(
\eta_\ell^k\eta_\chi +\eta_\ell-\eta_\chi-1\right)(\eta_\ell+\eta_\chi-1)  
+
\left(\eta_\ell^k(\eta_\ell+\eta_\chi-1) -
\eta_\chi\right)(\eta_\ell-\eta_\chi-1)  \right].
\end{aligned}
\end{equation}
With $\ell=2$, the limit $\chi \rightarrow \infty$ gives then the asymptotic behavior
\begin{equation}
\lim_{\chi\rightarrow \infty} S_2\left(\rho_{k,n}\right)_{\max} = \frac{1}{2^k} + \frac{2^k}{2^n} 
\end{equation}
Then, the Renyi-2 entropy is predominantly driven by the $k$-order considered.

\subsection{Variance}
\begin{figure}[htp]
\centering
\includegraphics[width=\textwidth]{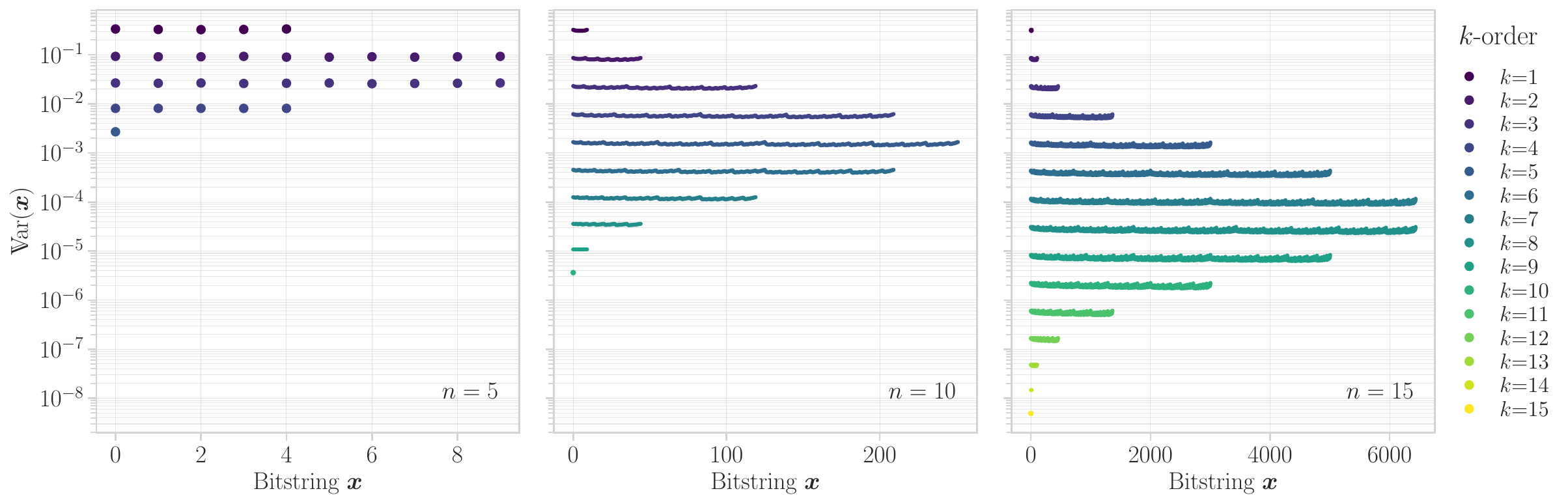}
\caption{\justifying\label{fig:additional_TN}\justifying\textbf{$\Var(\operatorname{Pr}^{(k)}(\bx))$ for general bit-string $\bx$.} In the main text we showed the result for $\bx=\bzero$, here we show that the scaling is the same for any bitstring and the vanishing dependence is still with $k$.}
\end{figure}

We start the proof of the variance for the case of a marginal of size $m$ and $k$-order equal to $m$. Such operator is defined as,
$$\Pi_{\vec x}=\bigotimes_{i=1}^n \Pi_i: \Pi_i= \begin{cases}\1 & \text{if } x_i=0 \quad \rightarrow T_{\1} \\ O =  \przero \text { or } \prone & \text{if } x_i=1 \quad \rightarrow T_{O}\end{cases}.$$

Such marginal is given by,
\begin{equation}
\Var\left(\Pi^{(m)}_{\vec{x}} \right)_{\max} = \Tr(\Pi^{(m)}_{\vec{x}}  U_\btheta \bprzero U_\btheta^\dagger),
\end{equation}
where the maximum values is taken over the configuration $\vec x$. $$\Var\left(\Pi^{(m)}_{\vec{x}} \right) = \E\left(\Pi^{(m)2}_{\vec{x}}\right) - \E\left(\Pi^{(m)}_{\vec{x}} \right)^2$$
In this case, as $O$ is a projector $\Tr(O^2) = \Tr(O) = 1$, and hence for \raisebox{-.1em}{\resizebox{.05\textwidth}{!}{\tikzset{every picture/.style={line width=0.75pt}} 

\begin{tikzpicture}[x=0.75pt,y=0.75pt,yscale=-1,xscale=1]

\draw    (41.1,28.1) -- (57.6,28.1) ;
\draw  [color={rgb, 255:red, 245; green, 166; blue, 35 }  ,draw opacity=1 ][fill={rgb, 255:red, 242; green, 206; blue, 133 }  ,fill opacity=1 ] (58.09,15.61) -- (93.28,15.61) -- (93.28,40.59) -- (58.09,40.59) -- cycle ;
\draw    (93.28,28.1) -- (109.51,28.1) ;

\draw (68.19,20) node [anchor=north west][inner sep=0.75pt]  [font=\large]  {$O$};

\end{tikzpicture}}} = \raisebox{-.6em}{\resizebox{.05\textwidth}{!}{\tikzset{every picture/.style={line width=0.75pt}} 

\begin{tikzpicture}[x=0.75pt,y=0.75pt,yscale=-1,xscale=1]

\draw    (86.06,18.94) -- (86.06,38.24) ;
\draw   (45.08,18.3) .. controls (46.71,22.04) and (48.27,25.6) .. (50.08,25.6) .. controls (51.89,25.6) and (53.45,22.04) .. (55.08,18.3) .. controls (56.71,14.56) and (58.27,11) .. (60.08,11) .. controls (61.89,11) and (63.45,14.56) .. (65.08,18.3) .. controls (66.71,22.04) and (68.27,25.6) .. (70.08,25.6) .. controls (71.89,25.6) and (73.45,22.04) .. (75.08,18.3) .. controls (76.71,14.56) and (78.27,11) .. (80.08,11) .. controls (81.89,11) and (83.45,14.56) .. (85.08,18.3) .. controls (85.21,18.61) and (85.35,18.91) .. (85.48,19.21) ;
\draw  [fill={rgb, 255:red, 0; green, 0; blue, 0 }  ,fill opacity=1 ] (82.76,18.94) .. controls (82.76,17.12) and (84.24,15.64) .. (86.06,15.64) .. controls (87.88,15.64) and (89.36,17.12) .. (89.36,18.94) .. controls (89.36,20.76) and (87.88,22.24) .. (86.06,22.24) .. controls (84.24,22.24) and (82.76,20.76) .. (82.76,18.94) -- cycle ;
\draw    (118.28,18.94) -- (86.06,18.94) ;
\draw  [color={rgb, 255:red, 245; green, 166; blue, 35 }  ,draw opacity=1 ][fill={rgb, 255:red, 242; green, 206; blue, 133 }  ,fill opacity=1 ] (68.69,38.01) -- (103.88,38.01) -- (103.88,62.99) -- (68.69,62.99) -- cycle ;

\draw (70.69,41.41) node [anchor=north west][inner sep=0.75pt]  [font=\Large]  {$\ket{O}$};

\end{tikzpicture}}}:

\setlength{\arraycolsep}{1.5em}
\begin{equation}
\begin{array}{cc}
\1\raisebox{-.1em}{\resizebox{.05\textwidth}{!}{\tikzset{every picture/.style={line width=0.75pt}} 

\begin{tikzpicture}[x=0.75pt,y=0.75pt,yscale=-1,xscale=1]

\draw    (41.1,28.1) -- (57.6,28.1) ;
\draw  [color={rgb, 255:red, 245; green, 166; blue, 35 }  ,draw opacity=1 ][fill={rgb, 255:red, 242; green, 206; blue, 133 }  ,fill opacity=1 ] (58.09,15.61) -- (93.28,15.61) -- (93.28,40.59) -- (58.09,40.59) -- cycle ;
\draw    (93.28,28.1) -- (109.51,28.1) ;

\draw (68.19,20) node [anchor=north west][inner sep=0.75pt]  [font=\large]  {$O$};

\end{tikzpicture}}}\1 = \frac{\ell\chi^2-1}{\ell\left(\ell^2 \chi^2-1\right)} & \1 \raisebox{-.1em}{\resizebox{.05\textwidth}{!}{\tikzset{every picture/.style={line width=0.75pt}} 

\begin{tikzpicture}[x=0.75pt,y=0.75pt,yscale=-1,xscale=1]

\draw    (41.1,28.1) -- (57.6,28.1) ;
\draw  [color={rgb, 255:red, 245; green, 166; blue, 35 }  ,draw opacity=1 ][fill={rgb, 255:red, 242; green, 206; blue, 133 }  ,fill opacity=1 ] (58.09,15.61) -- (93.28,15.61) -- (93.28,40.59) -- (58.09,40.59) -- cycle ;
\draw    (93.28,28.1) -- (109.51,28.1) ;

\draw (68.19,20) node [anchor=north west][inner sep=0.75pt]  [font=\large]  {$O$};

\end{tikzpicture}}} \F = \frac{\chi(\ell-1)}{\ell\left(\ell^2 \chi^2-1\right)} \\ [1.5em]
\F \raisebox{-.1em}{\resizebox{.05\textwidth}{!}{\tikzset{every picture/.style={line width=0.75pt}} 

\begin{tikzpicture}[x=0.75pt,y=0.75pt,yscale=-1,xscale=1]

\draw    (41.1,28.1) -- (57.6,28.1) ;
\draw  [color={rgb, 255:red, 245; green, 166; blue, 35 }  ,draw opacity=1 ][fill={rgb, 255:red, 242; green, 206; blue, 133 }  ,fill opacity=1 ] (58.09,15.61) -- (93.28,15.61) -- (93.28,40.59) -- (58.09,40.59) -- cycle ;
\draw    (93.28,28.1) -- (109.51,28.1) ;

\draw (68.19,20) node [anchor=north west][inner sep=0.75pt]  [font=\large]  {$O$};

\end{tikzpicture}}} \1 = \frac{\chi(\ell-1)}{\ell\left(\ell^2 \chi^2-1\right)} & \F \raisebox{-.1em}{\resizebox{.05\textwidth}{!}{\tikzset{every picture/.style={line width=0.75pt}} 

\begin{tikzpicture}[x=0.75pt,y=0.75pt,yscale=-1,xscale=1]

\draw    (41.1,28.1) -- (57.6,28.1) ;
\draw  [color={rgb, 255:red, 245; green, 166; blue, 35 }  ,draw opacity=1 ][fill={rgb, 255:red, 242; green, 206; blue, 133 }  ,fill opacity=1 ] (58.09,15.61) -- (93.28,15.61) -- (93.28,40.59) -- (58.09,40.59) -- cycle ;
\draw    (93.28,28.1) -- (109.51,28.1) ;

\draw (68.19,20) node [anchor=north west][inner sep=0.75pt]  [font=\large]  {$O$};

\end{tikzpicture}}} \F = \frac{\ell\chi^2-1}{\ell\left(\ell^2 \chi^2-1\right)}
\end{array}
\end{equation}
\setlength{\arraycolsep}{\defaultarraycolsep}
The change of variables,
\begin{equation}
\zeta = \frac{\ell\chi^2-1}{\ell\left(\ell^2 \chi^2-1\right)}, \quad \mu = \frac{\chi(\ell-1)}{\ell\left(\ell^2 \chi^2-1\right)}
\end{equation}
allows to calculate and simplify 
\begin{equation}
\Var\left(\Pi^{(m)}_{\vec{x}} \right)_{\max} = \raisebox{-.3em}{\resizebox{.4\textwidth}{!}{\tikzset{every picture/.style={line width=0.75pt}} 

\begin{tikzpicture}[x=0.75pt,y=0.75pt,yscale=-1,xscale=1]

\draw  [color={rgb, 255:red, 245; green, 166; blue, 35 }  ,draw opacity=1 ][fill={rgb, 255:red, 242; green, 206; blue, 133 }  ,fill opacity=1 ] (27.81,12.97) -- (63,12.97) -- (63,37.96) -- (27.81,37.96) -- cycle ;
\draw    (63.1,25.46) -- (79.6,25.46) ;
\draw  [color={rgb, 255:red, 245; green, 166; blue, 35 }  ,draw opacity=1 ][fill={rgb, 255:red, 242; green, 206; blue, 133 }  ,fill opacity=1 ] (80.09,12.97) -- (115.28,12.97) -- (115.28,37.96) -- (80.09,37.96) -- cycle ;
\draw    (115.28,25.46) -- (131.51,25.46) ;
\draw  [dash pattern={on 4.5pt off 4.5pt}]  (131.51,25.46) -- (182.93,25.46) ;
\draw    (182.93,25.46) -- (199.17,25.46) ;
\draw  [color={rgb, 255:red, 245; green, 166; blue, 35 }  ,draw opacity=1 ][fill={rgb, 255:red, 242; green, 206; blue, 133 }  ,fill opacity=1 ] (199.42,12.97) -- (234.6,12.97) -- (234.6,37.96) -- (199.42,37.96) -- cycle ;
\draw  [dash pattern={on 4.5pt off 4.5pt}]  (27.32,25.46) -- (11.7,25.46) ;
\draw  [color={rgb, 255:red, 84; green, 142; blue, 161 }  ,draw opacity=1 ][fill={rgb, 255:red, 186; green, 235; blue, 252 }  ,fill opacity=1 ] (251.81,12.97) -- (287,12.97) -- (287,37.96) -- (251.81,37.96) -- cycle ;
\draw    (287,25.46) -- (303.23,25.46) ;
\draw  [color={rgb, 255:red, 84; green, 142; blue, 161 }  ,draw opacity=1 ][fill={rgb, 255:red, 186; green, 235; blue, 252 }  ,fill opacity=1 ] (304.09,12.97) -- (339.28,12.97) -- (339.28,37.96) -- (304.09,37.96) -- cycle ;
\draw    (339.28,25.46) -- (355.51,25.46) ;
\draw  [dash pattern={on 4.5pt off 4.5pt}]  (355.51,25.46) -- (406.93,25.46) ;
\draw    (406.93,25.46) -- (423.17,25.46) ;
\draw  [color={rgb, 255:red, 84; green, 142; blue, 161 }  ,draw opacity=1 ][fill={rgb, 255:red, 186; green, 235; blue, 252 }  ,fill opacity=1 ] (423.42,12.97) -- (458.6,12.97) -- (458.6,37.96) -- (423.42,37.96) -- cycle ;
\draw    (234.93,25.46) -- (251.17,25.46) ;
\draw  [fill={rgb, 255:red, 0; green, 0; blue, 0 }  ,fill opacity=1 ] (70,25.46) .. controls (70,24.75) and (70.58,24.16) .. (71.3,24.16) .. controls (72.02,24.16) and (72.6,24.75) .. (72.6,25.46) .. controls (72.6,26.18) and (72.02,26.76) .. (71.3,26.76) .. controls (70.58,26.76) and (70,26.18) .. (70,25.46) -- cycle ;
\draw  [fill={rgb, 255:red, 0; green, 0; blue, 0 }  ,fill opacity=1 ] (19,25.46) .. controls (19,24.75) and (19.58,24.16) .. (20.3,24.16) .. controls (21.02,24.16) and (21.6,24.75) .. (21.6,25.46) .. controls (21.6,26.18) and (21.02,26.76) .. (20.3,26.76) .. controls (19.58,26.76) and (19,26.18) .. (19,25.46) -- cycle ;
\draw  [fill={rgb, 255:red, 0; green, 0; blue, 0 }  ,fill opacity=1 ] (124.25,25.46) .. controls (124.25,24.75) and (124.83,24.16) .. (125.55,24.16) .. controls (126.27,24.16) and (126.85,24.75) .. (126.85,25.46) .. controls (126.85,26.18) and (126.27,26.76) .. (125.55,26.76) .. controls (124.83,26.76) and (124.25,26.18) .. (124.25,25.46) -- cycle ;
\draw  [fill={rgb, 255:red, 0; green, 0; blue, 0 }  ,fill opacity=1 ] (188.25,25.46) .. controls (188.25,24.75) and (188.83,24.16) .. (189.55,24.16) .. controls (190.27,24.16) and (190.85,24.75) .. (190.85,25.46) .. controls (190.85,26.18) and (190.27,26.76) .. (189.55,26.76) .. controls (188.83,26.76) and (188.25,26.18) .. (188.25,25.46) -- cycle ;
\draw  [fill={rgb, 255:red, 0; green, 0; blue, 0 }  ,fill opacity=1 ] (241.7,25.46) .. controls (241.7,24.75) and (242.28,24.16) .. (243,24.16) .. controls (243.72,24.16) and (244.3,24.75) .. (244.3,25.46) .. controls (244.3,26.18) and (243.72,26.76) .. (243,26.76) .. controls (242.28,26.76) and (241.7,26.18) .. (241.7,25.46) -- cycle ;
\draw  [fill={rgb, 255:red, 0; green, 0; blue, 0 }  ,fill opacity=1 ] (294.65,25.46) .. controls (294.65,24.75) and (295.23,24.16) .. (295.95,24.16) .. controls (296.67,24.16) and (297.25,24.75) .. (297.25,25.46) .. controls (297.25,26.18) and (296.67,26.76) .. (295.95,26.76) .. controls (295.23,26.76) and (294.65,26.18) .. (294.65,25.46) -- cycle ;
\draw  [fill={rgb, 255:red, 0; green, 0; blue, 0 }  ,fill opacity=1 ] (349.4,25.46) .. controls (349.4,24.75) and (349.98,24.16) .. (350.7,24.16) .. controls (351.42,24.16) and (352,24.75) .. (352,25.46) .. controls (352,26.18) and (351.42,26.76) .. (350.7,26.76) .. controls (349.98,26.76) and (349.4,26.18) .. (349.4,25.46) -- cycle ;
\draw  [fill={rgb, 255:red, 0; green, 0; blue, 0 }  ,fill opacity=1 ] (411.9,25.46) .. controls (411.9,24.75) and (412.48,24.16) .. (413.2,24.16) .. controls (413.92,24.16) and (414.5,24.75) .. (414.5,25.46) .. controls (414.5,26.18) and (413.92,26.76) .. (413.2,26.76) .. controls (412.48,26.76) and (411.9,26.18) .. (411.9,25.46) -- cycle ;
\draw  [dash pattern={on 4.5pt off 4.5pt}]  (458.84,25.46) -- (474.75,25.46) ;
\draw  [fill={rgb, 255:red, 0; green, 0; blue, 0 }  ,fill opacity=1 ] (464.9,25.46) .. controls (464.9,24.75) and (465.48,24.16) .. (466.2,24.16) .. controls (466.92,24.16) and (467.5,24.75) .. (467.5,25.46) .. controls (467.5,26.18) and (466.92,26.76) .. (466.2,26.76) .. controls (465.48,26.76) and (464.9,26.18) .. (464.9,25.46) -- cycle ;

\draw (153.5,5.7) node [anchor=north west][inner sep=0.75pt]  [font=\large]  {$m$};
\draw (359.5,5.1) node [anchor=north west][inner sep=0.75pt]  [font=\large]  {$n-m$};
\draw (37.91,17.36) node [anchor=north west][inner sep=0.75pt]  [font=\large]  {$O$};
\draw (90.19,17.36) node [anchor=north west][inner sep=0.75pt]  [font=\large]  {$O$};
\draw (209.51,17.36) node [anchor=north west][inner sep=0.75pt]  [font=\large]  {$O$};
\draw (262.91,17.36) node [anchor=north west][inner sep=0.75pt]  [font=\large]  {$\mathds{1}$};
\draw (315.19,17.36) node [anchor=north west][inner sep=0.75pt]  [font=\large]  {$\mathds{1}$};
\draw (434.51,17.36) node [anchor=north west][inner sep=0.75pt]  [font=\large]  {$\mathds{1}$};

\end{tikzpicture}}} = \left(\begin{array}{ll}
1 & 1
\end{array}\right) T_{\mathbb F}^m \cdot T_{\mathds{1}}^{n - m}\binom{1}{1}
\end{equation}

The transition matrix in the projector case takes the form,

\begin{equation}
T_{O} = \left(\begin{array}{ll}
\zeta & \mu \\
\mu & \zeta
\end{array}\right) \implies
P_{O} = \left(\begin{array}{cc}
-1 & 1 \\
1 & 1
\end{array}\right),\,
D_{O} = \left(\begin{array}{cc}
\zeta-\mu & 0 \\
0 & \zeta+\mu
\end{array}\right),\,
P^{-1}_{O}=\frac{1}{2}\left(\begin{array}{rr}
-1 & 1 \\
1 & 1
\end{array}\right)
\end{equation}
therefore  
\begin{equation}
\begin{aligned}
\Var\left(\Pi^{(m)}_{\vec{x}}\right)_{\max}&=\left(\begin{array}{ll}
1 & 1
\end{array}\right) P_{O} D_{O}^m P_{O}^{-1} \, P_\mathds{1} D_{\mathds{1}}^{n-m} P_{\mathds{1}}^{-1}\binom{1}{1}
\\
&=\left(\begin{array}{ll} -(\zeta-\mu)^m & (\zeta+\mu)^m\end{array}\right) \frac{1}{2}\left(\begin{array}{cc}
-1 & 1-\frac{\eta_\chi}{\eta_\ell-1} \\
 \rule{0pt}{15pt} 
1 & 1+\frac{\eta_\chi}{\eta_\ell-1}
\end{array}\right)\binom{1-\frac{\eta_\chi}{\eta_\ell-1}}{\eta_\ell^{n-m}}
\\
&=\frac{1}{1+\chi^2}\!\!\left(\frac{(2 \chi-1)(1+\chi)}{(2(1-4 \chi))^2}\right)^m\!\!\left(\!2^m(1+\chi)(1+2 \chi)+2^n(\chi-1)(2 \chi-1)\left(\frac{\chi^2-1}{4 \chi^2-1}\right)^{n-m}\right).
\end{aligned}
\end{equation}

The behavior for high bond dimension is,
\begin{equation}
\lim_{\chi\rightarrow \infty} \Var\left(\Pi^{(m)}_{\vec{x}} \right)_{\max} = \frac{1}{2^{2m}}\left(1+ \frac{2^m}{2^n} \right)
\end{equation}

Now we turn the focus to the case of calculating the probability up to a given $k$-order correlator. Such probability is given by the sum of different $Z_{\bi}$, hence we are interested in  \raisebox{-.1em}{\resizebox{.05\textwidth}{!}{\tikzset{every picture/.style={line width=0.75pt}} 

\begin{tikzpicture}[x=0.75pt,y=0.75pt,yscale=-1,xscale=1]

\draw  [color={rgb, 255:red, 65; green, 117; blue, 5 }  ,draw opacity=1 ][fill={rgb, 255:red, 211; green, 244; blue, 180 }  ,fill opacity=1 ] (76.81,27.61) -- (112,27.61) -- (112,52.59) -- (76.81,52.59) -- cycle ;
\draw    (112,40.1) -- (128.23,40.1) ;
\draw    (59.93,40.1) -- (76.17,40.1) ;

\draw (85.91,32) node [anchor=north west][inner sep=0.75pt]  [font=\Large]  {$Z$};

\end{tikzpicture}}} = \raisebox{-.6em}{\resizebox{.05\textwidth}{!}{\tikzset{every picture/.style={line width=0.75pt}} 

\begin{tikzpicture}[x=0.75pt,y=0.75pt,yscale=-1,xscale=1]

\draw    (84.06,29.94) -- (84.06,49.24) ;
\draw   (43.08,29.3) .. controls (44.71,33.04) and (46.27,36.6) .. (48.08,36.6) .. controls (49.89,36.6) and (51.45,33.04) .. (53.08,29.3) .. controls (54.71,25.56) and (56.27,22) .. (58.08,22) .. controls (59.89,22) and (61.45,25.56) .. (63.08,29.3) .. controls (64.71,33.04) and (66.27,36.6) .. (68.08,36.6) .. controls (69.89,36.6) and (71.45,33.04) .. (73.08,29.3) .. controls (74.71,25.56) and (76.27,22) .. (78.08,22) .. controls (79.89,22) and (81.45,25.56) .. (83.08,29.3) .. controls (83.21,29.61) and (83.35,29.91) .. (83.48,30.21) ;
\draw  [fill={rgb, 255:red, 0; green, 0; blue, 0 }  ,fill opacity=1 ] (80.76,29.94) .. controls (80.76,28.12) and (82.24,26.64) .. (84.06,26.64) .. controls (85.88,26.64) and (87.36,28.12) .. (87.36,29.94) .. controls (87.36,31.76) and (85.88,33.24) .. (84.06,33.24) .. controls (82.24,33.24) and (80.76,31.76) .. (80.76,29.94) -- cycle ;
\draw    (116.28,29.94) -- (84.06,29.94) ;
\draw  [color={rgb, 255:red, 65; green, 117; blue, 5 }  ,draw opacity=1 ][fill={rgb, 255:red, 211; green, 244; blue, 180 }  ,fill opacity=1 ] (66.69,49.01) -- (101.88,49.01) -- (101.88,73.99) -- (66.69,73.99) -- cycle ;

\draw (68.69,52.41) node [anchor=north west][inner sep=0.75pt]  [font=\Large]  {$\ket{Z}$};

\end{tikzpicture}}}:

\setlength{\defaultarraycolsep}{\arraycolsep}
\setlength{\arraycolsep}{1.5em}
\begin{equation}\label{eq:array_Z}
\begin{array}{ll}
\1\raisebox{-.1em}{\resizebox{.05\textwidth}{!}{\tikzset{every picture/.style={line width=0.75pt}} 

\begin{tikzpicture}[x=0.75pt,y=0.75pt,yscale=-1,xscale=1]

\draw  [color={rgb, 255:red, 65; green, 117; blue, 5 }  ,draw opacity=1 ][fill={rgb, 255:red, 211; green, 244; blue, 180 }  ,fill opacity=1 ] (76.81,27.61) -- (112,27.61) -- (112,52.59) -- (76.81,52.59) -- cycle ;
\draw    (112,40.1) -- (128.23,40.1) ;
\draw    (59.93,40.1) -- (76.17,40.1) ;

\draw (85.91,32) node [anchor=north west][inner sep=0.75pt]  [font=\Large]  {$Z$};

\end{tikzpicture}}}\1 = \frac{-1}{\ell(\ell^2\chi^2-1)} & \1 \raisebox{-.1em}{\resizebox{.05\textwidth}{!}{\tikzset{every picture/.style={line width=0.75pt}} 

\begin{tikzpicture}[x=0.75pt,y=0.75pt,yscale=-1,xscale=1]

\draw  [color={rgb, 255:red, 65; green, 117; blue, 5 }  ,draw opacity=1 ][fill={rgb, 255:red, 211; green, 244; blue, 180 }  ,fill opacity=1 ] (76.81,27.61) -- (112,27.61) -- (112,52.59) -- (76.81,52.59) -- cycle ;
\draw    (112,40.1) -- (128.23,40.1) ;
\draw    (59.93,40.1) -- (76.17,40.1) ;

\draw (85.91,32) node [anchor=north west][inner sep=0.75pt]  [font=\Large]  {$Z$};

\end{tikzpicture}}} \F = \frac{-\chi}{\ell (\ell^2\chi^2-1)}\\[1.5em]
\F \raisebox{-.1em}{\resizebox{.05\textwidth}{!}{\tikzset{every picture/.style={line width=0.75pt}} 

\begin{tikzpicture}[x=0.75pt,y=0.75pt,yscale=-1,xscale=1]

\draw  [color={rgb, 255:red, 65; green, 117; blue, 5 }  ,draw opacity=1 ][fill={rgb, 255:red, 211; green, 244; blue, 180 }  ,fill opacity=1 ] (76.81,27.61) -- (112,27.61) -- (112,52.59) -- (76.81,52.59) -- cycle ;
\draw    (112,40.1) -- (128.23,40.1) ;
\draw    (59.93,40.1) -- (76.17,40.1) ;

\draw (85.91,32) node [anchor=north west][inner sep=0.75pt]  [font=\Large]  {$Z$};

\end{tikzpicture}}} \1 = \frac{\chi}{\ell^2\chi^2-1} & \F \raisebox{-.1em}{\resizebox{.05\textwidth}{!}{\tikzset{every picture/.style={line width=0.75pt}} 

\begin{tikzpicture}[x=0.75pt,y=0.75pt,yscale=-1,xscale=1]

\draw  [color={rgb, 255:red, 65; green, 117; blue, 5 }  ,draw opacity=1 ][fill={rgb, 255:red, 211; green, 244; blue, 180 }  ,fill opacity=1 ] (76.81,27.61) -- (112,27.61) -- (112,52.59) -- (76.81,52.59) -- cycle ;
\draw    (112,40.1) -- (128.23,40.1) ;
\draw    (59.93,40.1) -- (76.17,40.1) ;

\draw (85.91,32) node [anchor=north west][inner sep=0.75pt]  [font=\Large]  {$Z$};

\end{tikzpicture}}} \F = \frac{\chi^2}{\ell^2 \chi^2-1}
\end{array}
\end{equation}
\setlength{\arraycolsep}{\defaultarraycolsep}

\begin{equation}
\begin{aligned}
\Var\left(\operatorname{Pr} ^{(k)}_{\btheta}(\bx)\right) & = 
\frac{1}{2^n} \sum_{p=0}^k \sum_{\bullet \in\{\1,\F\}^n}\sum_{ \substack{\raisebox{-.15em}{\resizebox{.04\textwidth}{!}{\tikzset{every picture/.style={line width=0.75pt}} 

\begin{tikzpicture}[x=0.75pt,y=0.75pt,yscale=-1,xscale=1]

\draw    (43.1,23.7) -- (59.6,23.7) ;
\draw  [color={rgb, 255:red, 139; green, 87; blue, 42 }  ,draw opacity=1 ][fill={rgb, 255:red, 195; green, 174; blue, 161 }  ,fill opacity=1 ] (60.09,11.21) -- (95.28,11.21) -- (95.28,36.19) -- (60.09,36.19) -- cycle ;
\draw    (95.28,23.7) -- (111.51,23.7) ;

\draw (72.19,15.6) node [anchor=north west][inner sep=0.75pt]  [font=\large]  {$\boldsymbol{i}$};

\end{tikzpicture}}}\in \{\raisebox{-.15em}{\resizebox{.04\textwidth}{!}{\tikzset{every picture/.style={line width=0.75pt}} 

\begin{tikzpicture}[x=0.75pt,y=0.75pt,yscale=-1,xscale=1]

\draw  [color={rgb, 255:red, 84; green, 142; blue, 161 }  ,draw opacity=1 ][fill={rgb, 255:red, 186; green, 235; blue, 252 }  ,fill opacity=1 ] (56.81,22.01) -- (92,22.01) -- (92,46.99) -- (56.81,46.99) -- cycle ;
\draw    (92,34.5) -- (108.23,34.5) ;
\draw    (39.93,34.5) -- (56.17,34.5) ;

\draw (67.91,26.4) node [anchor=north west][inner sep=0.75pt]  [font=\Large]  {$\mathds{1}$};

\end{tikzpicture}}} , \raisebox{-.15em}{\resizebox{.04\textwidth}{!}{\tikzset{every picture/.style={line width=0.75pt}} 

\begin{tikzpicture}[x=0.75pt,y=0.75pt,yscale=-1,xscale=1]

\draw  [color={rgb, 255:red, 65; green, 117; blue, 5 }  ,draw opacity=1 ][fill={rgb, 255:red, 211; green, 244; blue, 180 }  ,fill opacity=1 ] (76.81,27.61) -- (112,27.61) -- (112,52.59) -- (76.81,52.59) -- cycle ;
\draw    (112,40.1) -- (128.23,40.1) ;
\draw    (59.93,40.1) -- (76.17,40.1) ;

\draw (85.91,32) node [anchor=north west][inner sep=0.75pt]  [font=\Large]  {$Z$};

\end{tikzpicture}}}\}^{\otimes n}:\\ |\bi_Z|=p}}\!\!\!\!\!\!\!\!\!\!\!\!\raisebox{-.3em}{\resizebox{.19\textwidth}{!}{\tikzset{every picture/.style={line width=0.75pt}} 

\begin{tikzpicture}[x=0.75pt,y=0.75pt,yscale=-1,xscale=1]

\draw  [color={rgb, 255:red, 139; green, 87; blue, 42 }  ,draw opacity=1 ][fill={rgb, 255:red, 195; green, 174; blue, 161 }  ,fill opacity=1 ] (51.81,17.27) -- (87,17.27) -- (87,42.26) -- (51.81,42.26) -- cycle ;
\draw    (87.1,29.76) -- (103.6,29.76) ;
\draw  [color={rgb, 255:red, 139; green, 87; blue, 42 }  ,draw opacity=1 ][fill={rgb, 255:red, 195; green, 174; blue, 161 }  ,fill opacity=1 ] (104.09,17.27) -- (139.28,17.27) -- (139.28,42.26) -- (104.09,42.26) -- cycle ;
\draw    (139.28,29.76) -- (155.51,29.76) ;
\draw  [dash pattern={on 4.5pt off 4.5pt}]  (155.51,29.76) -- (206.93,29.76) ;
\draw    (206.93,29.76) -- (223.17,29.76) ;
\draw  [color={rgb, 255:red, 139; green, 87; blue, 42 }  ,draw opacity=1 ][fill={rgb, 255:red, 195; green, 174; blue, 161 }  ,fill opacity=1 ] (223.42,17.02) -- (258.6,17.02) -- (258.6,42.01) -- (223.42,42.01) -- cycle ;
\draw  [dash pattern={on 4.5pt off 4.5pt}]  (51.32,29.76) -- (35.7,29.76) ;
\draw  [fill={rgb, 255:red, 0; green, 0; blue, 0 }  ,fill opacity=1 ] (94,29.76) .. controls (94,29.05) and (94.58,28.46) .. (95.3,28.46) .. controls (96.02,28.46) and (96.6,29.05) .. (96.6,29.76) .. controls (96.6,30.48) and (96.02,31.06) .. (95.3,31.06) .. controls (94.58,31.06) and (94,30.48) .. (94,29.76) -- cycle ;
\draw  [fill={rgb, 255:red, 0; green, 0; blue, 0 }  ,fill opacity=1 ] (43,29.76) .. controls (43,29.05) and (43.58,28.46) .. (44.3,28.46) .. controls (45.02,28.46) and (45.6,29.05) .. (45.6,29.76) .. controls (45.6,30.48) and (45.02,31.06) .. (44.3,31.06) .. controls (43.58,31.06) and (43,30.48) .. (43,29.76) -- cycle ;
\draw  [fill={rgb, 255:red, 0; green, 0; blue, 0 }  ,fill opacity=1 ] (148.25,29.76) .. controls (148.25,29.05) and (148.83,28.46) .. (149.55,28.46) .. controls (150.27,28.46) and (150.85,29.05) .. (150.85,29.76) .. controls (150.85,30.48) and (150.27,31.06) .. (149.55,31.06) .. controls (148.83,31.06) and (148.25,30.48) .. (148.25,29.76) -- cycle ;
\draw  [fill={rgb, 255:red, 0; green, 0; blue, 0 }  ,fill opacity=1 ] (212.25,29.76) .. controls (212.25,29.05) and (212.83,28.46) .. (213.55,28.46) .. controls (214.27,28.46) and (214.85,29.05) .. (214.85,29.76) .. controls (214.85,30.48) and (214.27,31.06) .. (213.55,31.06) .. controls (212.83,31.06) and (212.25,30.48) .. (212.25,29.76) -- cycle ;
\draw  [dash pattern={on 4.5pt off 4.5pt}]  (258.84,29.76) -- (274.75,29.76) ;
\draw  [fill={rgb, 255:red, 0; green, 0; blue, 0 }  ,fill opacity=1 ] (264.9,29.76) .. controls (264.9,29.05) and (265.48,28.46) .. (266.2,28.46) .. controls (266.92,28.46) and (267.5,29.05) .. (267.5,29.76) .. controls (267.5,30.48) and (266.92,31.06) .. (266.2,31.06) .. controls (265.48,31.06) and (264.9,30.48) .. (264.9,29.76) -- cycle ;

\draw (61.91,21.66) node [anchor=north west][inner sep=0.75pt]  [font=\large]  {$i_{1}$};
\draw (114.19,21.66) node [anchor=north west][inner sep=0.75pt]  [font=\large]  {$i_{2}$};
\draw (233.51,21.66) node [anchor=north west][inner sep=0.75pt]  [font=\large]  {$i_{n}$};

\end{tikzpicture}}} \\ &= \frac{1}{2^n} \sum_{p=0}^k \sum_{\bi \in \{\1,Z\}^{\otimes n}} \left(\begin{array}{ll}
1 & 1
\end{array}\right) T_{\bi}\binom{1}{1}
\end{aligned}
\end{equation} 

Where $T_{\bi} = \prod_{j=1}^n T_{i_j}$ where $i_j \in \{\1,Z\} $ and $T_Z = \frac{1}{\ell(\ell^2\chi^2-1)}\left(\begin{array}{ll} -1 & -\chi \\ \ell\chi & \ell\chi^2 \end{array}\right)$. Fig.~\ref{fig:additional_TN} shows how the variance for different $\bx$ behaves.

\section{Note I: Correlator selection within the MMD kernel}\label{app:anova}
In this section, we isolate interaction Orders via ANOVA Kernels in the MMD loss to demonstrate how, if the kernel is too simple, it can lead to misleading behaviours. Specifically, one may observe significant reduction in training loss and that the training process is improving. However, in reality, the favourable behavior is observed purely because the learning model is too simplified, and not all correlators are actually considered.

The explicit form in Eq.~\ref{eq:mmd_matrix} follows directly from expanding the implicit expectation-based definition in Eq.~\ref{eq:mmd_def}. For discrete distributions, expectations become weighted sums. For example,
\begin{equation}
\E_{\boldsymbol{x},\boldsymbol{x}' \sim p}\!\big[K(\boldsymbol{x},\boldsymbol{x}')\big]
= \sum_{\boldsymbol{x},\boldsymbol{x}'} p(\boldsymbol{x})\,p(\boldsymbol{x}')\,K(\boldsymbol{x},\boldsymbol{x}').
\end{equation}
Defining the kernel matrix $K$ with entries $K_{\boldsymbol{x},\boldsymbol{y}} = K(\boldsymbol{x},\boldsymbol{y})$, the three terms in Eq.~\ref{eq:mmd_def} can be written as
\begin{align}
\sum_{\boldsymbol{x},\boldsymbol{x}'} p(\boldsymbol{x})\,K_{\boldsymbol{x},\boldsymbol{x}'}\,p(\boldsymbol{x}') &= p^{\mathsf T} K p, \\
\sum_{\boldsymbol{x},\boldsymbol{y}} p(\boldsymbol{x})\,K_{\boldsymbol{x},\boldsymbol{y}}\,q(\boldsymbol{y}) &= p^{\mathsf T} K q, \\
\sum_{\boldsymbol{y},\boldsymbol{y}'} q(\boldsymbol{y})\,K_{\boldsymbol{y},\boldsymbol{y}'}\,q(\boldsymbol{y}') &= q^{\mathsf T} K q.
\end{align}
Substituting these expressions back into Eq.~\ref{eq:mmd_def} yields
\begin{equation}
\mathrm{MMD}^2(p,q) = p^{\mathsf T} K p - 2\,p^{\mathsf T} K q + q^{\mathsf T} K q,
\end{equation}
which simplifies to the compact quadratic form
\begin{equation}
\mathrm{MMD}^2(p,q) = (p-q)^{\mathsf T} K (p-q).
\end{equation}

\label{par:anova}

\begin{figure*}[htbp]
    \centering
    \includegraphics[width=\textwidth]{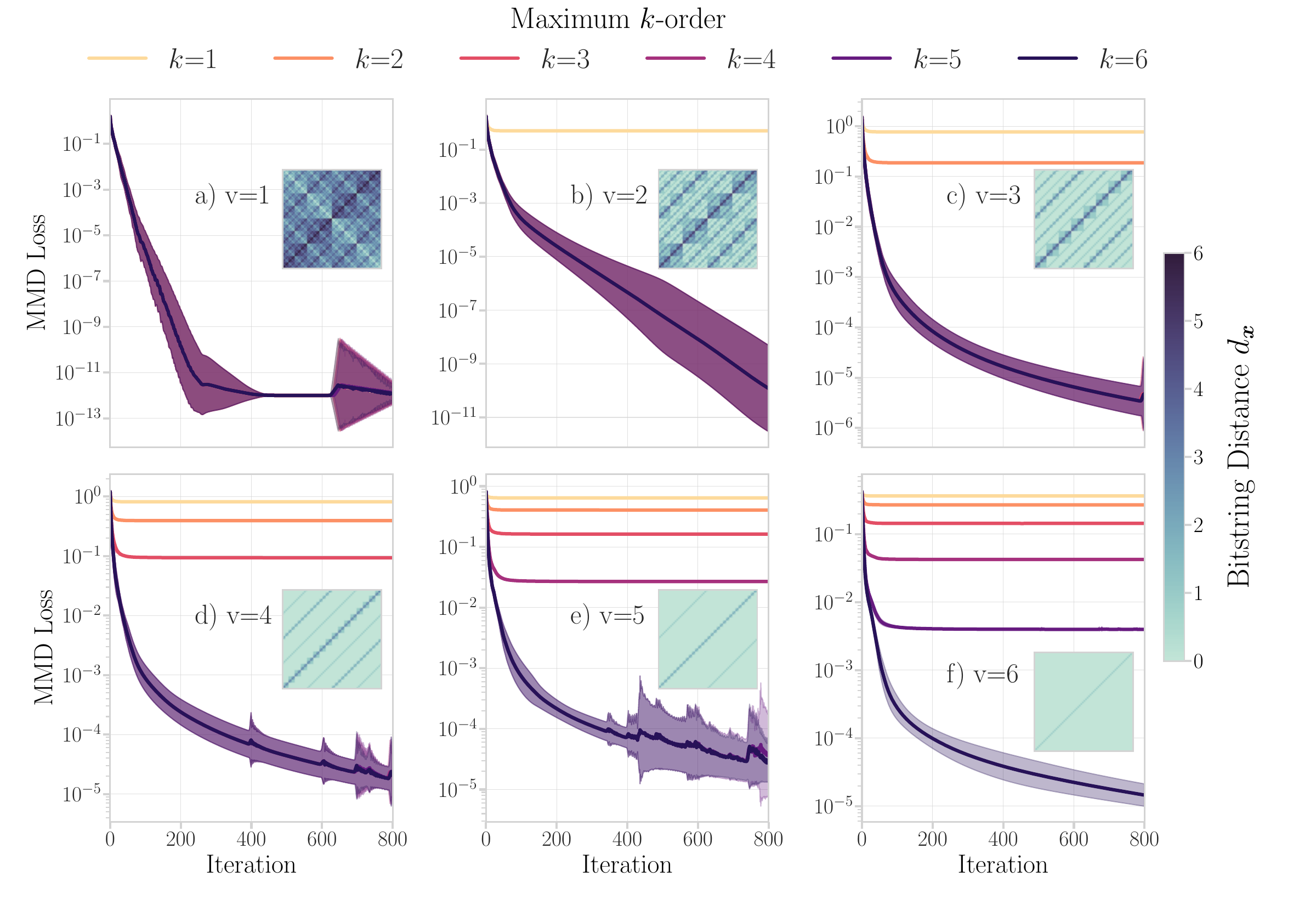}
    \caption{\justifying Effect of different ANOVA kernels in the distinction of truncated probability distributions for different maximum $k$-order of the same circuit with strongly entangling layers.}
    \label{fig:anova}
\end{figure*}

When choosing the kernel for the MMD loss, it is important that the loss renders a positive distance for all pairs of bitstrings. Moreover, different works have suggested and analyzed the potential use of a quantum kernel~\cite{coyle_quantum_2021,rudolph_trainability_2023}. We argue that as generative tasks use polynomial-sized batches to estimate the loss, we can mainly focus on classical-type kernels and analyze them from the perspective of $k$-order correlations. Truncating the Born distribution so that it captures only up to $k$-order correlations effectively restricts the loss to depend solely on those same $k$-order statistics. 

It is possible to have kernels that adapt~\cite{kurkin_note_2025}, however, one needs to be cautious to not be oversimplifying the problem and just training for an easy to train set of correlators. We exemplify this for MMD losses where we can manually choose the kernel such that the bitstring distance $d_{\bx}$ affects the $k$-orders it is accounting for. This can be achieved through the ANOVA (Analysis of Variance) kernel method.s
Given bitstrings $\boldsymbol{x}, \boldsymbol{y}\in {0,1}^n$, the general ANOVA kernel decomposes the similarity between bitstrings into contributions from interactions of distinct orders. Specifically, a kernel isolating $m$-th order interactions among subsets of bits can be expressed as:
\begin{equation}
K^{(v)}(\boldsymbol{x}, \boldsymbol{y}) = \sum_{u \subseteq [n],; |u|=v} K_u(\boldsymbol{x}_u, \boldsymbol{y}_u),
\end{equation}
where $u$ indexes subsets of bit positions, and $\boldsymbol{x}_u, \boldsymbol{y}_u$ are the corresponding sub-bitstrings. Typically, $K_u(\boldsymbol{x}_u, \boldsymbol{y}u)$ is an exponential function of the Hamming distance $d_{\boldsymbol{x}}$:
\begin{equation}
K_u(\boldsymbol{x}_u, \boldsymbol{y}_u) = e^{-\gamma d{\boldsymbol{x}}(\boldsymbol{x}_u, \boldsymbol{y}u)}, \quad \gamma > 0,
\end{equation}
with the Hamming distance defined as:
\begin{equation}
d{\boldsymbol{x}}(\boldsymbol{x}_u, \boldsymbol{y}_u) = \sum_{i\in u}\mathbf{1}[x_i \neq y_i].
\end{equation}

By varying the subset size $v$, the ANOVA kernel provides direct access to distinct orders of interactions between features, enabling precise characterization of higher-order correlations within data distributions.

The ANOVA kernel methodology systematically decomposes interactions among features into orthogonal contributions, facilitating focused analyses on interactions of specific cardinality. Given two bitstrings, $\boldsymbol{x}, \boldsymbol{y}\in{0,1}^n$, we introduce a general decomposition of an ANOVA kernel into terms of increasing order of interactions.

Formally, an ANOVA kernel up to order $m$ is defined as:
\begin{equation}
K_{\leq m}(\boldsymbol{x},\boldsymbol{y}) = \sum_{k=1}^{m}\sum_{u \subseteq [n], |u|=k} w_u k_u(\boldsymbol{x}_u, \boldsymbol{y}_u),
\end{equation}
where $u$ indexes subsets of coordinates, $w_u\ge0$ are weighting coefficients, and $k_u$ is a base kernel defined only on subset $u$. To isolate a specific order $m$, we set weights as $w_u=0$ unless $|u|=m$. The resulting $m$-th order ANOVA kernel then simplifies to:
\begin{equation}
K^{(m)}(\boldsymbol{x},\boldsymbol{y}) = \sum{u \subseteq [n], |u|=m} k_u(\boldsymbol{x}_u, \boldsymbol{y}_u).
\end{equation}

In our approach, we further specialize to contiguous subsets (substrings) of bits to leverage the inherent sequential structure of the data. Thus, we specifically consider substring kernels:

\begin{equation}
K_{\text{substring}}^{(m)}(\boldsymbol{x},\boldsymbol{y}) = \sum_{i=1}^{n-m+1} e^{-\gamma d_{\boldsymbol{x}}(\boldsymbol{x}_{i:i+m-1},\boldsymbol{y}_{i:i+m-1})},
\end{equation}
where the Hamming distance between substrings is given explicitly by:
\begin{equation}
d_{\boldsymbol{x}}(\boldsymbol{x}_{i:i+m-1},\boldsymbol{y}_{i:i+m-1}) = \sum_{\ell=i}^{i+m-1}\mathbf{1}[x_\ell \neq y_\ell].
\end{equation}

The parameter $\gamma$ controls the sensitivity of the kernel to differences between sub-bitstrings, adjusting the exponential decay of kernel contributions with increasing distance. For $m=1$, the ANOVA kernel captures first-order (single-bit) differences between strings, while higher values of $m$ systematically probe deeper, more complex interactions involving contiguous subsets of bits.

By individually computing kernels for multiple orders and subsequently combining them, we achieve flexible, interpretable representations of feature interactions across multiple scales.

A common question regarding the application of the MMD loss in quantum settings refers to the choice of the kernel. Taking the ANOVA method here described, we numerically show that if the kernel does not include the correlation of interest, then it will not be possible to distinguish between them. This is seen in Fig.~\ref{fig:anova}, each of the insets within the panels shows up to which correlation is considered in the kernel, with the respective kernel depiction in function of the bitstring distance $d_{\bx}$. We see that for a given $v$, any training loss which considers $k\geq v$ is resolved. Also, the kernel itself simplifies as $v$ increases, this is because when $v=n$ we are considering the whole window and hence all correlators are already present.

This numerical result points an important question when evaluating adaptive kernels~\cite{kurkin_note_2025}, if the kernel is not able to distinguish between orders we can still be under the illusion that the training is improving. We can see this by looking at Fig.~\figref[a)]{fig:anova} compared to Fig.~\figref[f)]{fig:anova}. We achieve a training error below the complete case with all the correlator orders collapsed into one unique curve. In summary, the modified kernel is  not calculating the distance up to the whole distribution but effectively only up to the set of selected correlators.

\section{Note II: Training iterations and correlator order}
In the following we study the case of having a PQC composed of locally scrambling layers. These distributions are invariant under single-qubit Clifford rotations and can be interpreted as being random locally in each qubit. They have the following orthogonal~\cite{angrisani_classically_2024} and mixing~\cite{huang_learning_2023} properties 
\begin{equation}
\Var_{\rm scr}[\langle Z_{\boldsymbol i}\rangle \langle Z_{\boldsymbol i'} \rangle] = \E[U_{\rm scr}^{\otimes 2\dagger} (Z_{\boldsymbol i}\otimes Z_{\boldsymbol i'})U^{\otimes 2}_{\rm scr}] = 0, \qquad \Var_{\rm scr} [\langle Z_{\boldsymbol i}\rangle^2] \leq \left(\frac{2}{3}\right)^{|\boldsymbol{i}|}
\end{equation}
Under the assumption that after each iteration step the variance does not increase, which in general may not necessarily be true, we derive the following theorem.
\begin{theorem}[Higher-Order Precision Bound for Scrambling Unitaries] On average, the number of training steps $T_{\boldsymbol{i}}$ required for the predicted correlator $\left\langle Z_{\boldsymbol{i}}\right\rangle$ to be within $\Delta_f$ of the exact correlator value $\left\langle Z_{\boldsymbol{i}}\right\rangle^*$ increases at least exponentially with the difference in order of the correlators. Specifically, given the mixing condition $
\operatorname{Var}_{\boldsymbol{\theta}}\!\left(\left\langle Z_{\boldsymbol{i}}\right\rangle\right) \leq \left(\frac{2}{3}\right)^{|\boldsymbol{i}|}$,
then for any pair of correlators $\boldsymbol{i}$ and $\boldsymbol{i}'$ with $\left|\boldsymbol{i}'\right|>|\boldsymbol{i}|$,
\begin{equation}
\mathbb{E}\!\left[T_{\boldsymbol{i}'}\right] \geq \left(\frac{3}{2}\right)^{\frac{\left|\boldsymbol{i}'\right|-|\boldsymbol{i}|}{2}} \mathbb{E}\!\left[T_{\boldsymbol{i}}\right].
\end{equation}
\end{theorem}
Hence, correlators of higher order require at least exponentially more training steps in the order difference. \hfill $\square$. This result allows us to understand why higher-order correlators become more important only when close to the global minimum when having srambling unitaries. 
\textit{Proof}. On average, for random initialization, the initial distance of any correlator can be assumed to be within $\Delta_0$ of its exact value,
\begin{equation}
\Delta_0 = \left|\left\langle Z_{\boldsymbol{i}}\right\rangle^*-\left\langle Z_{\boldsymbol{i}}\right\rangle_0\right| 
= \left|\left\langle Z_{\boldsymbol{i}'}\right\rangle^*-\left\langle Z_{\boldsymbol{i}'}\right\rangle_0\right|.
\end{equation}
Let $\Delta_t = \big|\left\langle Z_{\boldsymbol{i}}\right\rangle_{t+1} - \left\langle Z_{\boldsymbol{i}}\right\rangle_t\big|$ denote the expected change at iteration $t \to t+1$. Since the standard deviation gives the average variation of a random variable, we have
\begin{equation}
\mathbb{E}\!\left[\Delta_t\right] 
= \sqrt{\operatorname{Var}_{\boldsymbol{\theta}}\!\left(\left\langle Z_{\boldsymbol{i}}\right\rangle\right)} 
\leq \left(\frac{2}{3}\right)^{|\boldsymbol{i}|/2}.
\end{equation}
The expected distance covered to reach $\left\langle Z_{\boldsymbol{i}}\right\rangle_f$ satisfies
\begin{equation}
\Delta_f - \Delta_0 
\leq \mathbb{E}\!\left[T_{\boldsymbol{i}}\right]\,\mathbb{E}\!\left[\Delta_t\right],
\end{equation}
which gives the lower bound
\begin{equation}
\mathbb{E}\!\left[T_{\boldsymbol{i}}\right] 
\geq \frac{\Delta_f - \Delta_0}{\left(\frac{2}{3}\right)^{|\boldsymbol{i}|/2}} 
= (\Delta_f - \Delta_0)\left(\frac{3}{2}\right)^{|\boldsymbol{i}|/2}.
\end{equation}
Taking the quotient for correlators $\boldsymbol{i}'$ and $\boldsymbol{i}$, both with positive denominators, preserves the inequality,
\begin{equation}
\frac{\mathbb{E}\!\left[T_{\boldsymbol{i}'}\right]}{\mathbb{E}\!\left[T_{\boldsymbol{i}}\right]}
\geq
\left(\frac{3}{2}\right)^{\frac{\left|\boldsymbol{i}'\right|-|\boldsymbol{i}|}{2}}.
\end{equation}

\section{Note III: Dependencies between correlators}\label{app:dependencies}
As discussed, the variance between different correlators depends on the assumptions made. For Haar random distributions and scrambling unitaries we have a zero variance meaning zero dependency between correlators. For matchcircuits, the variance is non-zero when the orders are complementary, i.e. there is a correlator with order $k$ and another one with order $n-k$. 

\begin{figure*}[htbp]
    \centering
    \includegraphics[width=.7\textwidth]{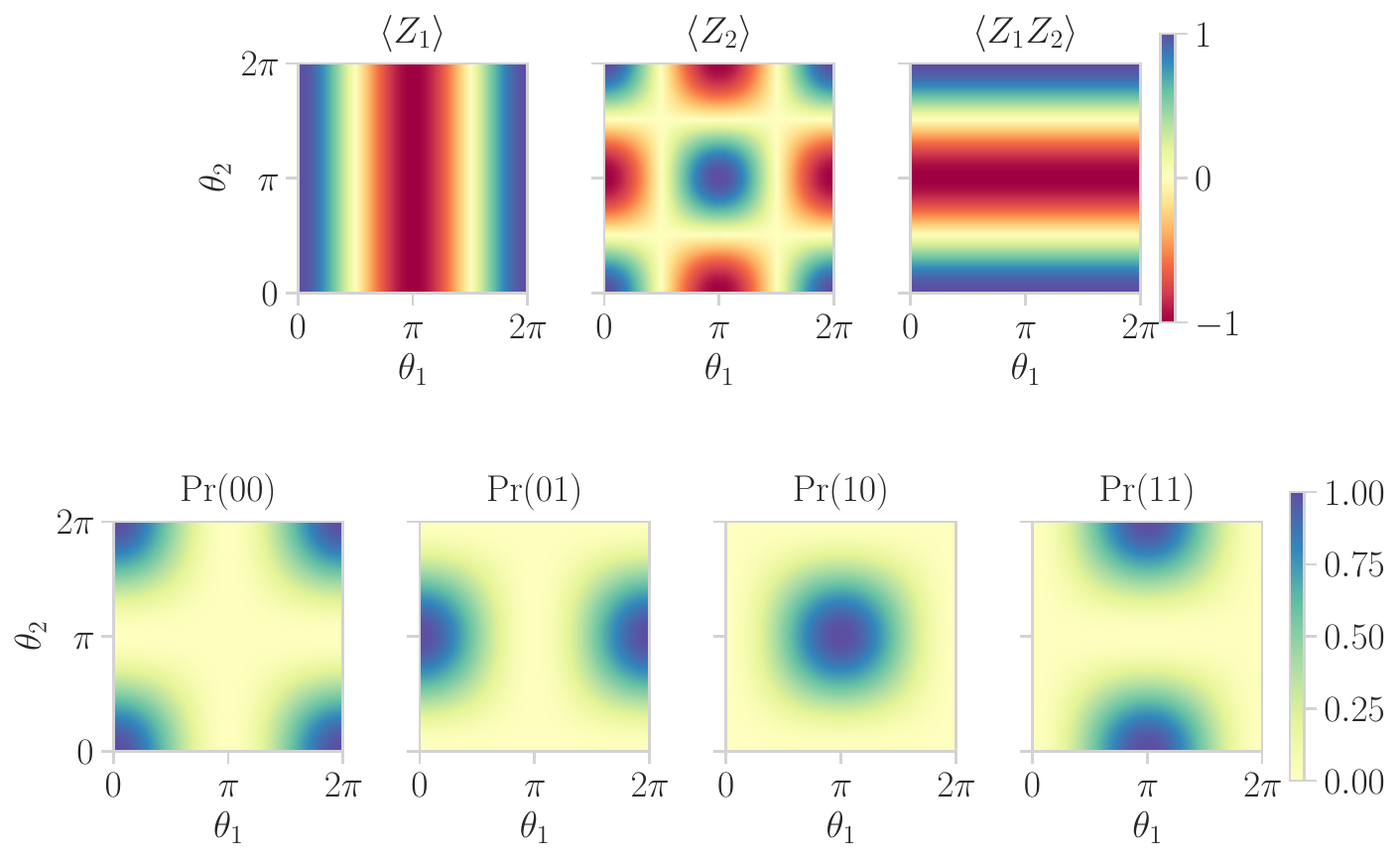}
    \caption{\justifying Simple example from Eq.~\ref{eq:simple_eg} showing how the different expectation values and probabilities depend on the parameters.}
    \label{fig:exps_probs_params}
\end{figure*}

For practical cases, there exists the framework of Fourier fingerprints~\cite{strobl_fourier_2025} developed for the supervised case. Here the dependencies between the Fourier modes are used as a metric of performance. In the correlators case we can also analyze the dependencies. For the simple example we show the analytical form of the correlators,
\begin{equation}\label{eq:simple_eg}
\begin{quantikz}
& \gate{RY(\theta_1)} & \ctrl{1} & \targ{} & \meter{}\\
& \gate{RY(\theta_2)} & \targ{}& \ctrl{-1} & \meter{}
\end{quantikz}
\quad
\begin{cases}
\hspace{.5em} \langle Z_1\rangle &= \cos(\theta_1)\\\hspace{.5em} \langle Z_2\rangle &= \cos(\theta_1)\cos(\theta_2) = \langle Z_1 \rangle \langle Z_1Z_2 \rangle\\
\hspace{.5em} \langle  Z_1Z_2\rangle &=  \qquad\quad\,\cos(\theta_2)\\
\end{cases}
\end{equation}
We see that the $\langle Z_2 \angle$ is entirely determined by the product of the $\langle Z_1\rangle$ and $\langle Z_12$.
If we do not focus in the overall landscape but rather in regions, we can see how important local variations are from Fig.~\ref{fig:exps_probs_params}.
Further, understanding how these fingerprints work for the QCBM may be of value for the performance in the deployment phase.

\section{Note IV: The Haldane algebra} 
\label{app_sec:Haldane_chain_proof}
To calculate the dimension of this algebra we use the Cartan decomposition~\cite{kokcu_fixed_2022}. We recall the Haldane Hamiltonian and the Haldane algebra,
\begin{equation}
    \mathcal{H} = -J \sum_{i=1}^{n-2} Z_i X_{i+1} Z_{i+2} 
    \;-\; h_1 \sum_{i=1}^{n} X_i \;-\; h_2 \sum_{i=1}^{n-1} X_i X_{i+1},
\end{equation}
is given by:
    \begin{equation}
    \begin{aligned}
        \g_{\text{Haldane}}\! = &\operatorname{span}\!\bigg(\bigg\{ 
        (\mathcal{YZ})_{\bj},
        (\mathcal{YZ})_{\bj}\mathcal{X}_{\boldsymbol{\noj}},
        \mathcal{X} \ \big|\ \bj = \bj_{\text{even}} \oplus \bj_{\text{odd}} : \\& \qquad \quad
         \left| \bj_{\text{even}} \right|\!,\!\left| \bj_{\text{odd}} \right|\! \in \! 2\mathbb{Z},  \, (\mathcal{YZ})_{\bj}\!\!=\!\!\{Y,Z\}_{\bj},\\ & \mathcal{X} \!\in\!\{X, I\}^{\otimes n},\mathcal{X}_{\boldsymbol{\noj}}\! \in\! \{X, I\}_{\boldsymbol{\noj}}
        \bigg\} \!\! \setminus \!\!
        \bigl\{\!X_o,
        X_e,
        X^{\otimes n}\!\bigr\}\bigg) .
    \end{aligned}
    \end{equation}
where we define $X_o := (\1 X)^{\lfloor n/2 \rfloor}\1^{(n \bmod 2)}$ and $X_e := (X\1)^{\lfloor n/2 \rfloor}X^{(n \bmod 2)}$. This expression is found by taking the Lie bracket, isolating each of the individual terms and then verifying it computationally.

A Cartan decomposition is the splitting of a Lie algebra (or Lie group) into a direct sum 
$\mathfrak{g}=\mathfrak{l}\oplus\mathfrak{m}$, where one part forms a subalgebra and the other transforms under it. This division arises from an involution that separates the elements into symmetric and antisymmetric components. The involutions of the Haldane algebra are $X_o,X_e,X^{\otimes n}$.

To determine the dimension $|\g_{\text{Haldane}}|$ we part from the $\mathfrak{su}(2^n)$ and decompose it with the involution  $X_o$, $\mathfrak{l} = span(g) : g = {\left[ g_i,X_o \right] = 0, \forall g_i \in \mathfrak{su}(2^n)}$. This decomposition is isomorphic to $\mathfrak{k} \cong \mathfrak{s} \mathfrak{u}\left(2^{n-1}\right) \oplus \mathfrak{s} \mathfrak{u}\left(2^{n-1}\right) \oplus \mathfrak{u}(1)$. We further decompose this subalgebra as $\mathfrak{l} = \mathfrak{l}^\prime \oplus \mathfrak{m'}$ with the involution $X_e$ such that
\begin{equation}
        \mathfrak{l}^\prime = \! = \operatorname{span}\!\bigg(\bigg\{ 
        (\mathcal{YZ})_{\bj},
        (\mathcal{YZ})_{\bj}\mathcal{X}_{\boldsymbol{\noj}},
        \mathcal{X} \ \big|\ \bj = \bj_{\text{even}} \oplus \bj_{\text{odd}} : 
         \left| \bj_{\text{even}} \right|\!,\!\left| \bj_{\text{odd}} \right|\! \in \! 2\mathbb{Z},  \, (\mathcal{YZ})_{\bj}\!\!=\!\!\{Y,Z\}_{\bj}, \mathcal{X} \!\in\!\{X, I\}^{\otimes n},\mathcal{X}_{\boldsymbol{\noj}}\! \in\! \{X, I\}_{\boldsymbol{\noj}}
        \bigg\} \bigg).
\end{equation}

This decomposition leaves the $\mathfrak{u}(1)$ component of $\mathfrak{k}$ unchanged. It acts only on each $\mathfrak{s}\mathfrak{u}\!\left(2^{n-1}\right)$ block independently, resulting in  
\begin{equation}
\mathfrak{k}' \cong 
\left(\mathfrak{s}\mathfrak{u}\!\left(2^{n-2}\right)\oplus \mathfrak{s}\mathfrak{u}\!\left(2^{n-2}\right)\oplus \mathfrak{u}(1)\right)
\oplus
\left(\mathfrak{s}\mathfrak{u}\!\left(2^{n-2}\right)\oplus \mathfrak{s}\mathfrak{u}\!\left(2^{n-2}\right)\oplus \mathfrak{u}(1)\right)
\oplus \mathfrak{u}(1).
\end{equation}

Hence, the dimension of $\mathfrak{k}'$ is  
\begin{equation}
|\mathfrak{k}'| = 4|\mathfrak{s}\mathfrak{u}\!\left(2^{n-2}\right)| + 3 = 4^{n-1} - 1.
\end{equation}

The only elements distinguishing the basis of $\mathfrak{k}'$ from that of the Haldane algebra $\mathfrak{g}$ are $X_o$, $X_e$, and $X^{\otimes n}$. Consequently, the dimension of the Haldane Hamiltonian algebra is  
\begin{equation}
|\mathfrak{g}\,(\text{Haldane})| = |\mathfrak{k}'| - 3 = 4^{n-1} - 4.
\end{equation}

This calculation is exactly analogous to that of the Heisenberg algebra~\cite{kokcu_fixed_2022}, where further details can be found, and both have the same exact dimension.

\end{document}